\documentclass[12pt]{article}
\pdfoutput=1
\usepackage{amsmath,amssymb,epsfig,amsfonts}
\usepackage{graphicx,subfigure}
\usepackage[usenames, dvipsnames]{color}
\usepackage[backref]{hyperref}
\usepackage[nosort]{cite}
\usepackage{lscape}
\usepackage{rotating}

\usepackage{extarrows}

\usepackage{verbatim} 

\makeatletter

\newtheorem{theorem}{Theorem}[section]
\newtheorem{fact}{Fact}[section]
\newtheorem{definition}{Definition}[section]

\newcommand{\be}{\begin{equation}}
\newcommand{\ee}{\end{equation}}
\newcommand{\ba}{\begin{aligned}}
\newcommand{\ea}{\end{aligned}}

\newcommand{\C}{\mathbb{C}}
\renewcommand{\P}{\mathbb{P}}

\newcommand{\nn}{\nonumber}
\newcommand{\bea}{\begin{eqnarray}}
\newcommand{\eea}{\end{eqnarray}}

\newcommand{\R}{{\mathbb R}}

\def\unit{{1\kern-.65ex {\rm l}}}
\def\1{{1\kern-.65ex {\rm l}}}

\newcount\hour \newcount\minute
\hour=\time \divide \hour by 60
\minute=\time
\def\now{%
\ifnum \hour<13
  \ifnum \hour=0 \advance \hour by 12 \number\hour:\else \number\hour:\fi%
     \ifnum \minute<10 0\fi%
     \number\minute%
\ A.M.%
\else \advance \hour by -12 \number\hour:%
  \ifnum \minute<10 0\fi%
  \number\minute%
  \ P.M.%
\fi%
}

\makeatother

\begin{document}

\baselineskip=18pt  
\numberwithin{equation}{section}  
\allowdisplaybreaks  


%
%


\thispagestyle{empty}

\vspace*{-2cm} 
\begin{flushright}
{\tt KCL-MTH-15-08} \qquad \qquad 
\end{flushright}

\vspace*{0.8cm} 
\begin{center}
{\Huge Box Graphs and Resolutions II:}\\
\bigskip

{\Large From Coulomb Phases to Fiber Faces}

 \vspace*{1.5cm}
Andreas P. Braun$\,^1$ and Sakura Sch\"afer-Nameki$\,^2$\\

 \vspace*{1.0cm} 
$^1$ {\it 
Rudolf Peierls Centre for Theoretical Physics, University of Oxford \\
Oxford, OX1 3NP, UK}\\
$^1$ {\it Mathematical Institute, University of Oxford, \\
Oxford, OX2 6GG, UK}\\

$^2$ {\it Department of Mathematics, King's College London \\
 The Strand, London WC2R 2LS, UK}\\
 {\tt {gmail:$\,$ andreas.braun.physics, sakura.schafer.nameki}}\\

\vspace*{0.8cm}
\end{center}
\vspace*{.1cm}

\noindent 
Box graphs, or equivalently Coulomb phases of three-dimensional $N=2$ supersymmetric gauge theories with matter, give a succinct, comprehensive and elegant  characterization of crepant resolutions of singular elliptically fibered varieties. 
Furthermore, the box graphs predict that the phases are  organized in terms of a network of flop transitions. 
The geometric construction of the resolutions associated to the phases is, however,  a difficult problem. Here, we identify a correspondence between box graphs for the gauge algebras $\mathfrak{su}(2k+1)$ with resolutions obtained using toric tops and generalizations thereof.  Moreover, flop transitions between different such resolutions agree with those predicted by the box graphs. Our results thereby provide explicit realizations of the box graph resolutions. 
 
\newpage
\setcounter{tocdepth}{2}
\tableofcontents



\section{Introduction}

Beyond its applications in the modeling of particle physics and classification of 6d superconformal field theories,  recent developments in F-theory
have led to tremendous progress in uncovering properties of higher-dimensional elliptically fibered complex varieties. 
Much of the progress has been made in particular in the study of crepant resolutions of singular elliptic fibers in higher dimensional varieties, i.e. resolutions that keep the canonical class unchanged. 
 
The canonical setup of interest in F-theory compactifications \cite{Vafa:1996xn}  is an elliptically fibered Calabi-Yau variety in dimension $3$ and $4$, which models $N=(1,0)$ six dimensional  or four-dimensional $N=1$ supersymmetric field theories, with gauge algebra $\mathfrak{g}$, and matter in the representations ${\bf R}_i$ of $\mathfrak{g}$. Four-folds in addition allow for codimension three singularities, where Yukawa couplings are realized. 
The F-theory limit is  obtained by taking the volume of the fiber to zero, and this singular limit is in fact not sensitive to which crepant resolution is used \cite{Morrison:1996na, Morrison:1996pp}. However, various more refined aspects of the F-theory compactification, such determining the $G_4$-flux, the possible $U(1)$ and discrete symmetries, make use of the singularity resolutions. 

By Kodaira's classification of singular fibers, one can associate a Lie algebra $\mathfrak{g}$ to an elliptic fibration. These are 
characterized in terms of an ADE type affine Dynkin diagram representing the dual graph to the intersection graph of the rational curves in the singular fiber. This classification holds for all singular fibers over codimension one loci in the base. In higher codimension, this classification ceases to be comprehensive, and additional structures emerge that are required in order to characterize how higher codimension singular fibers can occur, and what their characterization is. 

In \cite{Hayashi:2014kca} (see also \cite{Hayashi:2013lra, Braun:2014kla, Lawrie:2015hia, Esole:2014bka, Esole:2014hya}), inspired by the correspondence to classical Coulomb phases in 3d and 5d supersymmetric gauge theories \cite{Intriligator:1997pq, Aharony:1997bx, deBoer:1997kr, Diaconescu:1998ua, Grimm:2011fx, Hayashi:2013lra}, a proposal was put forward to systematically describe the distinct small resolutions of singular elliptic fibrations, including fibers in codimension two and three. In addition to a Lie algebra $\mathfrak{g}$, which characterizes the codimension one fibers, the codimension two fibers have a representation ${\bf R}$ of $\mathfrak{g}$ associated to them, and by 
\cite{Hayashi:2014kca}, the fibers can be obtained by a decorated representation graph, or box graph. Flops between distinct small resolutions are realized by the action of a quotiented Weyl group. Note that the box graphs are motivated from a dual M-theory compactification point of view and map the problem of small resolutions to Coulomb phases. However, as shown in \cite{Hayashi:2014kca}, the analysis applies directly in the cone of effective curves of the elliptic fibration, and does not require any reference to the gauge theory. Recently, this work was utilized in \cite{Lawrie:2015hia} to determine a classification of the fibers in codimension two with additional $U(1)$ symmetries, which geometrically are realized in terms of rational sections. This has led to a survey of all F-theory Grand Unified Theories (GUTs) with additional $U(1)$ symmetries, with interesting phenomenological implications \cite{Krippendorf:2015kta}. Thus the results on codimension two fibers are not merely of mathematical relevance, but indeed have far-reaching implications for the particle physics, in particular flavor properties,  of F-theory compactifications. 

Beyond this abstract characterization of elliptic fibrations, much progress has been made in the direct realization of elliptic curves 
in terms of hypersurfaces or complete intersections, for instance in toric varieties \cite{EY, MS,  Braun:2011zm, Krause:2011xj, Morrison:2011mb, Kuntzler:2012bu, Lawrie:2012gg, Braun:2014qka}. What is apparent from all these resolutions is that neither toric, nor algebraic resolutions necessarily yield the full set of possible fibers predicted in \cite{Hayashi:2014kca}. 
Concrete realizations of the complete set of distinct resolutions have indeed been determined for $\mathfrak{su}(5)$ in \cite{Hayashi:2013lra, Braun:2014kla,  Esole:2014hya}, with both fundamental and anti-symmetric matter, in terms of resolutions of the Tate model for a codimension one $I_5$ Kodaira fiber \cite{Bershadsky:1996nh, Katz:2011qp}. 

The purpose of the present work is to clarify the connection between toric and algebraic resolutions on the one hand, and the more general resolutions that are predicted by the box graphs, on the other. We will determine a characterization of all algebraic resolutions in terms of a subclass of box graphs, which have a simple combinatorial description. Furthermore, resolutions associated to triangulations of toric tops \cite{Candelas:1996su} are determined in terms of triangulations of a so-called {\it fiber face}. 
We then show how fiber face triangulations form a subset of the box graph resolutions and determine a one-to-one map between these for $\mathfrak{su}(n)$ gauge algebras with anti-symmetric representation. The fate of fiber components when approaching these matter loci can be easily read off the fiber face triangulation. 
The resulting correspondence also provides an identification  between the flops of the fiber face triangulations and single box sign changes in the box graphs.

Beyond the class of fiber face triangulations originating from toric tops, which are the subject of section \ref{sec:BoxToric}, we determine a class of resolutions realizing the fiber as a complete intersection. Likewise, these have a succinct characterization in terms of triangulations of what we call a {\it secondary fiber face}. This again has a simple characterization in terms of box graphs as shown in section \ref{sec:CompleteInt}. This structure is then extended to a third layer, and we conjecture that it persists for all remaining phases in section \ref{sec:Layersconj}.

The correspondence between box graphs and fiber face triangulations is exemplified in the context of $\mathfrak{su}(7)$ with anti-symmetric representation ${\bf R} = {\bf 21}$, where each of these box graph layers is discussed in detail and the corresponding resolutions (which in this case is the complete set) are determined in appendix \ref{app:SU7}. 
We conclude with a discussion of extensions and applications of our results in section \ref{sec:Discussion}.


\section{Box Graphs, Coulomb Phases and Fibers}
\label{sec:BoxGraphs}

Consider a singular elliptic fibration, with trivial canonical class, and a base of dimension at least two. Let $\mathfrak{g}$ be the Lie algebra associated to the singular fibers, i.e. the intersection graph of the  exceptional curves of the singular Kodaira fibers are given in terms of the affine Dynkin diagram of $\mathfrak{g}$.
The fibers in codimension two, associated to a representation ${\bf R}$ of $\mathfrak{g}$, can be characterized in terms of box graphs, introduced in
 \cite{Hayashi:2014kca}, which are a  combinatorial, graphic presentation of the codimension two fibers, which are based on the representation graph ${\bf R}$.

This section is a review of the results obtained in \cite{Hayashi:2014kca}, and developed further in \cite{Lawrie:2015hia}, with a focus on the anti-symmetric representation $\Lambda^2 V$ for $\mathfrak{su}(2k+1)$. The codimension one fibers for this setup are of Kodaira type $I_{2k+1}$, corresponding to an $\mathfrak{su}(2k+1)$ gauge algebra. In codimension two, the rational curves in the fiber intersect according to Kodaira type $I_{2k-3}^*$,  which realize matter in the anti-symmetric representation $\Lambda^2 ({\bf 2k+1})$. However, in this case there are inequivalent topological realizations. These are obtained by resolutions of Weierstrass or Tate models and, depending on which resolution is carried out, different components of the $I_{2k+1}$ fiber become reducible in codimension two. The box graphs provide an elegant characterization of all resolutions, but do not provide a constructive way to realize these geometrically. One of the goals of this paper is to determine the corresponding resolutions.

\subsection{Coulomb Phases for $\mathfrak{su}(2k+1)$ with $\Lambda^2V$ Matter}
\label{sec:BoxSULambda}

\begin{figure}
    \centering
    \includegraphics[width=7cm]{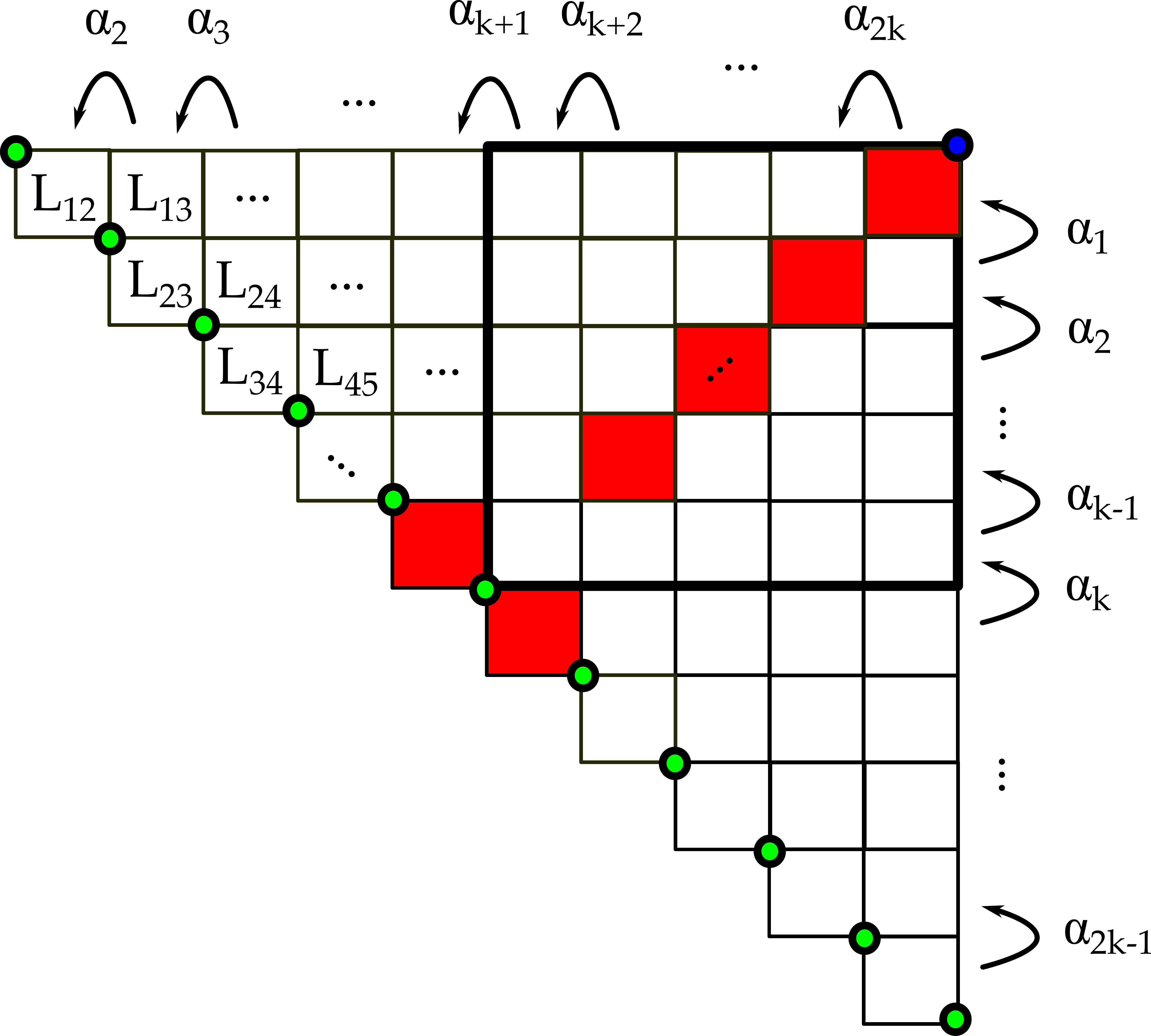}
\,    \includegraphics[width=7cm]{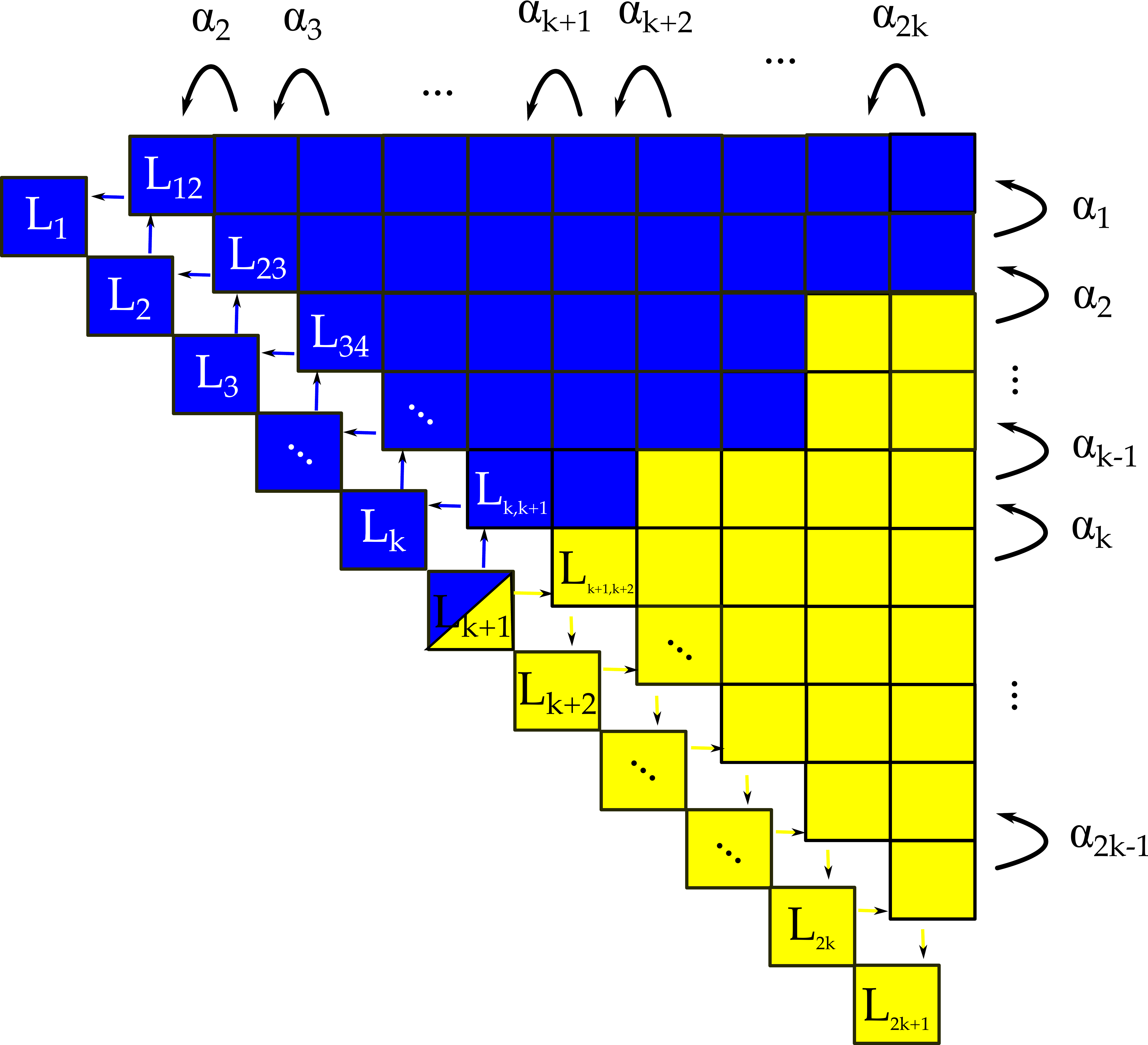}
    \caption{The left hand side shows the representation graph for the anti-symmetric representation of $\mathfrak{su}(2k+1)$ with weights $L_{ij} =L_i + L_j$ with $i<j$. The red boxes correspond to the diagonal $\mathcal{E}_{2k+1}$, defined in (\ref{diagdef}).
The right hand side shows a Box Graph for matter in the combined anti-symmetric  and fundamental representation of $\mathfrak{su}(2k+1)$, with $\pm$ shown in blue/yellow. The NW-SE additional diagonal corresponds to the box graph of the fundamental representation with weights $L_1$ to $L_{2k+1}$. The blue/yellow arrows indicate the flow rules between fundamental and anti-symmetric representation. For this box graph corresponding to the anti-symmetric representation, there are two box graphs consistent with the flow rules for the fundamental representation. These are distinguished by choosing $\epsilon(L_{k+1})= \pm$. 
    \label{fig:RepGraphA}}
\end{figure}

Let us begin with the discussion of (classical) Coulomb phases for $\mathfrak{su}(2k+1)$ with matter in the anti-symmetric representation and their succict characterization in terms of Box graphs. 
To begin with, let $\mathfrak{g} = \mathfrak{su}(2k+1)$ and let $L_i$, $i=1, \cdots, 2k+1$ be its fundamental weights.  With the constraint that $\sum_i L_i =0$, the simple roots can be represented as 
\be
\alpha_i = L_i- L_{i+1} \,,\qquad i= 1, \cdots , 2k\,.
\ee
The weights of the antisymmetric representation of dimension $(2k+1)k$ are 
\be
L_{ij} = L_i + L_j \,,\qquad  i<j \,.
\ee
The representation graph for a representation ${\bf R}$ is defined in terms of boxes, which correspond to the weights of ${\bf R}$. These are arranged in such a way that adjoining walls represent the action of simple roots within the representation. The representation graph for $\Lambda^2 ({\bf 2k+1}) = \Lambda^2V$ is shown in figure \ref{fig:RepGraphA}.  

The singular fibers in codimension two can be equally characterized in terms of the Coulomb branch phases of an $N=2$ supersymmetric gauge theory in $d=3$ (or $d=5$ depending on whether the elliptic fibration is a four-fold or three-fold) with chiral matter in the representation ${\bf R}$. Geometrically, this means that the singular fiber degenerates further in codimension two, and the singularity can be characterized in terms a higher rank Lie algebra $\widetilde{\mathfrak{g}}$. Higgsing the adjoint of this algebra ${\mathfrak{g}}$ gives rise to bifundamental matter\footnote{The case of a non-abelian commutant of $\mathfrak{g}$ in $\widetilde{\mathfrak{g}}$ was discussed also in \cite{Hayashi:2014kca}, and has very interesting properties. Here we are only interested in the case of an abelian commutant. }
\be\ba
\widetilde{\mathfrak{g}} \quad &\rightarrow\quad \mathfrak{g} \oplus \mathfrak{u}(1)\cr
\hbox{Adj} (\widetilde{\mathfrak{g}})  & \mapsto \quad (\hbox{Adj} (\mathfrak{g}) , {\bf 1}) \oplus ({\bf 1}, \hbox{Adj} (\mathfrak{u}(1))  ) \oplus ({\bf R},  +1) \oplus ( \overline{\bf R}, -1) \,.
\ea\ee

The key insight of \cite{Hayashi:2014kca} is that Coulomb phases, and thereby  singular fibers in codimension two, are characterized in terms of  box graphs $\mathcal{B}_{\epsilon}^{\bf R}$, i.e. a  sign-decorated representation graphs of ${\bf R}$, where the signs are given by a map 
\be
\epsilon: \hbox{Weights}({\bf R}) \quad \rightarrow \quad \{\pm\} \,, \qquad \epsilon(L_{ij}) = \pm \,,
\ee
satisfying a set of  conditions, which e.g. for $\mathfrak{su}(2k+1)$ with ${\bf R} = \Lambda^2 ({\bf 2k+1})$ are
\begin{itemize}
\item[(1.)] Flow rules for the anti-symmetric representation: \\
If $\epsilon(L_{ij}) = +$ then $\epsilon(L_{kl}) = +$ for all $(k, l)$ with $k\leq i$ and $l \leq j$.\\
If $\epsilon(L_{ij}) = -$ then $\epsilon(L_{kl}) = -$ for all $(k, l)$ with $k\geq i$ and $l \geq j$.
\item[(2.)] Trace condition for the anti-symmetric representation:\\
Let $\mathcal{E}_{2k+1} =\{ \epsilon (L_{1, 2k+1}),  \epsilon (L_{2, 2k }), \cdots,  \epsilon (L_{k-1, k+3}), \epsilon (L_{k, k+1}), \epsilon (L_{k+1, k+2})\}$. Then 
\be\label{diagdef}
\mathcal{E}_{2k+1} \not=( +,\cdots, +) \qquad\hbox{and}\qquad \mathcal{E}_{2k+1} \not=( -,\cdots, -)\,.
\ee
\end{itemize}
The flow rules ensure that if two weights are related by the action of a positive root, then their sign assignment needs to be the same. The trace condition says that the weights on the `diagonal' defined in terms of $\mathcal{E}_{2k+1}$ cannot all have the same sign. This ensures that we obtain an $\mathfrak{su}(2k+1)$ phase, rather than a $\mathfrak{u}(2k+1)$ one. The diagonal is shown in figure \ref{fig:RepGraphA} in terms of the red boxes. 

The sign assignment is uniquely characterized in terms of  the path separating the $+$ and $-$ signed boxes, starting at the upper right hand corner (blue point in figure \ref{fig:RepGraphA}), and ending on one of the points on the NW-SE diagonal (one of the green points in figure \ref{fig:RepGraphA}). These are so-called anti-Dyck path associated to the box graph. 
As an example, in figure \ref{fig:SU7FlopGraph} all the phases of $\mathfrak{su}(7)$ with the anti-symmetric representation ${\bf 21}$, including the anti-Dyck paths, are shown. 

Flop transitions between two phases are defined as single-box sign changes which map between two consistent phases, both satisfying (1.) and (2.). Geometrically, these correspond exactly to flop transitions in the codimension two fibers. One of the goals of this paper is to realize these concretely in a geometric setting, such as a toric realization of the singular fibers. The flop network for $\mathfrak{su}(7)$ is shown in figure \ref{fig:SU7FlopGraph}.


\subsection{Coulomb Phases for $\mathfrak{su}(2k+1)$ with $\Lambda^2V \oplus V$ Matter}

Although the main concern of this paper is the anti-symmetric representation, we will make several references to the Coulomb phases and box graphs for the fundamental representation as well. The weights for the ${\bf 2k+1}$ fundamental representation are $L_i$, $i=1, \cdots, 2k+1$, with $\sum L_i=0$. The phases can again be mapped to representation graphs with a sign decoration ${\epsilon}: {\bf R} \rightarrow \pm 1$, satisfying a set of flow rules and trace condition:
\begin{itemize}
\item[(1.)] Flow rules for the fundamental representation:\\
If $\epsilon (L_i) = + $ then $\epsilon (L_{j}) =+$ for all $j <i$.  \\
If $\epsilon (L_i) = -$ then $\epsilon (L_{j}) =-$ for all $j > i$.  
\item[(2.)] Trace condition for the fundamental representation: 
The signs cannot be all $+$ or all $-$. 
\end{itemize}
Furthermore, the phases for the combined anti-symmetric and fundamental representations are obtained by combining the phase of the  fundamental and anti-symmetric \cite{Hayashi:2014kca, Braun:2014kla} such that 
\begin{itemize}
\item[(AF0.)] The phases for each representation separately  are consistent $\mathfrak{su}(2k+1)$ phases.
\item[(AF1.)] Flow rules for combined anti-symmetric and fundamental representation: \\
\be\ba
\epsilon(L_i)=+ \quad  &\Rightarrow\quad  \epsilon(L_{i-1, i})= + \cr
\epsilon(L_{i, i+1})=+ \quad  &\Rightarrow\quad   \epsilon(L_{i})= +\cr
\epsilon(L_i)=-  \quad  &\Rightarrow\quad   \epsilon(L_{i, i+1})= - \cr
\epsilon(L_{i, i+1})=+ \quad  &\Rightarrow\quad   \epsilon(L_{i+1})= -\,.
\ea\ee
\end{itemize}
One can determine the corresponding box graphs by attaching the fundamental representation along the NW to SE diagonal to the anti-symmetric box graph. The resulting graph then needs to satisfying the flow rules, viewed as a  box graph for $\mathfrak{su}(2k+2)$ with the anti-symmetric representation. This is shown in figure \ref{fig:RepGraphA}.


\subsection{Fibers from Coulomb Phases/Box Graphs}

The box graphs give a succinct characterization of all the small resolutions of singular Weierstrass models. 
First we introduce the notion of a relative cone of effective curves (see e.g. \cite{Debarre}). Let $X$ be a projective variety. Then the group of Cartier divisors is
\be
N^1 (X) = \{D \hbox{ Cartier divisor in $X$}\}/\sim  \,,
\ee
where $\sim$ corresponds to numerical equivalence, i.e. 
\be
 D \sim D' \ \hbox{ if } \ D \cdot C = D' \cdot C \ \hbox { for all }\  C\in H_2(X, \mathbb{Z}) \,.
\ee
Two curves are numerically equivalent $C\sim C'$, if their intersections with any element in $N^1(X)$ agrees, and we correspondingly define  
$N_1 (X)$ as the quotient of all  (complex) 1-cycles by numerical equivalence. 
Any 1-cycle in $X$ can be written as a formal integral sum $\sum_i n_i C_i$, with  $n_i\in \mathbb{Z}$, where $C_i$ are integral curves in $X$ (i.e. actual subspaces of complex dimension $1$ in $X$). A curve is called {\it effective} if all coefficients $n_i$ are non-negative.

In $N_1(X)$ the effective curves form a convex cone, denoted by $NE(X)$. 
\begin{definition}
Let $X$ and $Y$ be two projective varieties and $\pi: X \rightarrow Y$ a morphism. Then the {relative cone of curves} $NE(\pi)$ of the morphism $\pi$ is the convex subcone of the cone of effective curves $NE(X)$,  generated by the curves that are contracted by $\pi$.
\end{definition}

Let $X$ be a smooth elliptically fibered Calabi-Yau variety of dimension $n$, with a section, and let 
\be
\pi: \quad X \ \rightarrow \ Y
\ee
be the contraction of all rational curves in the fiber which do not meet the zero section, so that $Y$ is the singular Weierstrass model associated to $X$. 
This definition of a singular limit \cite{Morrison:1996na, Morrison:1996pp} is the relevant one for F-theory. We can associate to a singular Weierstrass model with Kodaira fibers in codimension one in the base a Lie algebra $\mathfrak{g}\, $\footnote{We focus our attention here to the $I_n$ and $I_n^*$ as well as $II^*, III^*, IV^*$, with associated gauge algebras $\mathfrak{su}(n)$, $\mathfrak{so}(2n)$, and $\mathfrak{e}_n$.}. In codimension two, the singularity can enhance, which associates a representation ${\bf R}$ to the fibers.  
In \cite{Hayashi:2014kca, Lawrie:2015hia} it was shown that $NE(\pi)$ for this map $\pi$ can be constructed using the box graphs, and 
for a given singular Weierstrass model $Y$, all the smooth models $X_i$, with singular limits $\pi_i$, which are related by flop transitions, were determined: 
\begin{fact}[Box Graphs and Resolutions]
Let $Y$ be a singular Weierstrass model of dimension at least three, with codimension one singularity associated to a Lie algebra $\mathfrak{g}$ and codimension two singularities associated to a representation ${\bf R}$ of $\mathfrak{g}$.  
There is a one-to-one correspondence between box graphs $\mathcal{B}_{\epsilon^{(i)}}^{\bf R}$  -- associated to a representation ${\bf R}$ and a sign assignment $\epsilon^{(i)}: \hbox{weights} ({\bf R})\rightarrow \pm 1$ -- 
and a pair $(X_i, \pi_i)$ of  smooth elliptic Calabi-Yau varieties, with maps $\pi_i: X_i \rightarrow Y$. In particular, the cone $NE(\pi_i)$ can be characterized in terms of the box graphs as follows 
\be
NE (\pi_i) =\left\langle \{F_i,  i= 0,\cdots, \hbox{rank}(\mathfrak{g})\} \cup \{C_w\,, w \in {\bf R}: \ \exists j:\  \mathcal{B}_{\epsilon^{(j)}}^{\bf R} =\mathcal{B}_{\epsilon^{(i)}}^{\bf R}|_{\epsilon^{(j)}(w) =- \epsilon^{(i)}(w) }  \}\right\rangle_{\mathbb{Z}}\,.
\ee
\end{fact}
Here, $F_i$ are the rational curves associated to the simple roots of $\mathfrak{g}$, and $C_w$ are the rational curves associated to weights $w$ of the representation ${\bf R}$. 
The extremal generators of these cones, and flop transitions between two smooth models $(X_i, \pi_i)$, can be determined as follows, see Facts 2.2 and 2.3 in \cite{Lawrie:2015hia}: 
\begin{fact}[Flops and Extremal Rays]
Single box sign changes that map between box graphs $\mathcal{B}_{\epsilon^{(i)}}^{\bf R}$ correspond to flop transitions between the geometries $X_i$.
The convex cones $NE(\pi_i)$ can be written in terms of extremal rays 
\be
NE (\pi_i) = \bigoplus_k \mathbb{Z}^+ \ell_k \,,
\ee
where $\ell_k$ are the generators of the extremal rays, given by:
\begin{itemize}
\item[(1.)] $F_i$, associated to the simple roots of $\mathfrak{g}$ are extremal generators, is extremal if  the anti-Dyck path of $\mathcal{B}_{\epsilon^{(i)}}^{\bf R}$ 
 does not cross the horizontal or vertical lines in the box graph, which correspond to adding the simple root $\alpha_i$,
\item[(2.)] $\epsilon^{(i)}(w) C_w$ is extremal if  there exists  $j$ such that 
$\mathcal{B}_{\epsilon^{(j)}}^{\bf R} =\mathcal{B}_{\epsilon^{(i)}}^{\bf R}|_{\epsilon^{(j)}(w) =-\epsilon^{(i)}(w) }$.
\end{itemize}
\end{fact}
The condition $(1.)$ essentially states that $F_i$ is extremal if it stays irreducible in codimension two. The second condition states that a 
rational curve associated to a weight $w$ is extremal if it can be flopped, i.e. changing its sign gives rise to another consistent phase.
The extremal generators of $NE (\pi_i)$ correspond to the fiber components of the codimension two fiber, and we will explain the  construction of this when discussing the toric fibers.  In section \ref{sec:ToricTBG} we will provide more details on the precise identification of Coulomb Phases/Box Graphs, with fiber components. 

The characterization of crepant resolutions of elliptic Calabi-Yau varieties in terms of box graphs is very elegant and concise, however it does not give a constructive way of determining the resolutions $X_i$ of the singular Weierstrass models $Y$. The main purpose of this paper is to
show how such resolutions can be geometrically realized. We continue now with a  brief summary of various toric tools, which will be useful in this process.


\section{Toric Resolutions, Tops and Weighted Blowups}
\label{sect:gentoricstuff}

To keep this paper reasonably self-contained, we collect some background on the toric resolution techniques to be used below
and set up notations and conventions. A more in-depth treatment tailored to our needs can be 
found in \cite{Braun:2014kla}, see also 
\cite{danilov, Fultontoric, cox2011toric, Kreuzer:2006ax, Bouchard:2007ik} for basic definitions and
properties concerning toric varieties and their Calabi-Yau submanifolds.

Given a (appropriate) fan\footnote{As usual, we assume that one starts with dual lattices $N$ and $M$. The fan $\Sigma$ sits inside $N \otimes \R$ and is
rational (with respect to $N$), polyhedral, strongly convex and simplicial. If there is a strongly convex piecewise linear support function on $\Sigma$, the
corresponding toric variety is projective. See e.g. \cite{Fultontoric} for explanations of these terms.
As is customary in the literature, we denote the dual lattice to $N$ by $M$ and the product between elements of the two lattices
by $\langle M , N \rangle$.} 
$\Sigma$, there is an associated (smooth, projective) toric variety $T_\Sigma$. 
A special role is played by the generators of the rays $\rho_i$ (one-dimensional cones in $\Sigma$), which we denote by $v_i$. For every $v_i$ there
is an associated homogeneous coordinate $z_i$ and a toric divisor $D_i$. The fan $\Sigma$ encodes the linear relations
between the divisors $D_i$ as well as their intersections.

We may describe $T_\Sigma$ as the quotient
\begin{equation}\label{toricasquotient}
 T_\Sigma = \left( \C^{n+k} \setminus Z \right) / \left( (\C^*)^k \times G \right) \, .
\end{equation}
The Stanley-Reisner (SR) ideal $Z$ contains all collections of homogeneous coordinates $\{z_i\}$ for which the corresponding
rays $\{ \rho_i \}$ do not share a common cone in $\Sigma$. The weights $s_i$ of the $\C^*$ actions which are modded out can be found from
relations of the form
\begin{equation}
\sum  s_i v_i =0  \, .
\end{equation}
Finally, the finite group $G$ is isomorphic to the quotient $N / N_v$, where $N_v$ is the lattice spanned by all $v_i$ in $\Sigma$.

\subsection{Weighted Blowups} 

Refinements $\Sigma' \rightarrow \Sigma $ of the fan induce birational maps $T_{\Sigma'}\rightarrow T_\Sigma$, i.e. we may think of them as
(generalized) blowups. In particular, refinements in which we introduce a single new ray $v_E$ into $\Sigma$ correspond to weighted blowups
according to the following rules. Let us assume that $v_E$ sits in the interior of a $d$-dimensional cone $\sigma$, generated by
$\{ v_1 , \cdots, v_d \}$. The introduction of $v_E$ means we have to subdivide $\sigma$ into the cones
\begin{equation}
\langle v_1, \cdots, v_d \rangle \, \rightarrow  \langle v_1, \cdots v_E \rangle , \cdots , \langle v_E, \cdots v_n \rangle \,.
\end{equation}
For $d < n$, we also have to accordingly subdivide all higher-dimensional cones containing $\sigma$ as a face.
On the level of the description \eqref{toricasquotient}, the upshot of such a refinement is that the SR-ideal of $Z$ now contains the relation $z_1=\cdots = z_d =0$. Furthermore, $v_E$ being contained in the interior of $\sigma$ means that
we may write
\begin{equation}\label{eq:scalingE}
 \sum_i a_i v_i = a_E v_E \, ,
\end{equation}
so that there is a new $\C^*$ action with the corresponding weights in $T_{\Sigma'}$. If all of the weights $a_i=1$ and $a_E=1$,
this fan refinement is equivalent to a standard algebraic blowup $(z_1,\cdots,z_d;z_E)$, {where the notation means that the locus $z_1= \cdots = z_d=0$ gets resolved with new exceptional section $z_E$} (see section \ref{sec:SUOdd} for more details). In general, we can think of such a refinement as
a weighted blowup with weights $a_i$ and $a_E$.

\subsection{Toric Calabi-Yau Hypersurfaces}

The anti-canonical class of $T_\Sigma$ can be expressed as
\begin{equation}
-K = \sum_i D_i \, .
\end{equation}
A Calabi-Yau hypersurface is hence described by taking the zero locus of a section $P(z_i)$ of the corresponding line bundle. 
Calabi-Yau hypersurfaces in compact toric varieties can be described by means of pairs of reflexive polytopes, see
\cite{Kreuzer:2006ax} for a lightning review.
Here, all rays of $\Sigma$ are generated by vectors $v_i$ on the surface of an $N$-lattice polytope $\Delta^\circ$, which is called
reflexive if its polar dual $\Delta$, defined by
\begin{equation}
 \langle \Delta , \Delta^\circ \rangle \geq -1 \, ,
\end{equation}
is a lattice polytope as well (in the dual lattice $M$). While the $N$-lattice polytope $\Delta^\circ$ gives rise to the fa{\bf N}, the {\bf M}onomials of a generic hypersurface equation $P(z_i) = 0$ are determined by the $M$-lattice polytope $\Delta$. Every point $m$ on $\Delta$ gives rise to a monomial
\begin{equation}\label{eq:monos}
P(z_i) \supset c_m \prod_i z_i^{\langle m, v_i \rangle + 1} \, .
\end{equation}
This presentation allows for a convenient resolution of singularities: if we are given a singular Calabi-Yau hypersurface defined
by a set of monomials with generic coefficients, which lie on a (Newton) polytope $\Delta$, we automatically get a crepant
(partial) resolutions by performing toric resolutions for which all of the new rays in $\Sigma(1)$ are points on $\Delta^\circ$.

More generally, one may construct a maximal smooth ambient toric variety (and thereby a maximally smooth hypersurface) by 
considering a fine triangulation of $\Delta^\circ$ and simply taking all cones over the simplices on the boundary of 
$\Delta^\circ$. In this case, not all lattice points on $\Delta^\circ$ necessarily give rise to divisors on a Calabi-Yau hypersurface:
divisors corresponding to points interior to maximal-dimensional faces of $\Delta^\circ$ miss any smooth Calabi-Yau hypersurface.

\subsection{Tops and Elliptic Fibrations}

In the present context we are not interested in Calabi-Yau hypersurfaces per se, but rather elliptic Calabi-Yau manifolds
for which the elliptic fiber is described by a Tate model. This means that we can describe the elliptic fiber by a hypersurface
equation
\begin{equation}\label{eq:tatemodelgeneric}
{ y^2 +  yxw b_1 + yw^3 b_3 = x^3 + x^2 w^2 b_2 + xw^4 b_4 + w^6 b_6} \,,
\end{equation}
in the weighted projective space $\P_{123}$. The whole elliptic Calabi-Yau manifolds is then obtained by fibering $\P_{123}$
over a base such that the $b_n$ are sections of $- n K_{B}$. Different types of singular fibers can then be engineered by making 
the coefficients $b_n$ have appropriate vanishing degrees along a divisor $\zeta_0 = 0$ of the base. 

This presentation can be rephrased in terms of toric geometry by constructing a fan $\Sigma$ with $v_i$ given by
\be
\ba\label{eq:toppts}
v_x = \left(\begin{array}{c}
       -1 \\ 0 \\ 0
      \end{array}
\right)\, , \, \,
v_y = \left(\begin{array}{c}
       0 \\ -1 \\ 0
      \end{array}
\right)\, , \,\,
v_w = \left(\begin{array}{c}
       2 \\ 3 \\ 0
      \end{array}
\right)\,,  \,\,
v_{\zeta_0} = \left(\begin{array}{c}
       2 \\ 3 \\ 1
      \end{array}
\right)\,.
\ea
\ee
The fan $\Sigma$ contains the following three-dimensional cones 
\begin{equation}
 \langle v_x, v_y, v_{\zeta_0} \rangle \, ,  \langle v_x, {v_w}, v_{\zeta_0} \rangle\, , \langle {v_w}, v_y, v_{\zeta_0} \rangle \, .
\end{equation}
We may then capture the leading terms (in $\zeta_0$) in \eqref{eq:tatemodelgeneric} via \eqref{eq:monos} in terms of 
points on a Newton polyhedron $\Delta$.

This presentation allows for a straightforward application of the techniques discussed above to find all crepant weighted blowups.
If we perform a blowup associated with a refinement $\Sigma' \rightarrow \Sigma$, which introduces a single 
one-dimensional cone with generator $v_E$, the anticanonical class of $T_\Sigma$ receives the contribution
\begin{equation}
\delta K = \left(a_E - \sum_i a_i\right)D_E \, .
\end{equation}
This tells us that the above only is a crepant (partial) resolution of $X$ if its class after the proper transform is $-K_{X_\Sigma} - \delta K$. 
In other words, the proper transform must allow us to `divide out' the right power of the exceptional coordinate $z_E$ to make $P(z_i)$
acquire the weight $(-a_E + \sum_i a_i)$ under the $\C^*$ action \eqref{eq:scalingE}.

\begin{figure}
    \centering
    \includegraphics[width=6cm]{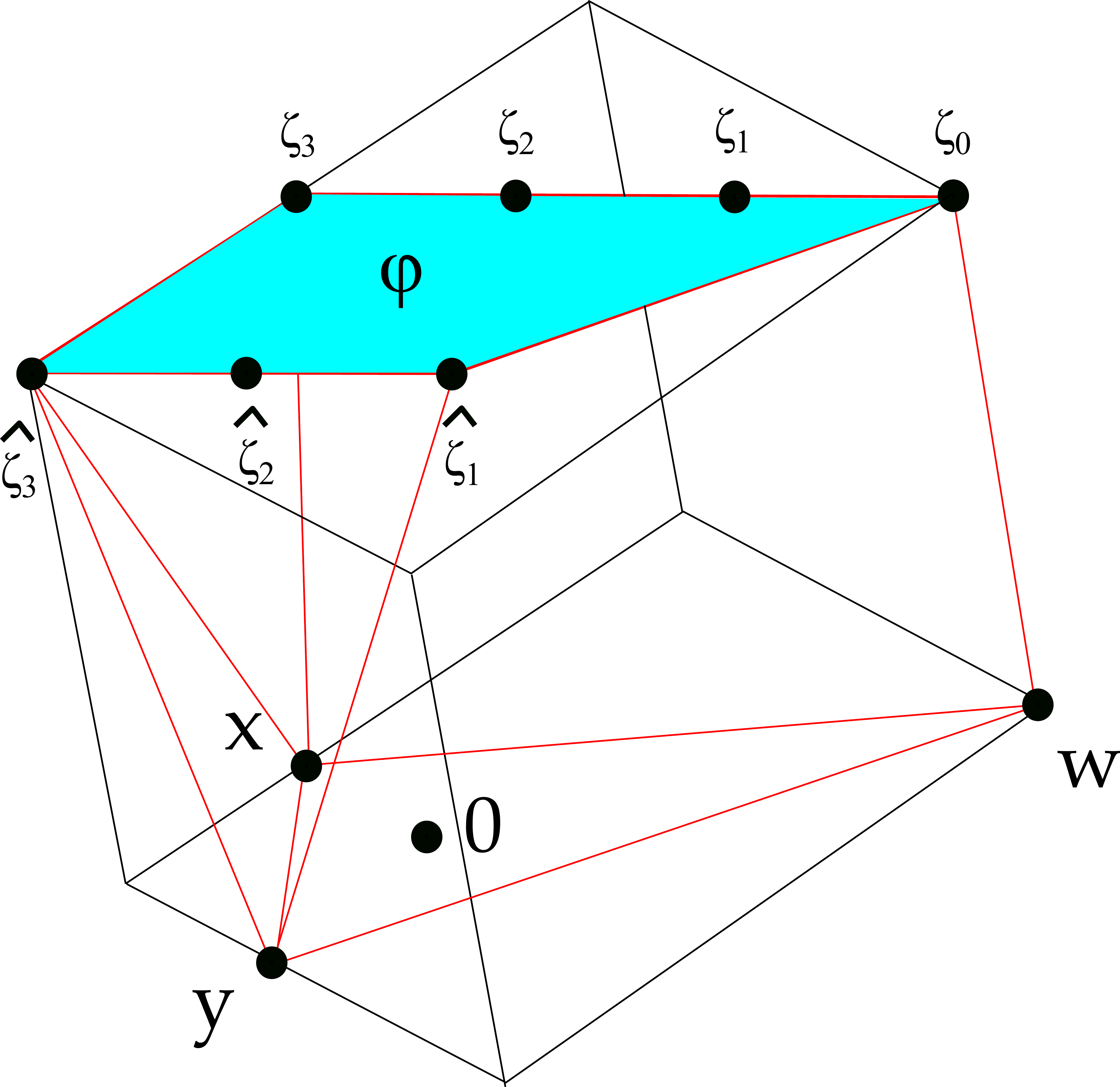}
    \caption{Toric top and fiber face $\varphi$ (in blue) for and $I_7$ Kodaira singular fiber, with the cone generators corresponding to $v_{\zeta_i}$ and $v_{\hat{\zeta}_i}$. The coordinates are summarized in (\ref{eq:toppts}).\label{fig:ToricFan}}
\end{figure}

A weighted blowup sends $z_i \rightarrow z_i z_E^{a_1/a_E}$. In order for such a blowup to be crepant, \eqref{eq:monos} must be divided by
$z_E^{(-a_E + \sum_i a_i)/a_E}$ under the proper transform. Using \eqref{eq:scalingE}, any monomial in \eqref{eq:monos} is 
then turned into
\begin{equation}
\ba\label{eq:monoscrep}
 &z_E^{(a_E - \sum_i a_i)/a_E} \prod_i z_i^{\langle m_j,v_i\rangle+1} z_E ^{\tfrac{1}{a_E}(a_i \langle m_j,v_i\rangle + a_i)} \cr
  =\, &z_E^{(a_E - \sum_i a_i)/a_E} z_E ^{\langle m_j,v_E\rangle + \sum_i a_i/a_E} \prod_i z_i^{\langle m_j,v_i\rangle+1} \cr
  =\,& z_E^{\langle m_j,v_E\rangle + 1 }  \prod_i z_i^{\langle m_j,v_i\rangle+1} \, ,
\ea
\end{equation}
i.e. we simply need to use \eqref{eq:monos} for the new coordinate $z_E$ as well. Note, however, that \eqref{eq:monoscrep}
is a holomorphic section if and only if 
\begin{equation}\label{constrv_E}
 \langle m_j,v_E  \rangle \geq -1 \quad\quad \mbox{for all}\quad m_j \,,
\end{equation}
and hence only blowups related to the introduction of new generators $v_E$ satisfying this relation can be crepant.
For a given singularity\footnote{We are only interested in singularities which can be resolved by refining the cone spanned by $v_x$, $v_y$ and $v_{\zeta_0}$.} in \eqref{eq:tatemodelgeneric}, this will single out a finite number of crepant weighted blowups. 
After performing such a weighted blowup (cone refinement), the set $m_j$ of monomials is not 
changed, i.e. at every step of a sequence of blowups we find the same condition \eqref{constrv_E} for the next step.
We hence learn that we can only use weighted blowups originating from the set of $v_E$ satisfying \eqref{constrv_E} in any 
step of a sequence of blowups. 

The finite number of points above the $v_x,v_y,{v_w}$ plane satisfying \eqref{constrv_E} form the 
tops \cite{Candelas:1996su,Perevalov:1997vw,Bouchard:2003bu} corresponding
to various degenerate fibers in Tate models. An example is shown in figure \ref{fig:ToricFan}.

Even though tops naturally appear in the study of toric hypersurfaces, they have
a more general applicability. The above argument shows that given any elliptic Calabi-Yau manifold for which the 
fiber is given by a Weierstrass model, and a singularity is engineered via assigning vanishing orders, we may use the corresponding 
top \eqref{constrv_E} to find all weighted crepant blowups for which the fiber persists to be embedded as a hypersurface.

\subsection{Triangulations of Tops and Fiber Faces}

As discussed in the last section, weighted blowups are crepant if the exceptional divisors correspond to lattice points
on the relevant top. However, performing resolutions through sequences of weighted blowups is inconvenient for two reasons:
First of all, we may end up with the same resolution although we have performed two different sequences of weighted blowups, see the figure
\ref{fig:busqvstr} and the related discussion for an example. Here, constructing the associated fan of the ambient space provides a
convenient way of identifying (in)equivalent resolutions. As we already know that the rays of this fan will be sitting on the relevant top,
each sequence of blowups will yield a triangulation of this top. Secondly, sequences of weighted blowups are not the most general resolutions
which can be conveniently described by toric methods. In fact, any refinement\footnote{In contrast to elementary blowups, we have to make sure 
the resulting variety is still projective. The condition for projectivity says that the simplices need to be images of faces of a higher-dimensional 
polytope, see e.g. \cite{Fultontoric}.} of a fan supplies us with a morphism which may be used
to construct a resolution \cite{Fultontoric}. In the case of tops, the fan refinements we are looking for are those associated with triangulations and it turns out that not all triangulations can be obtained through a series of weighted blowups, an easy example is given in figure \ref{fig:nobutriang}.

For these reasons, we can conveniently characterize different resolutions of elliptic singularities by considering different 
triangulations of the associated tops. Note that all of the corresponding models are described by the same hypersurface equation, 
which is essentially given by \eqref{eq:monos}, and only the SR-ideal changes when we consider different triangulations. This will
allow us to easily read off properties of the resolved geometries from triangulations.

Starting from a Weierstrass model, all singularities sit in the cone spanned by the rays $v_x$, $v_y$ and $v_{\zeta_0}$ before resolution. Consequently,
it is only this cone which is refined when performing a resolution. We can project the bouquet of cones sitting inside the cone 
$\langle v_x, v_y, v_{\zeta_0}\rangle$ after resolution to a plane resulting in a diagram showing which homogeneous coordinates
are allowed to vanish simultaneously. We call this type of diagram a {\it fiber face} and it will prove very useful to conveniently
read off which triangulation corresponds to which of the phases. An example is shown in figure \ref{fig:ToricFan}.


\subsection{Flops}

For a toric variety, we may perform a flop if there are cones in the associated fan which can be re-triangulated as shown in the following figure, with four
ray generators on a plane:
\be\label{AtiFlop}
    \includegraphics[width=10cm]{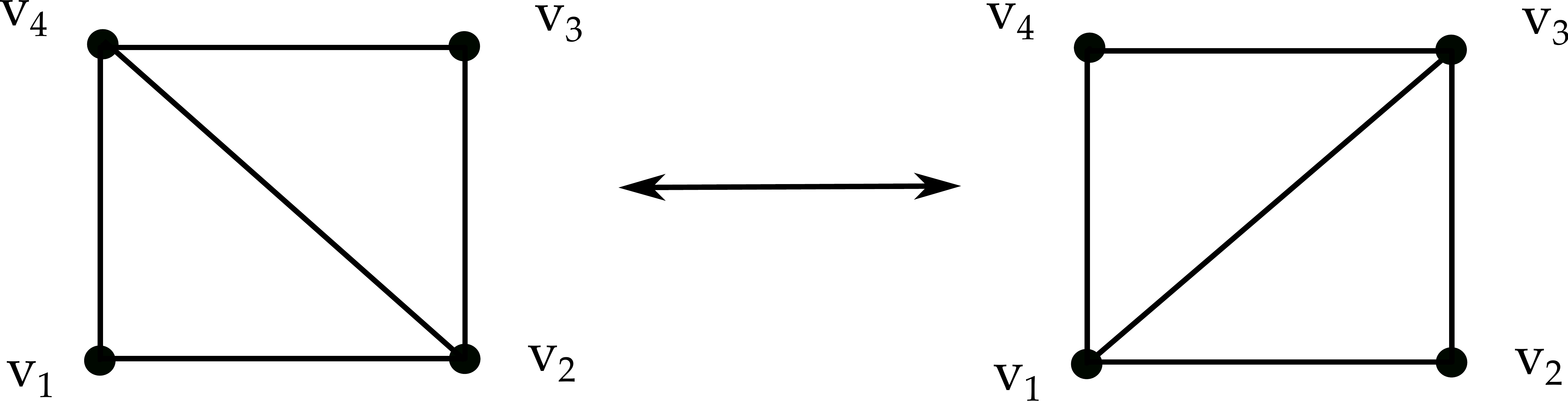}
\ee
We may understand this flop as a two-step process in which we first take out the cone $\sigma_{24}$ connecting $v_2$ and $v_4$, resulting in
a singularity, and then introduce the cone $\sigma_{13}$ connecting $v_1$ with $v_3$ to resolve. The cones $\sigma_{24}$ and $\sigma_{13}$
correspond to subvarieties of codimension two (intersection of two divisors) and each of these subvarieties have normal bundles {in the Calabi-Yau} $\mathcal{O}(-1)\oplus \mathcal{O}(-1)$.

For a Calabi-Yau hypersurface, or more generally complete intersection, embedded in a toric ambient space, performing a flop on the level of the ambient space induces a flop of the Calabi-Yau as well\footnote{Of course, this is only true if the relevant subvariety which is flopped in the ambient space also meets the embedded Calabi-Yau.}. 
The class of flops of the Calabi-Yau which descend from such flops of the ambient space can be conveniently described in terms of re-triangulations of tops. However, there are also other flops for which this is not the case. This stems from the fact that not all rational curves descend from rational curves in the ambient space. Flop transitions involving such curves are much harder to determine, and will be of consideration in the following.


\section{Fiber Faces and Box Graphs for $\mathfrak{su}(2k+1)$}
\label{sec:BoxToric}

We will now show that for elliptic fibrations with $I_{2k+1}$ singular fibers, corresponding to a gauge algebra $\mathfrak{su}(2k+1)$ with anti-symmetric matter, the algebraic resolutions as well as triangulations of the top/fiber faces yield (strict) subclasses of box graphs, and that there is a precise correspondence between the triangulations and the properties of the phases. The starting point for the toric resolutions is the Tate resolution (i.e. the resolution of the Tate model), which proceeds via a specific algebraic sequence of blowups, to be discussed in the next subsection. 
We then show how algebraic resolutions have a simple characterization in terms of specific box graphs, whose anti-Dyck path is a concatenation of corners $\includegraphics[width=.3cm]{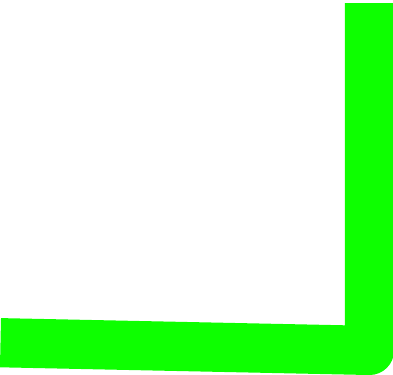}$ and  $\includegraphics[width=.3cm]{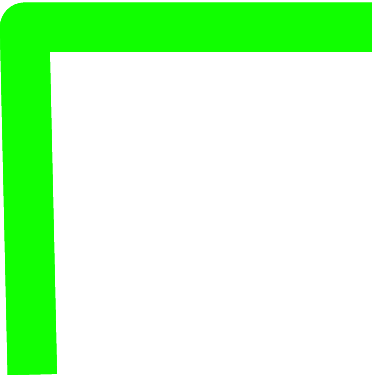}$. The toric resolutions obtained by top triangulations are explained in section \ref{sec:ToricResTring}. Finally the main argument identifying these with a sub-class of box graphs is given in section \ref{sec:ToricTBG}. 

\subsection{Tate Resolution}
\label{sec:SUOdd}

The gauge algebras $\mathfrak{su}(2k+1)$ are realized in F-theory in terms of singular fibers in codimension one of Kodaira type $I_{2k+1}$.
There are two matter loci of interest, corresponding to the fundamental representation of dimension $2k+1$ and the anti-symmetric $\Lambda^2 V$ of dimension $(2k+1)k$. 
The singular Tate form is \cite{Bershadsky:1996nh, Katz:2011qp}
\be\label{eq:su2kp1sing}
y^2 + b_1 x y + b_3 y \zeta_0^k  = x^3 + b_2 \zeta_0 x^2  + b_4 x \zeta_0^{k+1} + b_6 \zeta_0^{2k+1}  \,,
\ee
where $\zeta_0=0$ is the discriminant component for which the discriminant has vanishing order $\Delta = O(\zeta_0^{2k+1})$, and above which the singular $I_{2k+1}$ fiber is located. 
The two matter enhancements occur along the following loci
\be\ba
\Lambda^2 V: &\qquad b_1=0 \cr
V: & \qquad P_{V}= b_2b_3^2 - b_1 b_3 b_4 +b_1^2b_6 =0 \,.
\ea\ee
Resolutions of this class of models were described in \cite{Lawrie:2012gg} using algebraic blowups:
\be
\ba
(x, y, \zeta_i; \zeta_{i+1})&\,,\qquad i=0, \cdots k-1\cr
(y, \zeta_i; \hat\zeta_{i}) &\,,\qquad i= 1, \cdots, k \,.
\ea
\ee
Here the notation indicates that the singular locus $x=y=\zeta_i=0$ is blown up with new exceptional section $\zeta_{i+1}$. 
This can also be expressed in terms of the $\mathbb{C}^*$ scalings
\be
\begin{array}{ccccc}
x & y& \zeta_i & \zeta_{i+1}  & \hat\zeta_i\cr\hline
1& 1& 1& -1& 0 \cr
0& 1& 1& 0 &-1\cr
\end{array}
\ee
The resolved Tate model (in codimension one, two,  and for four-folds, three) is
\be\label{TateResolved}
\ba
&T_{2k+1}: \qquad y^2 B(\hat\zeta)\hat\zeta_k + b_1xy + b_3y\zeta_0^kB(\zeta\hat\zeta)C(\zeta \hat\zeta) \cr
&\  = x^3B(\zeta)A(\zeta \hat\zeta)\zeta_k^k\hat{\zeta}_k^{k-1} + b_2x^2\zeta_0B(\zeta)\zeta_k 	
			+ b_4x\zeta_0^{k+1}B(\zeta^2 \hat\zeta)C(\zeta\hat\zeta)\zeta_k 
			+ b_6\zeta_0^{2k+1}B(\zeta^3\hat\zeta^2)C(\zeta^2\hat\zeta^2)\zeta_k \,.
\ea			
\ee
where
\be
	A(z) = \prod_{i=2}^{k-1}z_{i}^{i-1} \,,\qquad 
	C(z) = \prod_{i=1}^{k-2}z_{i}^{k-(i+1)} \,,\qquad 
	B(z) = \prod_{i=1}^{k-1} z_i \,.
\ee
 The fibers above the codimension one locus are given by rational curves, and the associated exceptional divisors can be described in terms of the exceptional sections as follows:
  \be\label{CartanDivs}
\begin{array}{c|c|l}
\hbox{Simple root} & \hbox{Section} & \hbox{Equation in $Y_4$} \cr\hline
 \alpha_0&  \zeta_0  & 
 0=\left[\prod_{i=2}^{k}\hat{\zeta}_i\right] \left(y^2\hat{\zeta}_1 
 - x^3 A(\zeta)B(\zeta) \zeta_k^k \left[\prod_{i=2}^{k-1}\hat{\zeta}_i^{i-2}\right]\hat{\zeta}_k^{k-2} \right) + b_1 xy \\
 \alpha_{1\cdots k-1} &  \zeta_{1, \cdots , k-1 }  
  &0= y^2 B(\hat{\zeta})\hat{\zeta}_k   +b_1 xy \\
 \alpha_{k}& \zeta_k  & 0= yB(\hat{\zeta})\left( y \hat{\zeta}_k + b_3 \zeta_0^k B(\zeta)C(\zeta\hat{\zeta})\right)  +b_1 xy   \\ 
 \alpha_{k+1}  & \hat{\zeta}_{k}  &  0=   B(\zeta) \Big(  b_3 y \zeta_0^k B(\hat{\zeta})C(\zeta\hat{\zeta}) -b_2 x^2\zeta_0 \zeta_k  -b_4 x \zeta_0^{k+1} B(\zeta\hat{\zeta})C(\zeta\hat{\zeta}) \zeta_k \cr
 &&\qquad  -b_6 \zeta_0^{2k+1}B(\zeta^2\hat\zeta^2)C(\zeta^2\hat\zeta^2)\zeta_k  \Big) + b_1 xy   \\
 \alpha_{ k+2\cdots 2k-1}&  \hat{\zeta}_{k-1, \cdots ,2}  &  0= B(\zeta)\zeta_k b_2 x^2 \zeta_0 + b_1 xy   \\
 \alpha_{2k} &\hat{\zeta}_1&0= x^2 B(\zeta)\zeta_k \left(x A(\zeta\hat{\zeta})\zeta_k^{k-1}\hat{\zeta}_k^{k-1}+b_2\zeta_0 \right) - b_1 xy 
\end{array}
\ee
Here, the projective relations of the resolution were already used and the exceptional divisors, or Cartan Divisors, can be identified with the simple roots of $\mathfrak{su}(2k+1)$
\be\ba\label{eq:ptsvscartans}
 D_{\alpha_i} &= D_{\zeta_i} \qquad \qquad {\rm for} \quad i=0,\cdots, k \\
 D_{\alpha_i} &= D_{\hat{\zeta}_{2k+1-i}} \qquad {\rm for} \quad i=k+1, \cdots, 2k  \,.
\ea\ee
We will now consider various alternative resolutions, which will be shown to correspond to a subclass of box graphs.

\subsection{Algebraic Resolutions and Hypercubes}
\label{sec:AlgRes}

The first class of resolutions we will consider are algebraic resolutions, which were studied for $\mathfrak{su}(5)$ 
in \cite{EY, MS} and for general Tate models in \cite{Lawrie:2012gg}. The starting point is the codimension one resolved Tate 
model, i.e. (\ref{TateResolved}) with $\hat\zeta_i=1$. This has the form of a binomial form
\be
y \hat{y} = \left(\prod_{i=1}^{k} \zeta_i\right) \ P \,,
\ee
where 
\be
\ba
 \hat{y}&=  y  + b_1x  + b_3 \zeta_0^kB(\zeta)C(\zeta)  \cr
 P &=  x^3A(\zeta )\zeta_k^{k-1}
 		+ b_2x^2\zeta_0	
			+ b_4x\zeta_0^{k+1}B(\zeta )C(\zeta) 
			+ b_6\zeta_0^{2k+1}B(\zeta^2)C(\zeta^2) \,,
\ea
\ee
with the projective relations, obtained from the big resolutions $(x, y, \zeta_i; \zeta_{i+1})$. As we are interested in the case of $b_1=0$, i.e. matter in the anti-symmetric representation, the only relevant small resolutions are between $y$ and $\zeta_i$. 
The set of small resolutions is then
\be\label{AlgResDef}
\hbox{AlgRes}_\sigma: \qquad (y, \zeta_i; \hat\zeta_{\sigma (i)}), \qquad i=1, \cdots, k, \qquad \hbox{for a fixed }\sigma \in S_{k} \,.
\ee
Note that not all of these give inequivalent resolutions. 

\begin{figure}
    \centering
    \includegraphics[width=9cm]{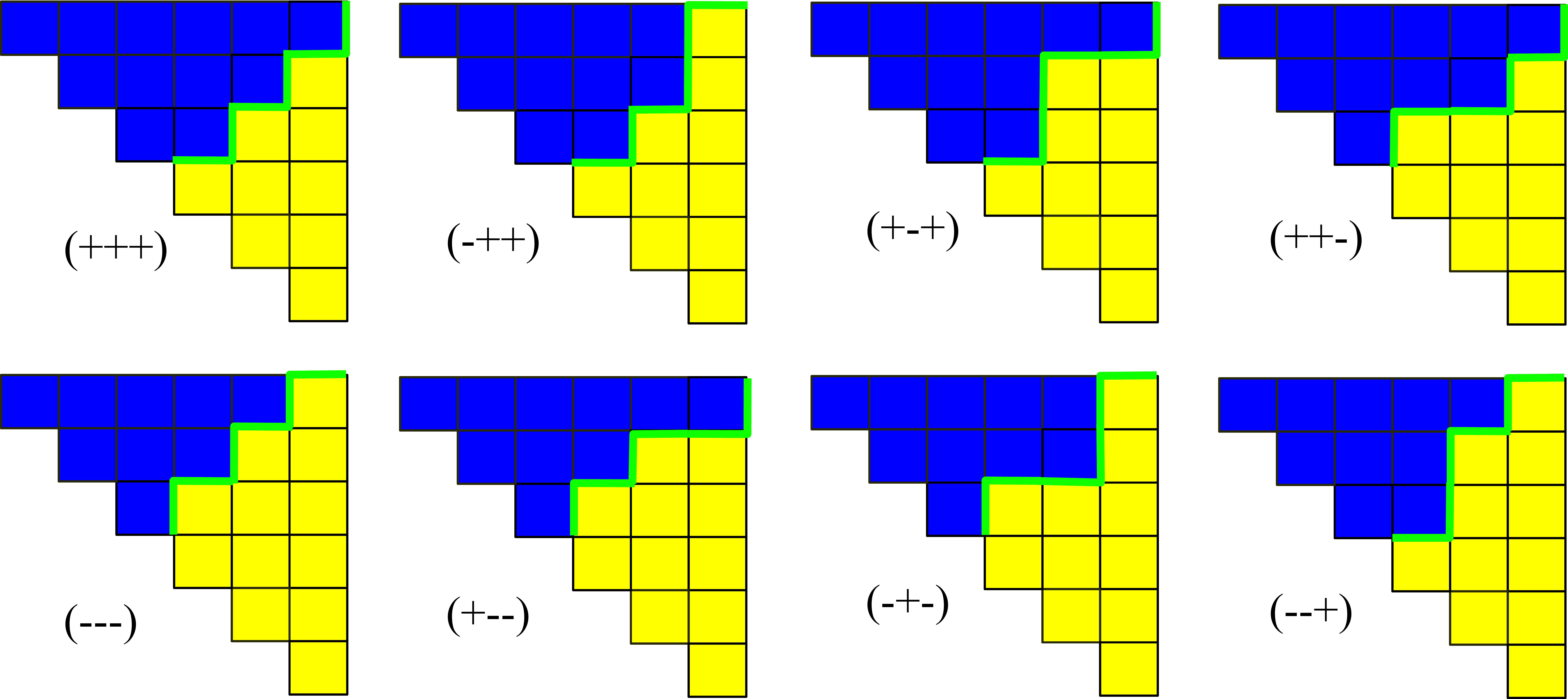}
    \caption{Example for $\mathfrak{su}(7)$: box graphs corresponding to algebraic resolutions, which are obtained as sequence of corners (\ref{AntiDyckCorners}).   \label{fig:AntiDyckAlg}}
\end{figure}

We can prove the following statement: 
The algebraic resolutions  (\ref{AlgResDef}) are exactly the box graphs, which have anti-Dyck paths that are concatenations of corners of the type
 \be\label{AntiDyckCorners}
  \includegraphics[width=1cm]{AntiDyck1.pdf}\qquad \hbox{and} \qquad 
  \includegraphics[width=1cm]{AntiDyck2.pdf}\,.
\ee
The resulting paths automatically satisfy the diagonal condition. For $\mathfrak{su}(7)$ the algebraic resolutions, and corresponding paths, are shown in figure \ref{fig:AntiDyckAlg}.

The network of flops between these algebraic resolutions for $\mathfrak{su}(2k+1)$ is a hypercube in {$k$ dimensions}, which follows straight forwardly from the decomposition into corners (\ref{AntiDyckCorners}): every anti-Dyck path, can be labelled by $(\pm1, \cdots, \pm 1)$, representing the decomposition into the two corners represented by $\pm 1$ in (\ref{AntiDyckCorners}). These are exactly points on a {$k$ dimensional} hypercube, so there are $2^k$ such phases/resolutions. A flop is a map   $\includegraphics[width=.3cm]{AntiDyck1.pdf}\leftrightarrow    \includegraphics[width=.3cm]{AntiDyck2.pdf}$, which in the hypercube corresponds to moving along one of the edges, which exchanges $+1 \leftrightarrow -1$. 

For $\mathfrak{su}(7)$, the 3d cube is shown in red in figure \ref{fig:SU7FlopGraph}, for $\mathfrak{su}(5)$, the flop diagram for algebraic resolutions of singular elliptic fibrations with {\bf 10} matter is a square \cite{Hayashi:2013lra}. 


\subsection{Fiber Face Triangulations}\label{sect:toric_triang}
\label{sec:ToricResTring}

In this section we will identify the precise correspondence between toric hypersurface\footnote{Here, of course, by hypersurface we always mean that the fiber is embedded as a hypersurface in a toric ambient space.} resolutions, which are characterized by  fiber face triangulations and a subclass of box graphs for $\mathfrak{su}(2k+1)$ with anti-symmetric matter.

\subsubsection{Top and Fiber Face}

We now discuss the resolutions of \eqref{eq:su2kp1sing} using the toric techniques discussed in section \ref{sect:gentoricstuff}.
As a first step, let us record the defining data. If the generators of the rays corresponding to ${x,y,w,\zeta_0}$ are fixed 
to be given by \eqref{eq:toppts}, the monomials in \eqref{eq:su2kp1sing} correspond to the following lattice points
\be 
\begin{array}{c|c|c|c|c|c|c|c}
\hbox{Monomial} &y^2 & b_1 xy & b_3 y \zeta_0^k & x^3& b_2 \zeta_0 x^2 & b_4 x \zeta_0^{k+1} &b_6 \zeta_0^{2k+1}  \cr\hline
&&&&&&&\cr 
\begin{array}{c}
\hbox{Lattice} \cr
\hbox{Point}
\end{array}
&	\left(\begin{array}{c}
       -1 \\ 1 \\ 0
      \end{array}\right)
      &
      \left(\begin{array}{c}
       0 \\ 0 \\ -1
      \end{array}\right) 
      &
      \left(\begin{array}{c}
       1 \\ 0 \\ k-3
      \end{array}\right) 
      &
      \left(\begin{array}{c}
       -2 \\ 1 \\ 0
      \end{array}\right) 
      &
      \left(\begin{array}{c}
       -1 \\ 1 \\ -1      \end{array}\right)
       &
       \left(\begin{array}{c}
       0 \\ 1 \\ k-3      \end{array}\right)
       &
       \left(\begin{array}{c}
       1 \\ 1 \\ 2k      \end{array}\right) 
       \cr
             &&&&&&
     \end{array}
\ee
%
Using \eqref{constrv_E}, this means that any crepant resolution obtained by subdividing the fan must only use the rays
\be \label{vzetadef}
\ba
v_{\zeta_i} &= (2-i,3-i,1) \,,\qquad  i= 1, \cdots, k  \cr
v_{\hat{\zeta}_i} &= (2-i,2-i,1) \,,\qquad  i= 1, \cdots, k \,.
\ea
\ee
An example of the top for $k=3$ can be found in figure \ref{fig:ToricFan}.
In fact, we could have already obtained this from the fact that all of the blowups discussed in the previous sections are crepant.
Translating these blowups into toric language shows that we need to subdivide the cone $\langle v_x,v_y,v_{\zeta_0}\rangle$ using the rays
generated by \eqref{vzetadef}. The algebraic resolutions discussed in section \ref{sec:AlgRes} are precisely those, in which we first subdivide
using the coordinates $\zeta_i$ for $i=1...k$ (in this order) and only then introduce the $\hat{\zeta}_i$ in an arbitrary order. 

\begin{figure}
    \centering
    \includegraphics[width=14cm]{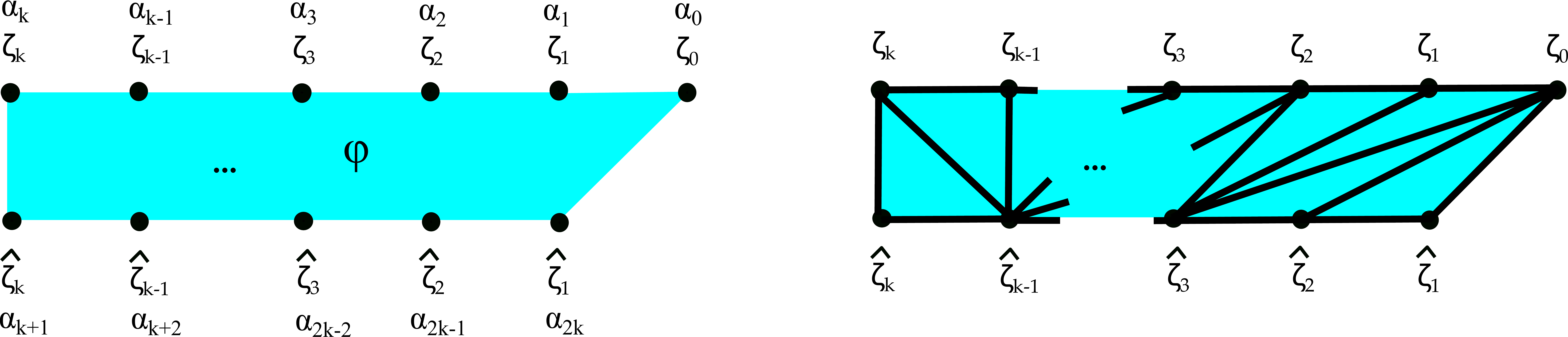}
    \caption{The picture on the left hand side shows the fiber face $\varphi$ for $\mathfrak{su}(2k+1)$ with vertices ${\zeta_i}$ and  ${\hat\zeta_i}$ defined in (\ref{vzetadef}).
    The label $\alpha_i$ correspond to the simple roots that each node is associated with. On the right a sample 
    triangulation of the fiber face is shown.\label{fig:ToricFanGen}}
\end{figure}

In general, we may of course subdivide the cone $\langle v_x,v_y,v_{\zeta_0}\rangle$ by introducing the points \eqref{vzetadef} in any order, or
more generally, consider an arbitrary fine triangulation of the corresponding top. Any triangulation will contain the cones $\langle v_x, v_{\zeta_i}, v_{\zeta_{i+1}}\rangle$ $\langle v_x, v_{\zeta_k}, v_{\hat{\zeta}_{k}}\rangle$, $\langle v_y, v_{\hat{\zeta}_i}, v_{\hat{\zeta}_{i+1}}\rangle$ and 
$\langle v_y, v_{\hat{\zeta}_1}, v_{\zeta_{0}}\rangle$, so that a triangulation is specified by giving the simplices on the face containing the points \eqref{vzetadef}. We can hence present a triangulation by drawing an image of what we call the fiber face, see figure \ref{fig:ToricFanGen}. 
Given such a toric resolutions, one has to check projectivity. This is already guaranteed for triangulations related to sequences of weighted blowups as this necessarily 
preserves projectivity. In the general case, we can argue like this. A toric variety is projective if there is a piecewise linear and strongly convex support function on the cones of its fan. This is equivalent to the simplices of our triangulation being the images of faces of a polytope 
embedded in a higher-dimensional space. In the present case, this can easily be seen to be true: for any triangulation, one may distribute
the $\zeta_i$ and $\hat{\zeta}_j$ along an arch such that all of the simplices become faces. In the present case any triangulation 
 gives rise to a projective toric ambient space.

Summarizing the above discussion, sequences of weighted blowups are a subclass of resolutions as toric hypersurfaces which in turn can be
constructed via triangulations. As shown in appendix \ref{sect:triang}, there are $\binom{2k-1}{k}$ such triangulations.
By construction, such resolutions will all lead to the same defining equation, \eqref{TateResolved}, and only differ in the SR ideal, 
which can be read off from the triangulation. 

Let us now discuss the structure of fiber components. At a generic point of the locus $\zeta_0$, the fiber will split into
$2k + 1$ components, as the proper transform for any resolution is $\zeta_0 \rightarrow  \zeta_0 \prod_i \zeta_i \hat{\zeta}_i$.
We can hence identify the points in \ref{fig:ToricFanGen} with the Cartan divisors. Over codimension one in the base, two such
divisors will only intersect if they are connected by a one-simplex of the triangulation along an edge of the top (see e.g. \cite{Candelas:1996su, Perevalov:1997vw}), i.e. we can identify 
\be\ba\label{eq:ptsvscartans2}
 D_{\alpha_i} &= D_{\zeta_i} \qquad \qquad {\rm for} \quad i=0,\cdots, k \\
 D_{\alpha_i} &= D_{\hat{\zeta}_{2k+1-i}} \qquad {\rm for} \quad i=k+1, \cdots, 2k  \,.
\ea\ee

Spelled out more explicitly, the expressions for the irreducible codimension-one fiber components are as in (\ref{CartanDivs})
and one may use this to check more explicitly that the identification \eqref{eq:ptsvscartans2} is sensible.

\subsubsection{Flops}

For two distinct triangulations which only differ by two simplices (and hence cones in the fan), 
\begin{equation}
\begin{aligned}
 \varphi_1 \supset&  \langle \zeta_k,\zeta_{k+1},\hat{\zeta}_l \rangle\, , \langle \zeta_{k+1},\zeta_{l},\hat{\zeta}_{l+1} \rangle \\
  \varphi_2 \supset&  \langle \zeta_k,\zeta_{k+1},\hat{\zeta}_{l+1} \rangle \, ,  \langle \zeta_{k},\hat{\zeta}_{l} ,\hat{\zeta}_{l+1} \rangle
\end{aligned}
\end{equation}
both fans can be seen as a subdivision of a fan containing the `fused' cone
\begin{equation}
 \varphi_{1 \cup 2} \supset \langle \zeta_k,\zeta_{k+1},\hat{\zeta}_{l} ,\hat{\zeta}_{l+1} \rangle \, .
\end{equation}
In other words, the fiber face contains four vertices which are positions as shown in (\ref{AtiFlop}).
Correspondingly, the geometrical transition between the two phases determined by triangulations $ \varphi_1$ and $  \varphi_2$ is a flop, both
at the level of the ambient space and the level of the embedded Calabi-Yau \eqref{TateResolved}. It is not hard to see that all triangulations of the fiber face are linked by passing through a number of transitions of this type. Hence all phases realized by triangulations are connected via flop transitions.


\subsubsection{Anti-Symmetric Representation}
\label{sect:anti-symsplitrules}

We now turn to the splitting of fiber components above $b_1=0$, corresponding to matter in the anti-symmetric $\Lambda^2 {\bf 2k+1}$ representation, where the fiber type enhances from {$I_{2k+1}$ to $I^*_{2k-3}$.}
This occurs over codimension two in the base, and 
thus, the `connections' along the fiber face  $\varphi$, i.e. the triangulation data, becomes relevant in characterizing the fibers. 
One-simplices connecting a divisor $\zeta_j$ with $\hat{\zeta}_l$ (for $j,l \neq k$) indicate that the two divisors intersect 
along a codimension two locus in the base. As such pairs are never neighbouring Cartan divisors, this can only happen if the two divisors
share a common component, which means there is a component of multiplicity at least two over the corresponding locus.
The one-simplices, which connect the $\zeta_i$ with $\hat{\zeta}_i$ hence gives us information
relevant to the phase with respect to the antisymmetric representation. Let us discuss this in a bit more detail by analysing the 
behaviour of the different Cartan divisors over $b_1=0$ in turn, which will then enable us to identify the corresponding box graphs. 

\begin{itemize}
\item[{$\alpha_0$}:]
Over $b_1=0$, the number of irreducible components $\zeta_0$ splits into depends on how many of the coordinates
$\hat{\zeta}_i$, $i=2..k$ are allowed to vanish simultaneously with $\zeta_0$. In toric language, this means we have to
count the number of one-simplices of the considered triangulation $\varphi$, which contain $v_{\zeta_0}$ and one of the
$v_{\hat{\zeta}_i}$, $i=2, \cdots, k$. Note that this number can be zero, depending on the triangulation.
There is always at least one component over $\zeta_0 = \left(y^2\hat{\zeta}_1 
 - x^3 A(\zeta)B(\zeta) \zeta_k^k \left[\prod_{i=2}^{k-1}\hat{\zeta}_i^{i-2}\right]\hat{\zeta}_k^{k-2} \right)=0$.
As there is always a one-simplex connecting $v_{\zeta_0}$ with $v_{\hat{\zeta}_1}$, we can summarize
the splitting rule of $\zeta_0$ by saying that the number of components it splits into is equal to the number of one-simplices
connecting $v_{\zeta_0}$ with any of the $v_{\hat{\zeta}}$.

\item[{$\alpha_{i\not=0}$}:]

Considering $\zeta_{i}=0$ for $i=1,\cdots, k-1$, the number of components over $b_1=0$ is determined by the number of factors
of $B(\hat{\zeta})\hat{\zeta}_k = \prod_{j=1}^{k} \hat{\zeta}_i$ that $\zeta_{i}$ is allowed to vanish simultaneously with.
Again, this directly translates into the number of one-simplices connecting $v_{\zeta_i}$ with any of the $v_{\hat{\zeta}_j}$.
Note that any triangulation will at least contain one such one-simplex.
\end{itemize}

\noindent Continuing in this fashion, one may easily see that all of the splittings over $b_1=0$ may be elegantly summarized by
the simple rule: 
\begin{theorem}\label{thm:splittingrules}
Each fiber component $F_\ell$ corresponds to a root $\alpha_\ell$ and a homogeneous coordinate
according to the table above. Let $Z=\{\zeta_i\, |\, i =0, \cdots, k\} $ and 
$\hat{Z} = \{ \hat{\zeta}_i\,|\, i=1, \cdots, k\}$. Above $b_1=0$, the rational curve 
 $F_\ell$ corresponding to the section $\zeta_i\in Z$ splits into $n_\ell$ components, where 
 \be
 n_\ell = \# \hbox{connections between $\zeta_i$ and any element in $\hat{Z}$}.
 \ee
Likewise, if $F_\ell$ corresponds to $\hat\zeta_i$, then the number of splitting components is the number of connections between $\hat{\zeta}_i$ and any element of $Z$. 
\end{theorem}
Let us now see how many resolutions can be obtained in the way outlined above for $\mathfrak{su}(2k+1)$ with ${\Lambda^2}{V}$. 
As any two such {triangulations of the fiber face $\varphi$}
determine a different phase, this question is equivalent to determining the number of triangulations of {$\varphi$}. Using \eqref{eq:Tmn} derived in appendix \ref{sect:triang},
we find that this number is given by
\begin{equation}\label{NumberToric}
 \# \hbox{Triangulations}\ = 2\,  \binom{2k-1}{k}\, .
\end{equation}
The factor of 2 arises as we get two phases from each triangulation by reordering the simple roots.
Note that we can also easily reproduce the total number of fiber components (counted with multiplicities) over the $b_1=0$ locus. 
From the above discussion it follows that we simply need to count the number of one-simplices connecting the two sides of the
fiber face, as each gives rise to two components over $b_1=0$. For any triangulation, there are $2k$ such one-simplices, so that
we find a total of $4k$ components which matches with the $ (2k + 2 - 4)\cdot 2 + 4 = 4k $ components expected for a fiber of type 
$I^*_{2k-3}$.

\subsubsection{Fundamental Representation}

Let us now discuss which fiber component splits over the matter curve carrying the fundamental representation, i.e. over 
$P = b_1^2 b_6 - b_2 b_3^2 + b_1 b_3 b_4 = 0$. Consider the fiber component corresponding to the root $\alpha_{k+1}$.
Over $b_1^2 b_6 - b_2 b_3^2 + b_1 b_3 b_4 = 0$, it splits into the two components
\be
\ba
0 &= b_1 x +  b_3 B(\zeta\hat{\zeta}) C(\zeta\hat{\zeta}) \zeta_0^k  \cr
0 &= b_1^2 y -b_1 b_2 x  B(\zeta)  \zeta_0 \zeta_k + b_2 b_3 B(\zeta^2\hat{\zeta}) C(\zeta\hat{\zeta}) \zeta_0^{k+1} \zeta_k - 
  b_1 b_4 B(\zeta^2\hat{\zeta}) C(\zeta\hat{\zeta}) \zeta_0^{k+1} \zeta_k \,.
\ea
\ee
Note that this statement is completely independent of which triangulation we have choosen, so that we conclude that
{\it all models} in which the fiber is realized as toric hypersurface are in the same phase with respect to  the fundamental representation.
Similarly, one easily convince oneself that all other fiber components stay irreducible over the matter curve related to the 
fundamental representation.


\subsection{Coulomb Phases/Box Graphs for Triangulations of Tops}
\label{sec:ToricTBG}

\begin{figure}
    \centering
    \includegraphics[width=8cm]{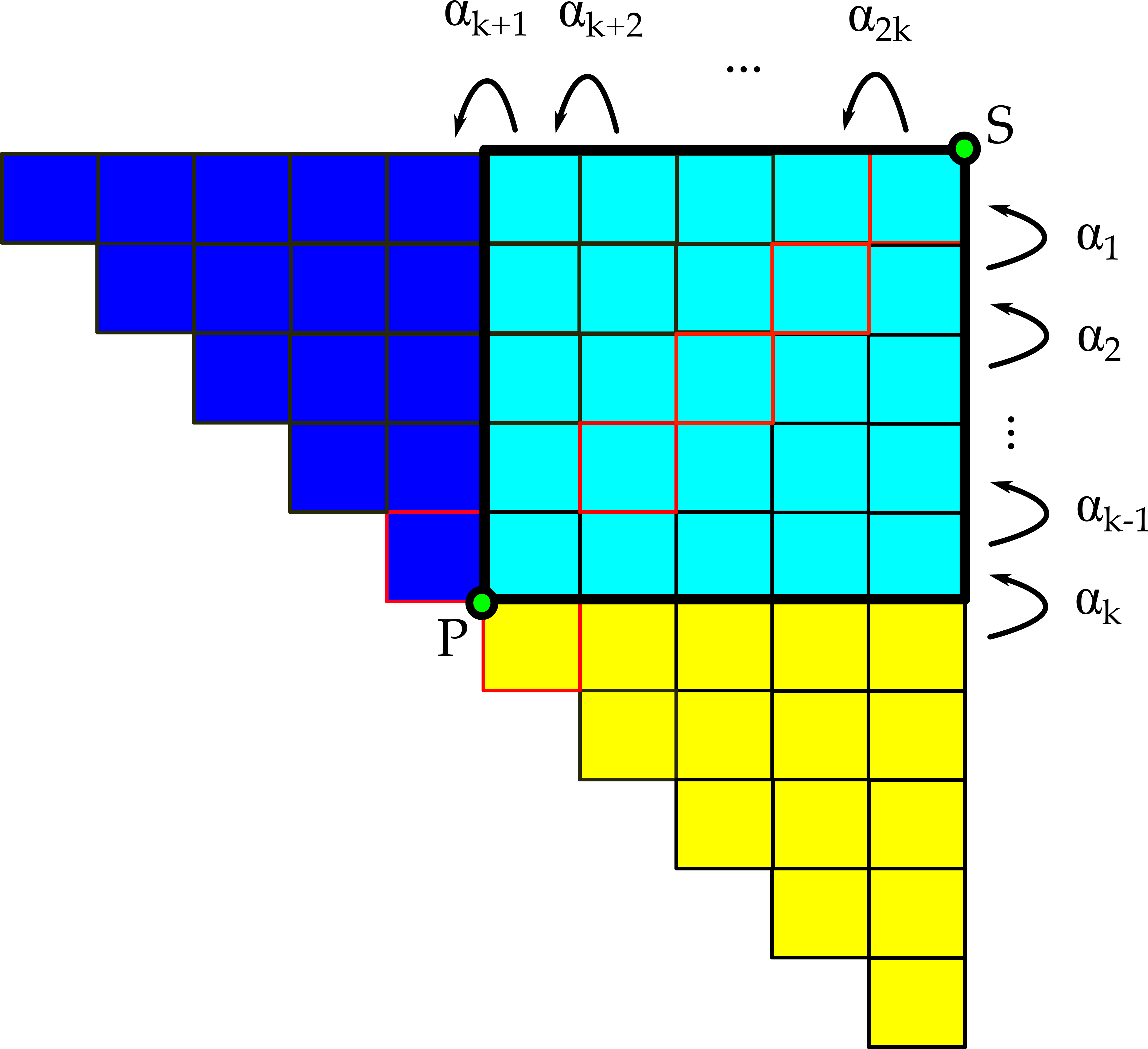}
    \caption{Box Graphs corresponding to fiber face triangulations. Blue/yellow are fixed $+/-$ sign assignments, 
    and each sign assignment/coloring of the turquois region (satisfying the consistency requirements, i.e. trace condition and flow 
    rules for box graphs) corresponds to a  triangulation of a fiber face.  
    \label{fig:ToricBox}}
\end{figure}

We now turn to the alternative description of the fiber face triangulations in terms of Coulomb phases, or equivalently box graphs. 
The fiber face triangulations correspond to a sub-class of box graphs which can be characterized as follows. 
\begin{theorem}\label{thm:ToricBox}
There is a one-to-one correspondence between fiber face triangulations (\ref{vzetadef}) for an $I_{2k+1}$ fiber in codimension one with enhancement to $I_{2k-3}^*$ (or $\mathfrak{so}(4k+2)$) along  the codimension one locus $b_1=0$, and the box graphs, which correspond to the following decorations of the representation graph of $\Lambda^2({\bf 2k+1})$.
\begin{itemize}
\item[(a)] The  weights $L_i+ L_j$ with $i\in[1, k]$ and $j\in[2, k+1]$  are assigned $+$ (i.e. the boxes are colored blue)
\item[(b)]  The  weights $L_i+ L_j$ with $i\in[k+1, 2k-1]$ and $j\in[k+2, 2k]$  are assigned $-$ (i.e. the boxes are colored yellow)
\item[(c)] Any sign assignments in the remaining $k\times k$ square in the representation graph with weights $L_{i}+ L_j$, $i\in [1, k]$ and $j\in[k+2, 2k+1]$, which obeys the flow rules then defines a consistent box graph, and corresponds to exactly one fiber face triangulation. 
\end{itemize}
Equivalently, the anti-Dyck paths starting at the point $S$ and ending at $P$, as marked in figure \ref{fig:ToricBox}, are one to one with toric fiber face triangulations. 
\end{theorem}

We have shown the structure of the toric box graphs in figure \ref{fig:ToricBox}, where the turquois colored region can be filled with any sign assignment which satisfies the flow rules. The + (blue) and - (yellow) colorings in the remaining triangles defined by (a) and (b) in the theorem, respectively, are fixed. Any sign changes in those regions will correspond to deviations from fiber face triangulations.

Before we prove the theorem, we recall how box graphs encode various properties of the codimension two fiber. A box graph for the $\Lambda^2V$ representation determines a specific fiber by providing the extremal generators of the cone of effective curves along the codimension two locus $b_1=0$ in the Tate model, and their intersections. The central tool for that are the splitting rules, which specify how irreducible fiber components in codimension one split along the $b_1=0$ locus. 

The {\bf splitting rules} \cite{CLSSN} applied to the current problem of $\Lambda^2V$ for $\mathfrak{su}(2k+1)$ state:
Given a box graph or equivalently anti-Dyck path, it can be decomposed into horizontal and vertical segments, separated by the corners of the path. We will denote these lines by $H^i$ and $V^i$, when associated to horizontal or vertial lines in the box graph, which correspond to adding $\alpha_i$. 
Recall that each vertical and horizontal wall in the box graph corresponds to a simple root $\alpha_i = L_i - L_{i+1}$, and whenever the anti-Dyck path crosses such a wall, the curve $F_i$ labeled by the corresponding root splits along $b_1=0$. 

\begin{figure}
    \centering
    \includegraphics[width=13cm]{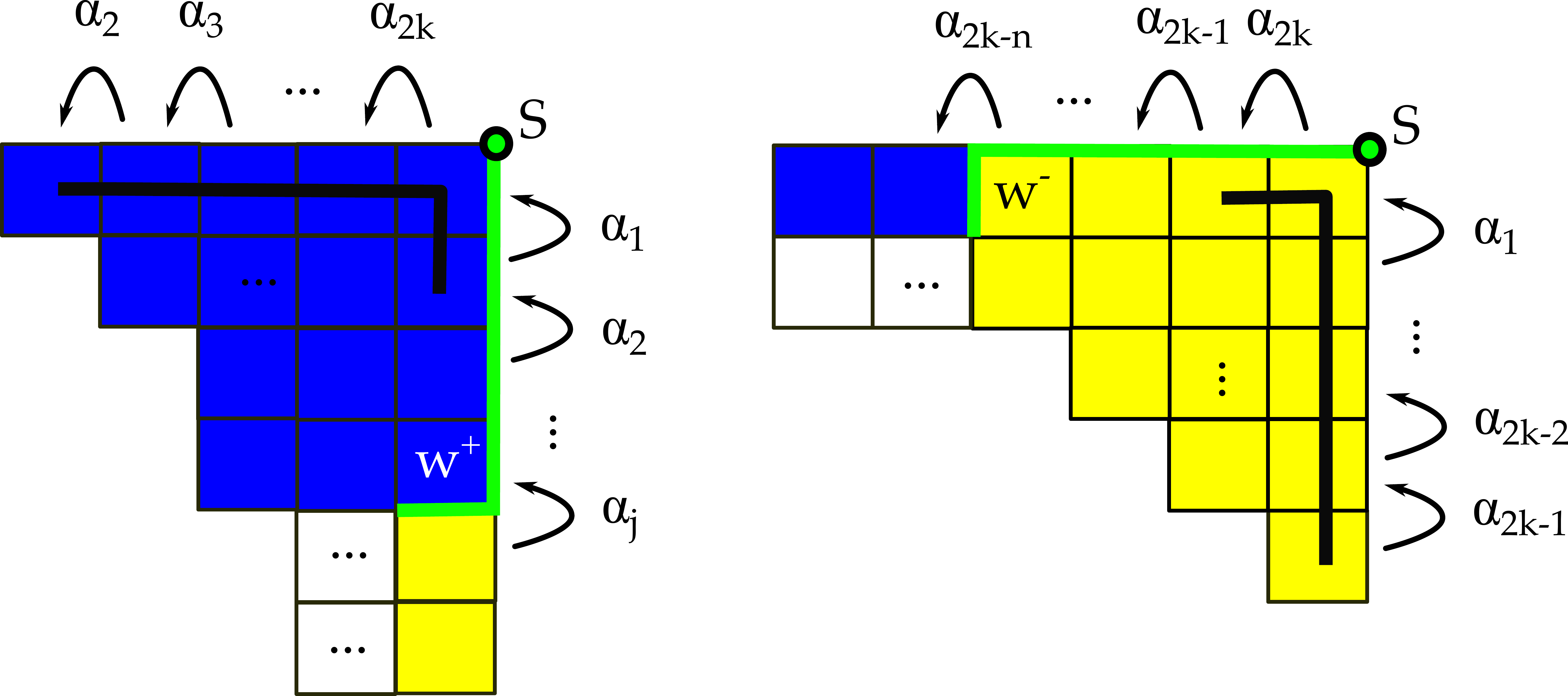}
    \caption{$F_0$ splitting rule: The $F_0$-hook is shown in bold-face black. Whenever a phase contains either of these, then $F_0$ splits. 
    \label{fig:F0Split}}
\end{figure}
\begin{figure}
    \centering
    \includegraphics[width=15cm]{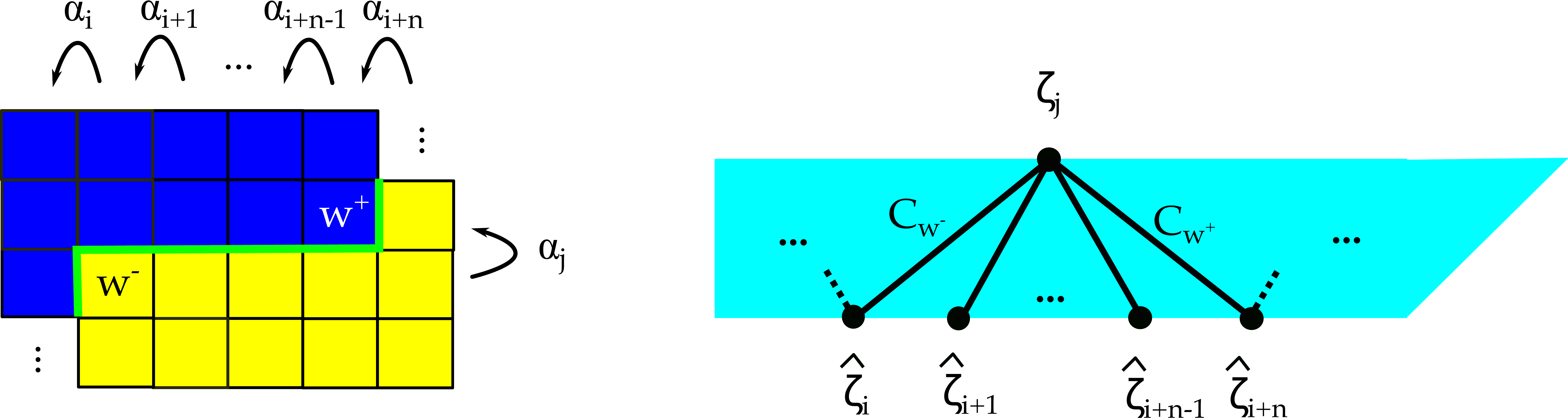}
    \caption{Splitting rule in terms of box graphs and corresponding 1-simplices in the toric fiber face triangulation. 
    \label{fig:ToricBoxProof1}}
\end{figure}

\begin{itemize}
\item $F_0$ splitting:\\
The affine node splits whenever the box graph contains the so-called ``$F_0$-hook", i.e. a 
path through the box graph, which crosses all $\alpha_i$-lines without changing the sign of the weights,
 I.e. whenever $F_0= -\sum F_i$ `fits' into the box graph.
Equivalently, this can be characterized by the anti-Dyck path starting at the point $S$ to move at least 
two boxes vertically down or at least two boxes horizontally to the left.
In figure \ref{fig:F0Split} we have shown such paths, with the black line indicating the $F_0$-hook, 
for which the splitting is 
\be\label{F0SplittingRule}
\ba
\hbox {LHS figure \ref{fig:F0Split}:}:& \qquad  F_0 \quad \rightarrow  \quad C_{j, 2k+1}^+ + F_{j-1}+ \cdots + F_2 + \tilde{F}_0 \,,\qquad \ \quad \tilde{F}_0= C^{-}_{1,2}  \cr
\hbox {RHS figure \ref{fig:F0Split}:}:& \qquad  F_0 \quad \rightarrow  \quad C_{1, n+1}^- + F_{2k-n-1} + \cdots + F_{2k-1} + \tilde{F}_0 \,,\qquad \tilde{F}_0 = C^{+}_{2k-1, 2k}\,.
\ea
\ee
Here $C^{\pm}$ are the curves corresponding to the extremal weight at the first 
corner of the anti-Dyck path that starts at $S$. $\tilde{F}_0$ is the affine node of the codimension two fiber, in particular it is not effective in the relative Mori cone. 

\item $F_{i\not=0}$ splitting:\\
For the splitting of the $F_i$
consider first a horizontal segment of the anti-Dyck path, along the horizontal line $H^j$ labeled by the simple root $\alpha_j$, 
bounded by the vertical lines $V^i$ and $V^{i+n}$, that correspond to adding $\alpha_i$ and $\alpha_{i+n}$, 
as shown on the left of figure \ref{fig:ToricBoxProof1}. 
Then the curve corresponding to $\alpha_j$ splits as follows 
\be\label{SplitProof}
F_j \qquad \rightarrow\qquad   C_{j-1, i+1}^-  + F_{i+1} + \cdots + F_{i+n-1} + C_{j, i+n}^+ \,.
\ee
Likewise a vertical segment of the anti-Dyck path between $H^i$ and $H^{i+j}$, along $V^n$, results in the splitting of the curve associated to $V^n$ into
\be\label{SplitV}
F_n \qquad \rightarrow\qquad  C_{i-1, n+1}^- + F_{i+1} + \cdots + F_{i+j-1} + C_{i+j, n}^+ \,.
\ee

\end{itemize}

{\bf Proof of Theorem \ref{thm:ToricBox}.}
{The idea of the proof is to systematically derive the splitting from the box graphs, and to map this to a triangulation of the fiber face. This is done inductively, by starting at the point $P$ of the box graphs, and determining the implied splitting from the anti-Dyck path. Roughly speaking one can think of each (horizontal or vertical) segment of the anti-Dyck path as specifying the 1-simplices that emanate from one of the vertices of the fiber face. }

To prove the theorem, note first that any  box graph defined by the rules (a)-(c) automatically is a consistent $\mathfrak{su}(2k+1)$ box graph, as the flow rules are satisfied and  the signs $\epsilon(L_k+ L_{k+1}) =+$ and $\epsilon(L_{k+1} + L_{k+2}) =-$ (which follow from (a) and (b)) guarantee, irrespective of the remaining signs in the region defined in (c), that the diagonal condition is satisfied. 

A fiber face triangulation can be specified by the splitting of the fiber components along the codimension two locus $b_1=0$, which introduces 1-simplices (lines in the fiber face diagram), connecting the sections $\zeta_j$ with the sections $\hat\zeta_i$, which share common components. 
We now show that a given box graph of the type specified in the theorem yields a fine triangulation of the fiber face (or top) shown in figure \ref{fig:ToricFanGen} and defined in (\ref{vzetadef}). 

\begin{figure}
    \centering
    \includegraphics[width=15cm]{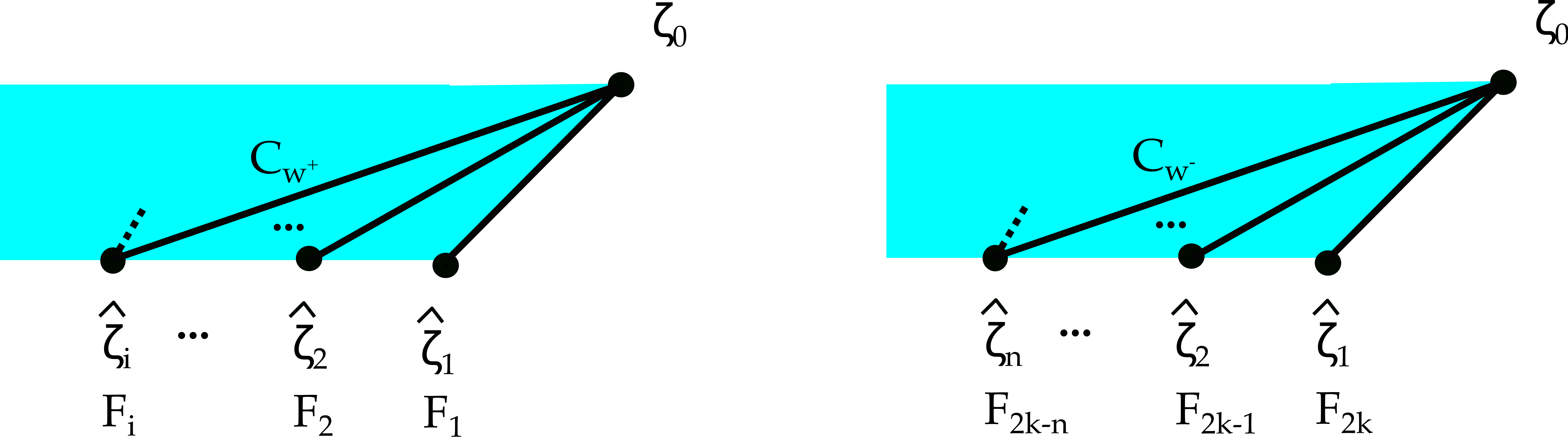}
    \caption{Splitting rule for $F_0$ as shown on the LHS/RHS of figure \ref{fig:F0Split} implies the 1-simplices as shown here on the LHS/RHS. Depending on the initial splitting of $F_0$, which is given by $\zeta_0=0$, the assignement of simple roots $\alpha_i$ and associated curves $F_i$ is determined in the diagram.   \label{fig:ToricF0Split}}
\end{figure}

The box graph defines an anti-Dyck path, which starts at $S$ and ends at $P$ (which is the intersection of the vertical line $V^{k+1}$ and horizontal line $H^k$). Starting at $S$, if the path proceeds horizontally/vertically, and turns at $V^{2k}$ ($H^1$), $F_0$ does not split and there is no additional 1-simplex attached to the node $\zeta_0$. Else, the path will turn at $V^n$ or $H^j$, in which case the curve $F_0$ splits as in (\ref{F0SplittingRule}). This implies the 1-simplices shown in figure \ref{fig:ToricF0Split}. 
Furthermore, this initial segment (and the thereby resulting splitting of $F_0$) determines the identification between $\zeta_i$, $\hat\zeta_i$ with the simple roots $\alpha_i$: 
\begin{itemize}
\item Dyck path segment starting at $S$ is vertical: then for $i=1, \cdots, k$
\be\label{Fzeta1}
F_i \leftrightarrow \{\hat\zeta_i =0\} \qquad 
F_{k+i} \leftrightarrow \{\zeta_{2k-i} =0 \}\,.
\ee
\item Dyck path segment starting at $S$ horizontal: then for $i=1, \cdots, k$
\be\label{Fzeta2}
F_i \leftrightarrow \{\zeta_i =0\} \qquad 
F_{k+i} \leftrightarrow \{\hat\zeta_{2k-i} =0 \}\,.
\ee
\end{itemize}

The remaining 1-simplices for the triangulation are introduced by considering alternatingly the horizontal and vertical segments of the path. 
Consider first a horizontal segment along the line labeled by the simple root $\alpha_j$\footnote{Without loss of generality, we consider the identification (\ref{Fzeta2}), which can be easily mapped to the identification of the sections with the roots should the splitting of $F_0$ imply the alternative identification (\ref{Fzeta1}).  }, bounded by the vertical lines that correspond to adding $\alpha_i$ and $\alpha_{i+n}$, as shown on the left hand side of figure 
\ref{fig:ToricBoxProof1}. The anti-Dyck paths for fiber face triangulations are specified as starting at $S$ and ending at the point $P$,  therefore $j< i$. 
The splitting rules imply the following splitting along $b_1=0$
\be\label{SplitProof2}
F_j \qquad \rightarrow\qquad   C_{j-1, i+1}^-  + F_{i+1} + \cdots + F_{i+n-1} + C_{j, i+n}^+ \,.
\ee
Monotony of the anti-Dyck path implies that the path will not intersect the corresponding vertical lines again, and thus $F_{i+1}, \cdots , F_{i+n-1}$ are irreducible along $b_1=0$. 
The remaining components are the curves from the endpoints of this segment (which are the extremal generators of the cone of curves, and can be flopped).  The curves $F_i$ and $F_{i+n}$ are also reducible, with a component $C_{j-1, i+1}^-$ and $C_{j, i+n}^+$, with the remaining components being determined by the next (vertical segment) of the anti-Dyck path. 
The splitting (\ref{SplitProof2}) implies that there are 1-simplices in the fiber face triangulation, which connect 
$\zeta_j$  with each of the vertices $\hat{\zeta}_{i},\cdots, \hat{\zeta}_{i+n}$.
Furthermore irreducibility of $F_{i+1}, \cdots , F_{i+n-1}$ implies that these are the only 1-simplices that end on $\hat{\zeta}_{i+1}, \cdots, \hat{\zeta}_{i+n-1}$, which are shown in the corresponding triangulation on the RHS of figure \ref{fig:ToricBoxProof1}. 
Monotony of the path implies that there is no 
crossing of 1-simplices, which would render the triangulation inconsistent.  

Likewise a vertical segment, (\ref{SplitV}) implies the 1-simplices connecting 
$\hat{\zeta}_n$ with $\zeta_i, \cdots, \zeta_{i+j}$, where $\zeta_{i+1}, \cdots, \zeta_{i+j-1}$ are irreducible (which implies again due to the monotony of the path that these will only have 1-simplices connecting them to $\hat{\zeta}_n$), and $\zeta_{i}$ and $\zeta_{i+j}$ split along the adjacent horizontal lines as described above. Iterating this process results in a fine triangulation of the toric top.

Let us conclude with a simple counting argument of these box graphs. We can characterize these by monotonous staircase paths, starting at $S$ and ending at $P$, which form a $k\times k$ grid. Note that the trace condition is already automatically satisfied for any sign assignment in the box graphs of the type in figure \ref{fig:ToricBox}, and thus, the paths are only required to satisfy the flow rules, which translates into monotony. The number of such paths is 
\be
\# \hbox{Box graphs of the type in figure \ref{fig:ToricBox}} =  {2k \choose k} \,,
\ee
which agrees with the result in from the fiber face triangulations (\ref{NumberToric}).

\section{Secondary Fiber Faces and Complete Intersections}
\label{sec:CompleteInt}

In the last section, we have shown how to construct all resolutions of $\mathfrak{su}(2k+1)$ fibrations
for which the fiber is embedded as a toric hypersurface, and the starting point was a 
 singular Weierstrass or Tate model. 
In terms of the box graphs this corresponded to anti-Dyck paths starting at $S$ and ending at $P$ in figure \ref{fig:ToricBox} (or $P_1$ in figure \ref{fig:HatZetaDelta}). 
In this section, we show how resolutions corresponding to paths ending at $P_2$ in figure \ref{fig:HatZetaDelta} can be obtained from fibers embedded as complete intersections. 
 They can be reached from the phases already considered
via flops, and thus a straight-forward identification of their box graphs is possible. 
However, these 
generalized, so-called secondary fiber face  triangulations, only realize a sub-class of the remaining phases. We discuss in section \ref{sec:Layers} how this decomposition of box graphs in terms of paths with varying endpoints can be emulated by embedding the fiber in an increasingly complex way.

\subsection{Blowdowns and Elementary Flops}
\label{sect:elemtflopssu2kp1}

Phases that are beyond those corresponding to fiber face triangulations can be reached by chains of elementary flops, which map out of the class of box graphs in figure \ref{fig:ToricBox}. 
Starting with the resolutions discussed in the last section, this will lead to geometries realized as complete intersections. Before discussing the general class of such resolutions, which will be done in section \ref{sec:RedFibFace}, we first consider elementary flops, obtained by blowdowns of toric divisors. We blow down a single coordinate from the ambient space and construct a new resolution, which cannot 
be realized as a hypersurface. The emerging structure is most easily seen by writing the 
resolved Tate model \eqref{TateResolved} in the two forms
\be\label{eq:su2kp1factforms1}
T_{2k+1}\qquad \Leftrightarrow \qquad y \hat{y} = \prod_{i=1}^k \zeta_i \   P 
\ee
and
\be\label{eq:su2kp1factforms2}
T_{2k+1}\qquad \Leftrightarrow \qquad x W =  \prod_{i=1}^{k-1} \hat\zeta_i \   S \,,
\ee
where we defined
\be\ba
 \hat{y}&=  y B(\hat\zeta)\hat\zeta_k + b_1x  + b_3 \zeta_0^kB(\zeta\hat\zeta)C(\zeta \hat\zeta)  \cr
 P &=  x^3A(\zeta \hat\zeta)\zeta_k^{k-1}\hat{\zeta}_k^{k-1} 
 		+ b_2x^2\zeta_0	
			+ b_4x\zeta_0^{k+1}B(\zeta \hat\zeta)C(\zeta\hat\zeta) 
			+ b_6\zeta_0^{2k+1}B(\zeta^2\hat\zeta^2)C(\zeta^2\hat\zeta^2) \cr
W &= 	-b_1 y + x^2 B(\zeta)A(\zeta \hat\zeta)\zeta_k^k\hat{\zeta}_k^{k-1}+ b_2x \zeta_0B(\zeta)\zeta_k 	
		+ b_4\zeta_0^{k+1}B(\zeta^2 \hat\zeta)C(\zeta\hat\zeta)\zeta_k \cr
S &=y^2 \hat\zeta_k + b_3y\zeta_0^kB(\zeta)C(\zeta \hat\zeta) 
			- b_6\zeta_0^{2k+1}B(\zeta^3\hat\zeta)C(\zeta^2\hat\zeta^2)\zeta_k  \,.
\ea			
\ee
The relevance of these forms is that they anticipate the conifold-like singularities, which may 
arise once one of the $\zeta_i$ or $\hat{\zeta}_i$ is blown down. Of course, as long as we 
use a fine triangulation of the top, we have resolved all singularities 
in codimensions one, two and three over the base and the factorized forms of 
\eqref{eq:su2kp1factforms1} and \eqref{eq:su2kp1factforms2} can never lead to a 
singularity. At a technical level, this happens because the coordinate $y$ may never 
vanish simultaneously with any one of the coordinates $\zeta_i$,  and the 
coordinate $x$ may never simultaneously vanish with any of the $\hat{\zeta}_i$ for $i = 1, \cdots, k-1$.

In toric language, a blowdown corresponds to a projection 
$\pi: \Sigma \rightarrow \Sigma'$ which maps every cone of $\Sigma$ (in)to a cone in $\Sigma'$. In other
words we can think of $\Sigma'$ as arising by appropriately gluing together cones of $\Sigma$. 
Blowing down a coordinate $z$ hence means that we have to glue cones such that the corresponding ray generated by
$v_z$ is not present in $\Sigma'$. Conversely, we may get back to $\Sigma$ by blowing up $\Sigma'$
via reintroducing $v_z$. 

In the case at hand, we can only have a situation in which $y$ can simultaneously vanish with
$\zeta_i$ if we blow down $v_{\hat{\zeta}_i}$: as $v_{\hat{\zeta}_i} = v_{\zeta_i} + v_y$, it follows that
$v_{\hat{\zeta}_i}$ sits in the middle between $v_{\zeta_i}$ and $v_y$ (a cone spanned by $v_{\zeta_i}$ and $v_y$ 
contains $v_{\hat{\zeta}_i}$).  Similarly, $x$ can only simultaneously vanish with $\hat{\zeta}_i$ (for any $i = 1 \cdots k-1$)
if we blow down $\zeta_{i+1}$ as $v_x + v_{\hat{\zeta}_i} = v_{\zeta_{i+1}}$.

We will use the notation $(z_1 , \cdots,z_n | z_e)$ to indicate a blowdown which can be undone by
a (weighted) blowup at $z_1 = \cdots = z_n = 0$ introducing the new coordinate $z_e$. 
We now discuss the various possible blowdowns and flops in turn.

\begin{figure}
    \centering
    \includegraphics[width=15cm]{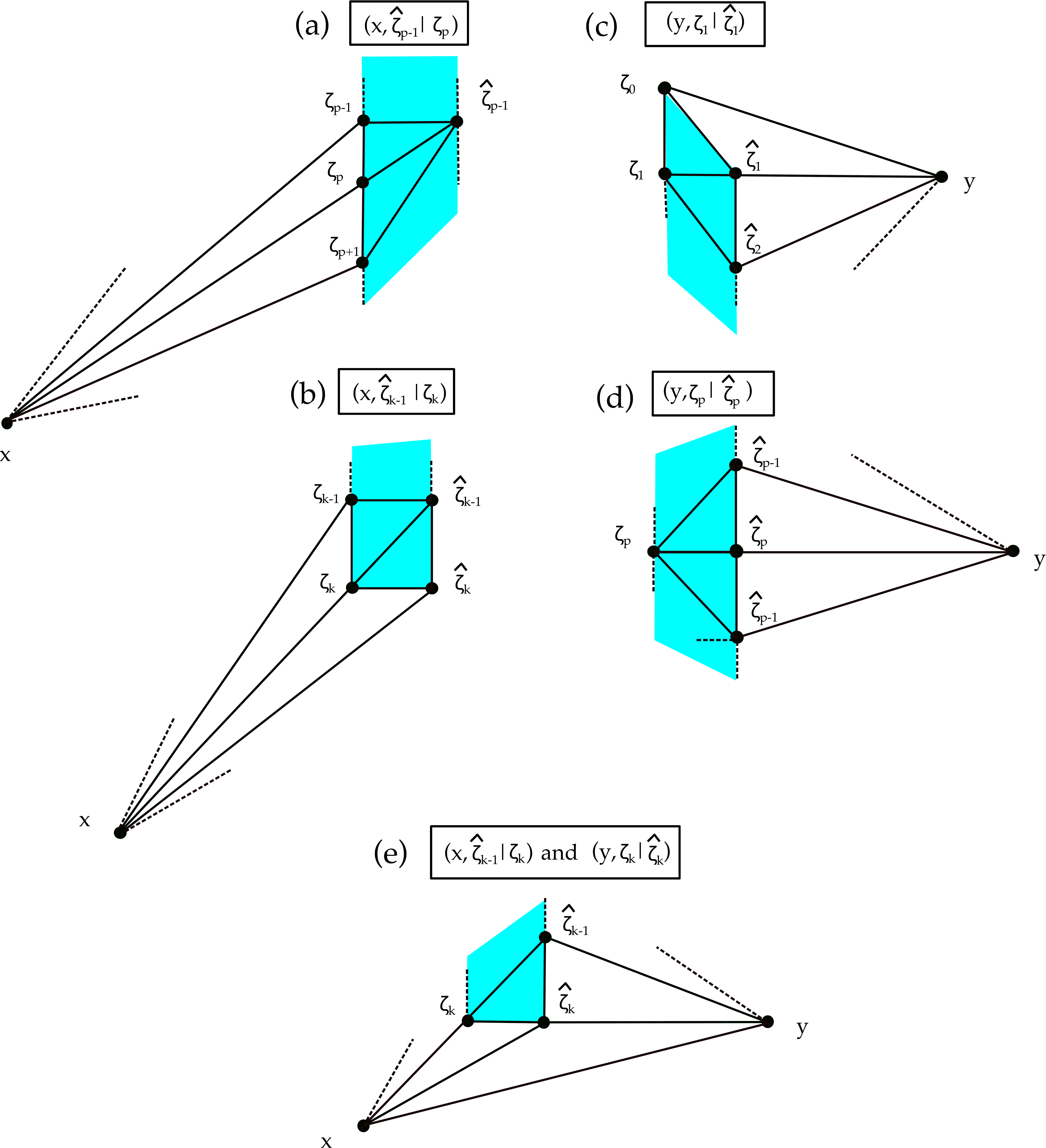}
    \caption{Toric triangulations for blowdowns and flops for $I_n$ fibers. 
    The notation is $(a, b|c)$, that we blow down $c$, which connects $a$ 
    and $b$.\label{fig:ToricTri}}
\end{figure}

\subsubsection{Flops based on $(y, \zeta_1| \hat{\zeta}_1)$ Blowdowns}

Let us start by investigating blowdows of $\hat{\zeta}_1$. For such a blowdown to be possible,
the triangulation of the fiber face in the vicinity of $v_{\hat{\zeta}_1}$ must be as shown in figure
\ref{fig:ToricTri} (c). After the blowdown the four cones $\langle v_{\zeta_0},v_{y},v_{\hat{\zeta}_1}\rangle, 
\langle v_{\zeta_0},v_{\zeta_1},v_{\hat{\zeta}_1}\rangle,\langle v_{\hat{\zeta}_2},v_{y},v_{\hat{\zeta}_1}\rangle,
\langle v_{\hat{\zeta}_2},v_{y},v_{\hat{\zeta}_1}\rangle$ are glued to 
$\langle v_{\zeta_0},v_{y},v_{\zeta_1}\rangle$  and $\langle v_{\zeta_1},v_{y},v_{\hat{\zeta}_2}\rangle$.
Correspondingly, there is now a singularity at $y = \hat{y} = \zeta_1 = P = 0$ which implies $b_1=0$. We have hence blown down 
a fiber component over the $\Lambda^2V$ matter curve.

We may perform a different resolution by blowing up along $y = P = 0$. To achieve this, we first introduce a new coordinate $\pi$
and a new equation $\pi = P$. After this we may perform a small resolution $(y,\pi;\delta)$ resulting in
\be\label{eq:su2kp1ysideblowup1}
\ba
 y\hat{y} &= \prod_{i=1}^k \zeta_i \   \pi \cr
 \delta \pi &= P \,.
\ea 
\ee
Let us now see how this has altered the splitting of fiber components over the $\Lambda^2V$ matter curve at $b_1 = 0$.
Note that in the phase before the blowdown, $D_{\alpha_{2k}}$ necessarily splits into two 
components, see figure \ref{fig:ToricTri} (c) and use the general rule formulated in theorem \eqref{thm:splittingrules}. 
After the resolution, the association of fiber components has changed, we now have
\be
\ba
 D_{\alpha_{1}} :\qquad & \zeta_1 = \delta y B(\hat\zeta)\hat\zeta_k + b_1x = 0 \\
 D_{\alpha_{2k}}:\qquad & \zeta_1 = y = 0 \, .
\ea
\ee
Hence $D_{\alpha_{2k}} $ is now irreducible over any locus in the base. Over $b_1=0$,
the fiber component $D_{\alpha_1}$ loses one component (the coordinate $\hat{\zeta_1}$ no longer appear in 
$B(\hat\zeta)$ after the blowdown) but gains the two components at $y=0$ and at $\delta = 0$. Hence we see that the 
total number of fiber components over $b_1 = 0$ stays constant: $D_{\alpha_{2k}}$ loses a component (so that 
it become irreducible over $b_1=0$) whereas $D_{\alpha_1}$ gains a component. 

As such a flop is possible whenever the triangulation in the vicinity of $\zeta_1$ is as shown in figure \ref{fig:ToricTri} (c),
the number of which is given by 
\be
T_{k,k-1} = \binom{2k-3}{k-1} \,.
\ee

\subsubsection{The Blowdowns $(y, \zeta_p| \hat{\zeta}_p)$ for $0 < p < k$}

Similarly, we may blow down any of the coordinates $\hat{\zeta}_p$ if we are in a phase with triangulation shown
in figure \ref{fig:ToricTri} (d). Note that this means that the fiber component associated with $\hat{\zeta}_p$ stays
irreducible over $b_1 = 0$. After the blowdown, we expect a singularity at $y = \hat{y} = \zeta_p = P = 0$.
Again $y= \hat{y} = 0$ implies $b_1=0$, but now $P = \hat{\zeta}_p = 0$ implies $ b_2 x^2 b\zeta_0 = 0$.
As both $x$ and $\zeta_0$ cannot vanish at the same time as $y$ and $\zeta_p$, this implies $b_2 = 0$ and
we conclude that this blowdown can never affect the splitting of fiber components over any of the matter curves.
One also easily finds that performing a flop as in \eqref{eq:su2kp1ysideblowup1} does not alter the phase.
It is not hard to see that the ambient space stays smooth after the blowdown as well.

\subsubsection{Flops based on $(y, \zeta_k| \hat{\zeta}_k)$ Blowdowns}

The blowdown $(y, \zeta_p| \hat{\zeta}_p)$, which can be performed when the triangulation is as shown in figure 
\ref{fig:ToricTri} (e), leads to a singularity at 
\be
y = \zeta_k = b_1x  + b_3 \zeta_0^kB(\zeta\hat\zeta)C(\zeta \hat\zeta) =
b_2x^2\zeta_0 + b_4x\zeta_0^{k+1}B(\zeta \hat\zeta)C(\zeta\hat\zeta) + b_6\zeta_0^{2k+1}B(\zeta^2\hat\zeta^2)C(\zeta^2\hat\zeta^2) = 0 \,.
\ee
These equations only have a common solution in the homogeneous coordinates $[x:B(\zeta \hat\zeta)C(\zeta\hat\zeta)]$ 
if we are over the matter curve of the fundamental representation, $P = b_2 b_3^2 - b_1 b_3 b_4 + b_1^2 b_6=0$. We hence expect 
the flop \eqref{eq:su2kp1ysideblowup1} to have no effect on the splitting over the ${\Lambda^2}{V}$ matter, but only to affect the matter
in the fundamental representation. After the blowdown, the divisor $D_{\zeta_k}$ becomes reducible and contains the fiber components
\be
\ba
D_{\alpha_{k+1}}: & \quad \zeta_k = y = 0 \cr
D_{\alpha_{k}}:& \quad \zeta_k = \hat{y} = 0 \,.
\ea
\ee
The fiber component corresponding to $D_{\alpha_{k+1}}$ stays irreducible over $P=0$ in the flopped phase \eqref{eq:su2kp1ysideblowup1},
whereas $D_{\alpha_{k}}$ splits into two components there.

There are 
\be
T_{k,k-1} = \binom{2k-3}{k-1} 
\ee
cases, in which such a flop is possible.

\subsubsection{Flops based on $(x, \hat{\zeta}_{k-1}| {\zeta}_k)$ Blowdowns}

This blowdown is possible if the triangulation is as shown in figure \ref{fig:ToricTri} (b). Setting $x = \hat{\zeta}_{k-1} = W= S = 0 $ implies
$b_1 = 0$ and $y^2 \hat\zeta_k + b_3y\zeta_0^kB(\zeta)C(\zeta \hat\zeta)= 0$, so that there is now a singularity at this locus.
The relevant exceptional divisors after the flop become
\be\ba
 D_{\alpha_{k}} &: \quad \hat{\zeta}_{k-1} = x = 0 \\
 D_{\alpha_{k+2}}&:\quad  \hat{\zeta}_{k-1} = W = 0 \, .
\ea\ee
Note that now $ D_{\alpha_{k}} $, which was splitting into two components over $b_1 = 0$ has become irreducible. $ D_{\alpha_{k+2}}$ has
gained this component: over $b_1=0$, $W=\hat{\zeta}_{k-1}=S=0$ implies that 
\be\ba
 b_2 \delta x \prod_{i=1}^{k-1} \zeta_i = 0 \\
 y^2 \hat\zeta_k + b_3y\zeta_0^kB(\zeta)C(\zeta \hat\zeta) =0
\ea\ee
While the component corresponding corresponding to a common solution of $\hat{\zeta}_{k-1} $ with $\zeta_k$ (this coordinate no longer exists) 
is lost, $D_{\alpha_{k+2}}$ has gained two more components at $x=0$ and $\delta=0$ over $b_1=0$. Such a flop can be performed in 
\be
T_{k,k-1} = \binom{2k-3}{k-1}
\ee
cases.

\subsubsection{The Blowdowns $(x, \hat{\zeta}_{p-1}| {\zeta}_p)$ for $1 < p < k$}

This type of blowdown is possible if the triangulation in the vicinity of $\zeta_p$ is as shown in figure \ref{fig:ToricTri} (a).
When we blow down $\zeta_p$, we expect a singularity over $x = \hat{\zeta}_{p-1} = W = S = 0$.
Setting at $x = \hat{\zeta}_{p-1}= W=S=0$ implies $b_1 = 0$ and $y^2 \hat{\zeta}_k=0$. As $v_{\hat{\zeta}_{p-1}}$ never shares 
a cone with $v_{\hat{\zeta}_{k}}$ and there is also never a common cone for $v_x,v_y$ and $v_{\hat{\zeta}_{p-1}}$, 
we conclude that no singularity arises in this blowdown, and the ambient space stays smooth. Hence any blowdown $(x, \hat{\zeta}_{p-1}| {\zeta}_p)$, $1 < p < k$ can never lead to a flop/change of phase.


\subsection{Complete Intersections and Secondary Fiber Faces} 
\label{sec:RedFibFace}
\begin{figure}
    \centering
    \includegraphics[width=8cm]{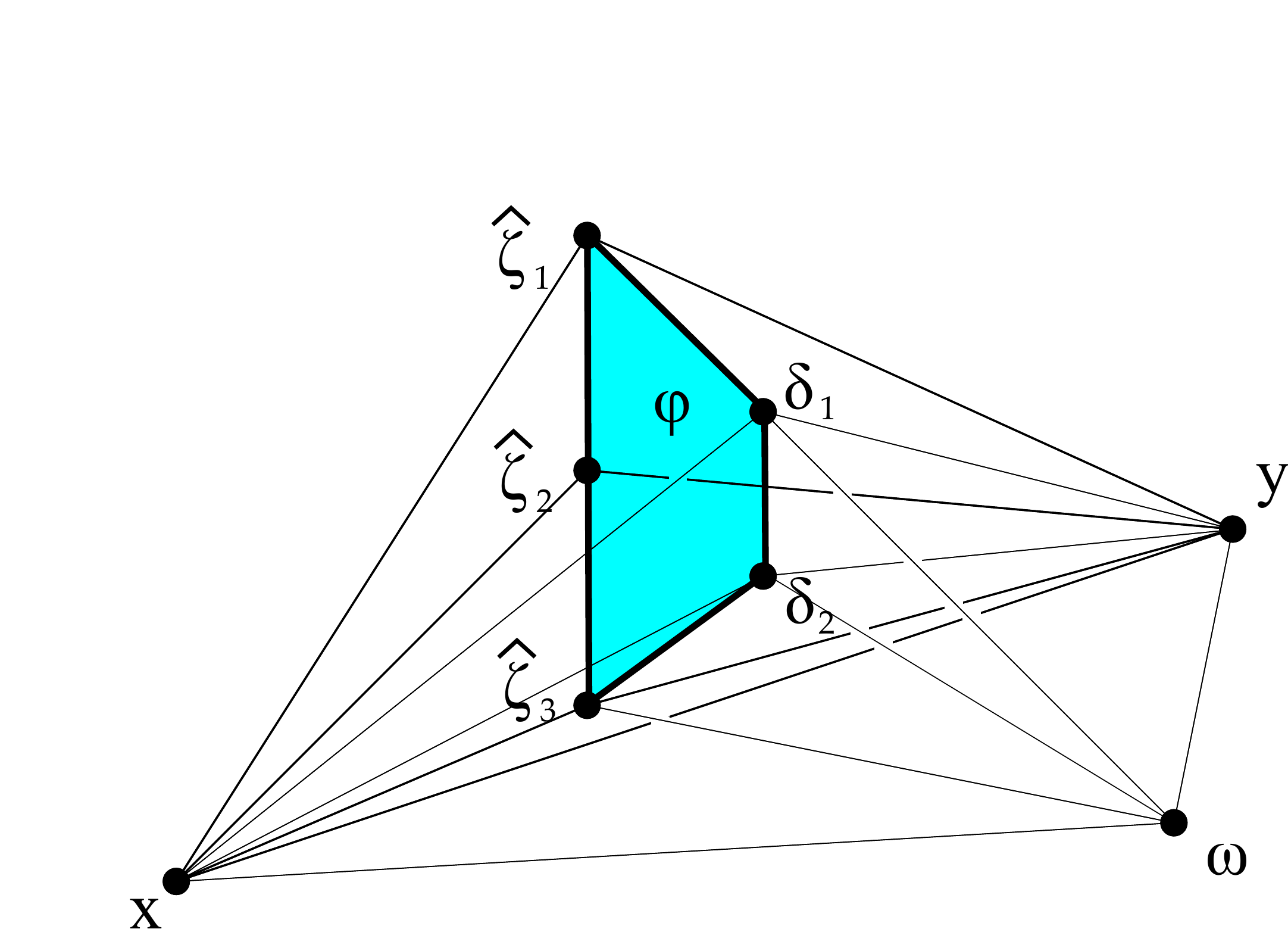}
    \caption{The secondary fiber face  $\varphi$ for the case of  $\mathfrak{su}(7)$.   \label{fig:SU7RedFibFace}}
\end{figure}

\begin{figure}
    \centering
    \includegraphics[width=5cm]{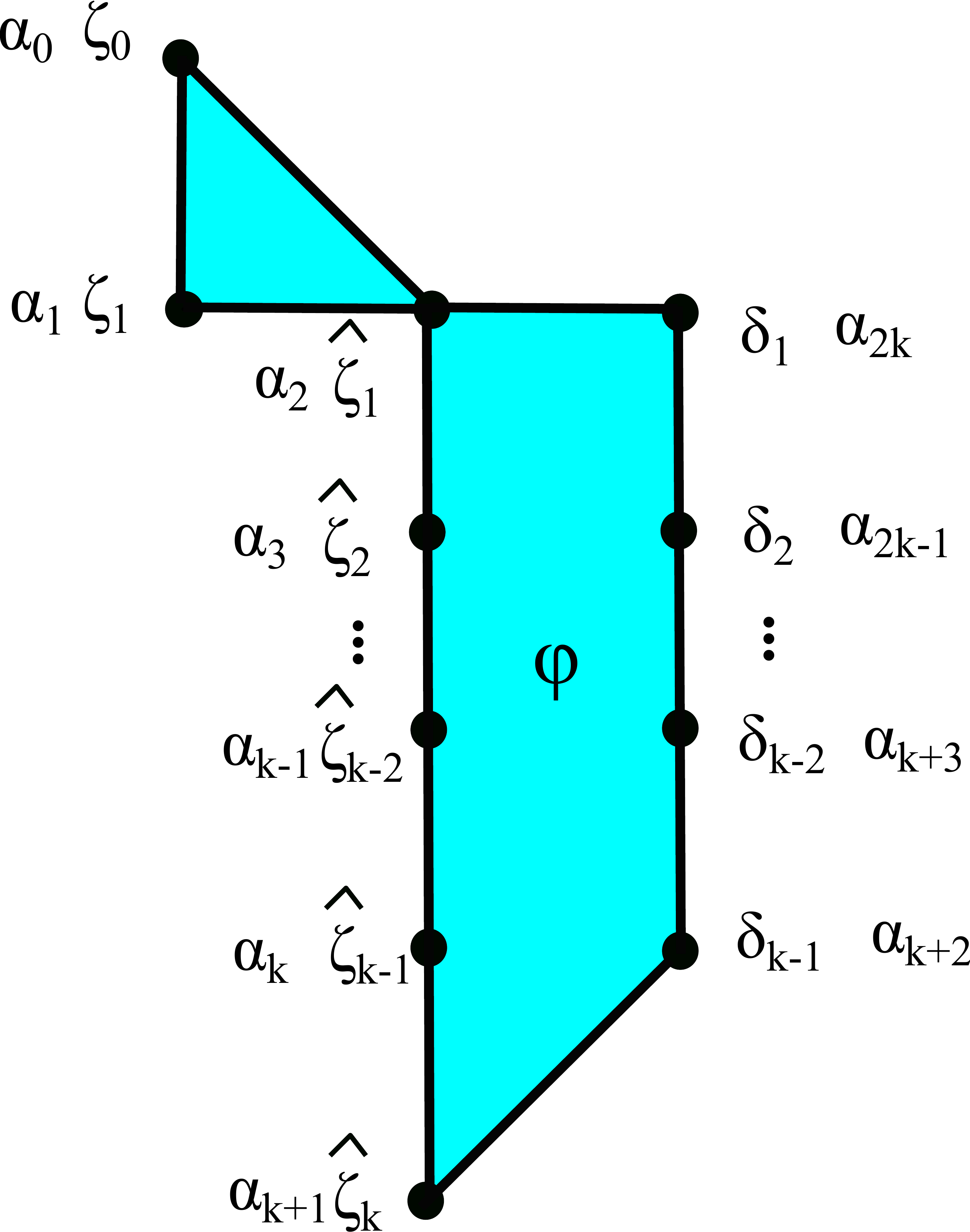}
    \caption{Secondary fiber face  $\varphi$ for the equations (\ref{Ell2Eq}). Triangulations $\rho$ of the  fiber face $\varphi$ correspond to resolutions that are characterized by the box graphs shown in figure \ref{fig:HatZetaDelta}. The labels indicate the simple root $\alpha_i$, as well as the section $\hat\zeta$ or $\delta$, associated to each node.  \label{fig:HatZetaDeltaFF}}
\end{figure}

In this section, we generalize the construction above by blowing down more than just a single toric divisor. It turns out that 
blowing down all coordinates $\zeta_i$ for $i= 2, \cdots, k$ allows us to access a new class of resolutions, which go beyond the standard toric tops, and originate from box graphs which do not fall into the toric class figure \ref{fig:ToricBox}. 
In the following, we will  work with the form (\ref{eq:su2kp1factforms2}), i.e. 
\be\label{AltForm}
\ba
x \omega &=  \prod_{i=1}^{k-1} \hat\zeta_i \   S \cr
\omega &=  -b_1 y + x^2 \zeta_1 A( \hat\zeta) \hat{\zeta}_k^{k-1}
		+ b_2x \zeta_0	\zeta_1
		+ b_4\zeta_0^{k+1} \zeta_1^{k}B(\hat\zeta) C(\hat\zeta) \,,
\ea
\ee
where $\omega$ is now a new coordinate. Torically, we enlarge the ambient space of the fan by one dimension, and associate the ray generated by $(0,0,0,1)$ with $\omega$ and lift all other cones of the fan. We give concrete description of this for $\mathfrak{su}(7)$ in (\ref{eq:gen356}). 
When all of the $\zeta_i$ (except $\zeta_0$ and $\zeta_1$) are blown-down, in particular, the corresponding cones are glued together, and then the resulting singularity is resolved by the set of resolutions 
\be
{(\hat\zeta_i, \omega;\delta_i) \,,\qquad i= 1, \cdots, k-1} \,,
\ee
we obtain
\be\label{Ell2Eq}
\ba
x\omega &=\left( \prod_{i=1}^{k-1} \hat\zeta_i \right) 
			\left(y^2 \hat\zeta_k + b_3y\zeta_0^k \zeta_1^{k-1} C(\hat\zeta \delta) 
			- b_6\zeta_0^{2k+1} \zeta_1^{2k-1}B(\hat\zeta\delta)C(\hat\zeta^2\delta^2)\right)   \cr
\left(\prod_{i=1}^{k-1} \delta_i \right) \omega &= 
		-b_1 y + x^2 \zeta_1 A( \hat\zeta\delta) \hat{\zeta}_k^{k-1}
		+ b_2x \zeta_0	\zeta_1
		+ b_4\zeta_0^{k+1} \zeta_1^{k}B(\hat\zeta\delta) C(\hat\zeta\delta) \,.
\ea
\ee
Alternative resolutions of the form (\ref{AltForm}) are obtained by similar blowups, introducing the same coordinates $\delta_i$, which however differ in the SR ideal, but not in the defining equation -- much like in the case of the Tate resolution discussed earlier. 
As before, distinct resolutions are characterized in terms of triangulations of a face spanned by
$\{\hat\zeta_1, \cdots, \hat\zeta_k, \delta_{k-1}, \cdots, \delta_1\}$, which we 
refer to  as the  {\it secondary fiber face } $\varphi$. This is shown in figure \ref{fig:SU7RedFibFace} for $\mathfrak{su}(7)$, including the remaining coordinates $x, y, \omega$, as well as figure \ref{fig:HatZetaDeltaFF}, which shows the secondary fiber face for $\mathfrak{su}(2k+1)$.

A triangulation $\rho$ of the secondary fiber face  $\varphi$ gives rise to a fan with cones as summarized in the following:
\be
\ba
&\langle x, y, \omega, \hat\zeta_k\rangle \,,\quad 
\langle x, \omega, \hat\zeta_k, \delta_{k-1}\rangle \,,\quad 
\langle y, \omega, \hat\zeta_k, \delta_{k-1}\rangle\cr
&\langle \delta_i, \delta_{i+1}, \omega, x \rangle\,,\quad
\langle \delta_i, \delta_{i+1}, \omega, y\rangle\,,\cr
&\langle \rho, x\rangle \,,\quad
\langle \rho ,y\rangle \,,\cr
& \langle  \zeta_0, \zeta_1, x, \omega \rangle \,,\quad 
\langle  \zeta_0, \zeta_1, \hat\zeta_1, \delta_1 \rangle \,,\quad 
\langle  \zeta_0, \zeta_1, \omega, \delta_1  \rangle \,,\quad 
\langle  \zeta_0, \hat\zeta_1, y, \delta_1 \rangle \cr
&\langle  \zeta_0, \omega, y, \delta_1 \rangle \,,\quad 
\langle  \zeta_1, \hat\zeta_1, x, \delta_1 \rangle \,,\quad 
\langle  \zeta_1, x, \omega, \delta_1 \rangle \,.
\ea
\ee

\begin{figure}
    \centering
    \includegraphics[width=6cm]{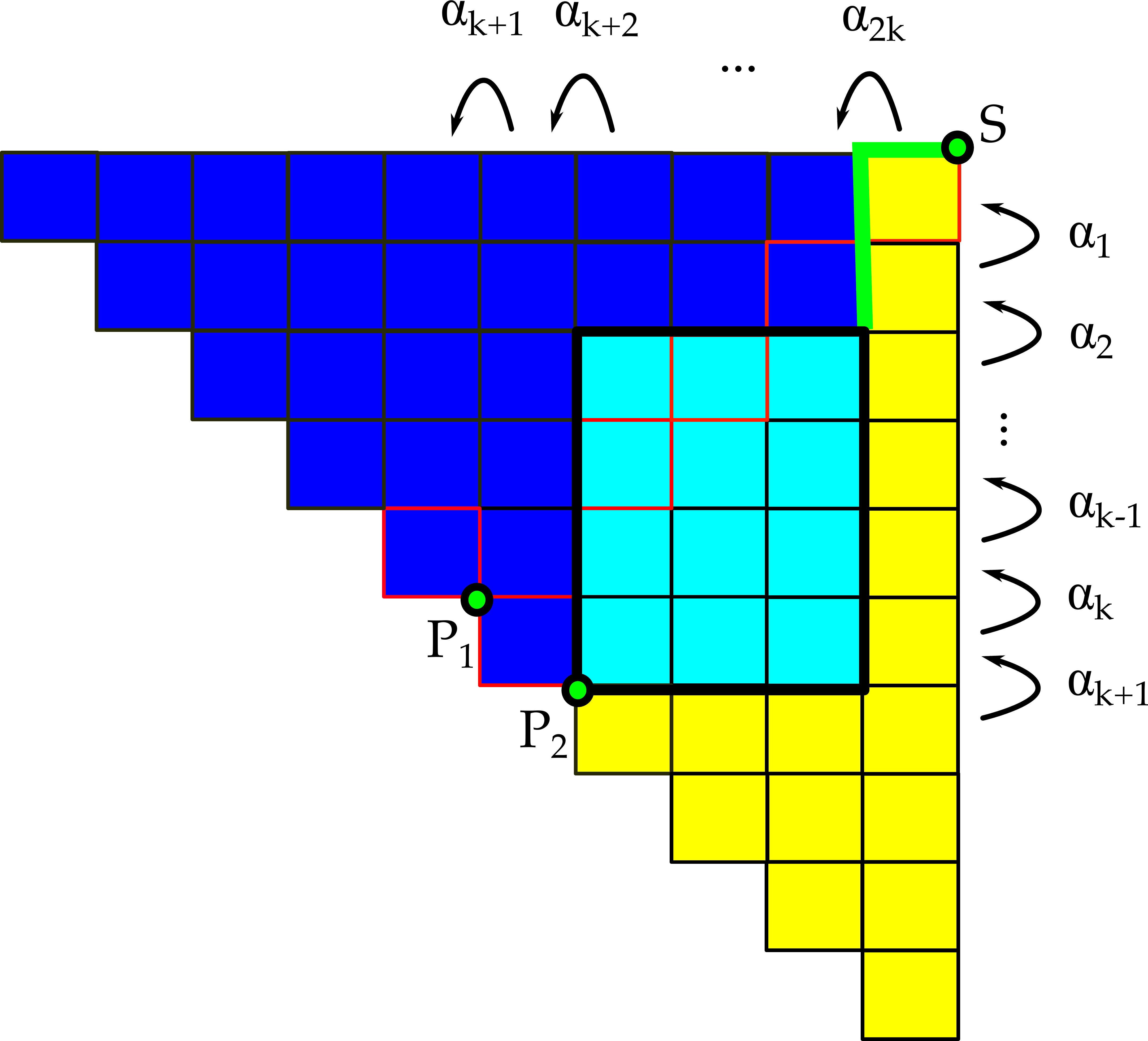}
    \caption{Box graphs realizing Coulomb phases, that correspond to the resolutions in  (\ref{Ell2Eq}). The fixed +/- sign assignments are shown in blue/yellow, whereas the possible triangulations are obtained by consistent sign assignments for  the  turquoise region, which satisfy the flow rules as well as the trace condition for $\mathfrak{su}(2k+1)$. The resulting geometries have a characterization in terms of triangulations $\rho$ of the secondary fiber face  $\varphi$ in figure \ref{fig:HatZetaDeltaFF}. \label{fig:HatZetaDelta}}
\end{figure}

To determine the fibers, first consider codimension one, where the $I_{2k+1}$ fiber components are identified with the sections as follows:
\be\label{eq:redfibfacefibercomp}
\begin{array}{c|c|l}
\hbox{Simple root} & \hbox{Section} & \hbox{Equations in $Y_4$} \cr\hline
 	\alpha_0&  \zeta_0  	&   x\omega = y^2  \prod_{i=1}^{k} \hat\zeta_i  \,,\quad 
						  \omega \prod_{i=1}^{k-1} \delta_i   = -b_1 y + x^2\zeta_1 A( \hat\zeta\delta) \hat{\zeta}_k^{k-1}	
						\cr
	\alpha_1&\zeta_1 & 	    x\omega = y^2  \prod_{i=1}^{k} \hat\zeta_i   \,,\quad 
						  \omega \prod_{i=1}^{k-1} \delta_i   = -b_1 y 
					 	\cr
  \alpha_2 & \hat\zeta_1 & x=0 \,,\quad 
 		\omega\prod_{i=1}^{k-1} \delta_i = 
		-b_1 y \cr
  \alpha_{3,\cdots, k} &  \hat\zeta_{2, \cdots , k-1 }  &  x =0 \,,\quad 
  		\omega\prod_{i=1}^{k-1} \delta_i  = 	-b_1 y 
			\cr 
 	\alpha_{k+1}  & \hat\zeta_{k}  & 
	x\omega =\left( \prod_{i=1}^{k-1} \hat\zeta_i \right) \zeta_1^{k-1} \zeta_0^k  C(\hat\zeta \delta) 
			\left( b_3y
			- b_6\zeta_0^{k+1} \zeta_1^{k}B(\hat\zeta\delta)C(\hat\zeta\delta)\right)\cr
	&&\omega\prod_{i=1}^{k-1} \delta_i  = 
		-b_1 y + b_2x \zeta_0	\zeta_1
		+ b_4\zeta_0^{k+1} \zeta_1^{k}B(\hat\zeta\delta) C(\hat\zeta\delta)
	\cr
	\alpha_{k+2} & \delta_{k-1} & 
			x\omega =\left( \prod_{i=1}^{k-1} \hat\zeta_i \right) 
			\left(y^2 \hat\zeta_k + b_3y\zeta_0^k \zeta_1^{k-1} C(\hat\zeta \delta) \right)   \cr
 			&&0= 	-b_1 y + b_2x \zeta_0	\zeta_1 
	  \cr
 	\alpha_{k+3,\cdots ,2k-1}& \delta_{k-2, \cdots, 2} &  
	x\omega = y^2 \prod_{i=1}^{k} \hat\zeta_i   \,,\quad 
		0=		-b_1 y 		+ b_2x \zeta_0	\zeta_1
	\cr
	\alpha_{2k} & \delta_1 & 
	x\omega = y^2 \prod_{i=1}^{k} \hat\zeta_i   \,,\quad 
	0= 	-b_1 y + x^2 \zeta_1 A( \hat\zeta\delta) \hat{\zeta}_k^{k-1}
		+ b_2x \zeta_0	\zeta_1
	\end{array}
\ee
This identification in codimension one is independent of the triangulation of the fiber face. 
With this data, one may again work out how the various fiber components split over the $\Lambda^2V$ matter curve.
With the notation
\be
X_\Delta (\zeta, \xi) = \#\hbox{ connections between $\zeta$ and $\xi$ in the triangulation}\,,
\ee
we can summarize the splitting rules along $b_1=0$ as follows:
\be
\begin{array}{l|c|c}
\hbox{Section} & \hbox{Equation along $b_1=0$} & \hbox{Number of components}\cr\hline
\zeta_0 & (\ref{Ell2Eq})& 1 \cr\hline
\zeta_1 & \delta_1=0 & 1 \cr\hline
\hat\zeta_1 & x=\prod_{i=1}^{k-1}\delta_i =0 & \sum_{i=1}^{k-1} X_\Delta(\hat\zeta_1, \delta_i) \cr\hline
\hat{\zeta}_j,\ j=3, \cdots, k-1 & x= \prod_{i=1}^{k-1} \delta_i =0 & \sum_{i=1}^{k-1} X_\Delta (\hat\zeta_j, \delta_i)   \cr\hline
\hat{\zeta}_k & (\ref{Ell2Eq}) & { \sum_{i=1}^{k-1} X_\Delta (\hat\zeta_k, \delta_i)}  \cr\hline
\delta_{k-1}  & x= (y \hat\zeta_k + b_3 \zeta_0^k \zeta_1^{k-1} C (\hat\zeta \delta))\prod_{i=1}^{k-1} \hat\zeta_i  =0& \sum_{i=1}^{k-1} X_\Delta (\delta_{k-1}, \hat\zeta_i) \cr\hline
\delta_j,  \ j= 2, \cdots, k-2 & x=\prod_{i=1}^k \hat\zeta_i=0 & \sum_{i=1}^k X_\Delta (\delta_j, \hat\zeta_i) \cr\hline
\delta_1  	& x= \prod_{i=1}^k \hat\zeta_i =0 & \sum_{i=1}^k X_\Delta (\delta_1, \hat\zeta_i)\cr
			& \zeta_1=x\omega - y^2 \prod_{i=1}^{k} \hat\zeta_i =0 & 1 \cr
			& x A( \hat\zeta\delta) \hat{\zeta}_k^{k-1}
		+ b_2\zeta_0  =  x\omega - y^2 \prod_{i=1}^{k} \hat\zeta_i   =0 & 1 
\end{array}
\ee
These splittings are in one-to-one correspondence with the splittings given in the box graphs of figure \ref{fig:HatZetaDelta}. 
The case of $\mathfrak{su}(7)$ with all the possible triangulations of $\varphi$ is shown in figure \ref{fig:SU7Sigma}.
Note that the splitting rules follow a similar pattern to the fiber face triangulations. However, 
 $\zeta_0, \zeta_1, \delta_1$ play a special role, which will also be clear from the splitting of $\alpha_0, \alpha_1, \alpha_{2k}$ in the associated box graphs, see figure \ref{fig:HatZetaDelta}.

\subsection{Coulomb Phases/Box Graphs for Secondary Fiber Faces}

The Coulomb phases associated to the secondary fiber face  triangulations $\rho_i$, i.e. corresponding to the equations \eqref{Ell2Eq}, are characterized in terms of box graphs, as shown in figure \ref{fig:HatZetaDelta}, where the blue/yellow colourings are fixed, and the only freedom in sign assignments  (compatible with the flow rules) is in the turquoise box, bounded by the vertical lines $V^{k+2}$ and $V^{2k}$, and horizontal lines 
$H^{2}$ and $H^{k+1}$. This implies in particular that $F_{2k}$ is always reducible, and splits off one $F_1$. Furthermore, $F_1$
is irreducible. The sign assignment in the region bounded by these lines is only constrained by the flow rules, as the trace 
condition is already automatically satisfied ($\epsilon(L_{1,2k+1})= -$ and $\epsilon(L_{2,2k})= +$).
Note also that we require at least one of the signs $\epsilon (L_{k+1, i})$, $i = k+2, \cdots, 2k+1$ to be positive, as otherwise the resulting box graphs already have a description in terms of standard toric 
 top triangulations, which we already discussed. By the flow rules 
\be
\epsilon (L_{k+1, k+2}) =+  \qquad \Rightarrow \qquad \epsilon (L_{i,k+2}) =+ \,,\ i=1, \cdots, k+1 \,.
\ee
This implies that $F_{k+1}$ is also reducible in codimension two.  
Following a similar reasoning to section \ref{sec:ToricTBG}, each sign assignment within this region results in a triangulation. This reduced set of box graphs is characterized in terms of the sign assignments as in  figure \ref{fig:HatZetaDelta}. 
Again applying similar arguments to the ones in section \ref{sec:ToricTBG}, we can map these one-to-one to triangulations of the secondary fiber face  $\varphi$. The number of such box graphs is again the number of monotonous lattice paths in a $(k-1) \times (k-2)$ grid, which is given by 
\be
\# \hbox{Box graphs as in figure \ref{fig:HatZetaDelta} } =  {2k-3 \choose k-1} \,. 
\ee
This agrees with the number of triangulations of the secondary fiber face  $\varphi$, as determined in appendix \ref{sect:triang} 
\be
\# \hbox{Triangulations of $\varphi$} = T_{k,k-1} = {2k-3 \choose k-1}\,.
\ee
Finally, it should be remarked that the toric hypersurface and complete intersection resolutions realize a subclass of the complete set 
of small resolutions. It is tantalizing to think that this process of blowdown and flop can be systematically generalized to cover all small resolutions that the box graphs predict. We will discuss this in appendix \ref{app:SU7} in detail for the case of $\mathfrak{su}(7)$, which has  (up to reordering) one additional phase, that does not fall into the category of resolutions discussed thus far. 


\section{Generalized Fiber Faces from Box Graph Layers}
\label{sec:Layers}

All of the phases discussed so far had a simple description matching that of the Coulomb phases/Box graphs, and furthermore all flops were realized by modifying the toric ambient space.
This approach is convenient,  as we can identify curves of the geometry as 2-dimensional cones, or equivalently 1-simplices on faces. Starting from box graphs, this gives a clear strategy for blow-downs or, more generally, flops. Unfortunately, at least in the present description, this structure does not persist to all Coulomb phases.

To conclude the general analysis of flops we now discuss how to realize the phases that go beyond the 
fiber face and secondary fiber face triangulations discussed so far.
The next layer in the box graph description corresponds to changing signs outside the turquoise region in figure \ref{fig:HatZetaDelta}, and require flopping the curve $C^{-}_{k+2, k+3}$.  The phase and fiber face triangulation, from which we start in order to access the next layer in the box graph is shown in figure \ref{fig:NextLayer}. In this case only two of the curves corresponding 
to the roots $\alpha_i$ split over the matter curve $b_1=0$, they are 
\begin{equation}
\begin{aligned}
F_{2k} & \ \rightarrow\  C^+_{k+1, 2k} + \sum_{i=1}^{k} F_i + C^-_{1, 2k+1}  \cr
F_{k+1}& \ \rightarrow\ C^+_{k+1, 2k} + \sum_{i=k+3}^{2k-1} F_i+  C^-_{k+2,k+3}  \,.
\end{aligned}
\end{equation}
Correspondingly, we can write the expression for $D_{\alpha_{k+1}}$ over $b_1 = 0$ (see \eqref{eq:redfibfacefibercomp}) as
a matrix equation
\be
M \left(\begin{array}{c}
 x \\ \prod_{i=1}^{k-2} \delta_i
\end{array}\right) = 0 \,.
\ee
The components of $D_{\alpha_{k+1}}$ are now found by setting either $x=\delta_i=0$ or $\det M =0$. The first group of components are the ones shared with the $D_{\alpha_i}$, for $i = k+3, \cdots, 2k$, and $C^{-}_{k+2, k+3}$ is identified as the component for which $\det M =0$. Hence it
cannot be identified as a stratum descending from the ambient space and we cannot flop it by re-triangulating the ambient space.

\subsection{Flops to the next Layer}
\label{sec:ThirdLayer}

In order to flop the curve $C^{-}_{k+2, k+3}$  we take a more pedestrian approach in this section. For this consider the equations (\ref{Ell2Eq})
in the patch where $\omega\not=0$\footnote{This assumption is without loss of generality, as none of the curves involved in the splittings have a component given by $\omega=0$, as can be readily checked.}. We can then solve the first equation for $x$ and insert into the second equation, which yields again a hypersurface. 
To blow down the curve $C_{k+2, k+3}^-$, which is a component of $\hat\zeta_{k}=0$ not shared with any of the $\delta_i$, we note that a good coordinate on this curve is given by $\delta_{k-2}$. More precisely, we can define the coordinates
\be\label{eq:defsvariables}
\ba
s_1 &= \hat{\zeta}_{k-1} \cr
s_2 &=  \delta_{k-2} \hat{\zeta}_{k-1} = \delta_{k-2} s_1 \cr 
s_3&=
	- 2 A'(\delta\hat\zeta) b_6 B'(\delta\hat\zeta) {C'}^3(\delta\hat\zeta) \left(\prod_{i=1}^{k-2} \hat\zeta_i^2\right) s_1 y \zeta_0^{3k+1}
\zeta_{1}^{3k-1} \hat{\zeta}_k^{k-1} s_2^k  \cr
	& + A'(\delta\hat\zeta) b_6^2 {B'}^2 (\delta\hat\zeta) {C'}^4(\delta\hat\zeta)  \left(\prod_{i=1}^{k-2} \hat\zeta_i^2\right) \zeta_0^{4k+2} \zeta_{1}^{4k-1} \hat{\zeta}_k^{k-1} s_2^{k+2} \cr 
	& - a A'(\delta\hat\zeta) b_6 B'(\delta\hat\zeta) {C'}^2(\delta\hat\zeta) \left(\prod_{i=1}^{k-2} \hat\zeta_i^2\right) s_1^2 y^2 \zeta_0^{2k+1} \zeta_1^{2k} \hat\zeta_k^{k} s_2^{k-1} \cr 
	& - b_2 b_6 B'(\delta\hat\zeta) {C'}^2(\delta\hat\zeta) \left(\prod_{i=1}^{k-2} \hat\zeta_i\right) s_2^2 \omega \zeta_0^{2k+2} \zeta_1^{2k} + b_4 B'(\delta\hat\zeta) {C'}(\delta\hat\zeta) s_2 \omega^2 \zeta_0^{k+1} \zeta_1^k - \omega^3 \delta_{k-2} \prod_{i=1}^{k-3} \delta_i 
\cr 
s_4&= \delta_{k-2} s_3 \,,
\ea
\ee
where we used the modified products (where all the $\delta_{k-2}$ and $\hat{\zeta}_{k-1}$ dependence is factored out)
\be
\ba
A(\delta\hat{\zeta}) = s_2^{k-3} s_1 A'(\delta\hat\zeta) \,,\qquad 
B(\delta\hat{\zeta}) = s_2 B'(\delta\hat{\zeta})  \,,\qquad 
C(\delta\hat{\zeta}) = \delta_{k-2} C'(\delta\hat{\zeta})  \,.
\ea
\ee
The hypersurface equation can then be written in the following way
\be
\ba
s_4= & -b_1 y \omega^2 + A'(\delta\hat{\zeta})  {C'}^2(\delta\hat\zeta) b_3^2  \left(\prod_{i=1}^{k-2} \hat\zeta_i\right) s_1 y^2 \zeta_0^{2k} \zeta_1^{2k-1} \hat{\zeta}_{k}^{k-1} s_2^{k-1} \cr
& + 2b_3  A'(\delta\hat{\zeta})  {C'}(\delta\hat\zeta)  \left(\prod_{i=1}^{k-2} \hat\zeta_i^2\right) s_2^2 y^3 \zeta_0^k \zeta_1^k \hat\zeta_k^k s_2^{k-2} +  A'(\delta\hat{\zeta})    \left(\prod_{i=1}^{k-2} \hat\zeta_i^2\right)\zeta_1 s_1^3 y^4 \hat\zeta_k^{k+1} s_2^{k-3} \cr 
&+ b_2 b_3  {C'}(\delta\hat\zeta)  \left(\prod_{i=1}^{k-2} \hat\zeta_i\right) s_2 y \omega \zeta_0^{k+1} \zeta_1^k 
+ b_2 \zeta_0 \zeta_1 \left(\prod_{i=1}^{k-2} \hat\zeta_i\right) \hat\zeta_k s_1 y^2 \omega \,,
\ea
\ee
with the additional constraint that the new coordinates $s_i$ need to satisfy the conifold equation:
\be\label{eq:conifoldinsvars}
s_1 s_4 = s_2 s_3 \,.
\ee

\begin{figure}
    \centering
    \includegraphics[width=12cm]{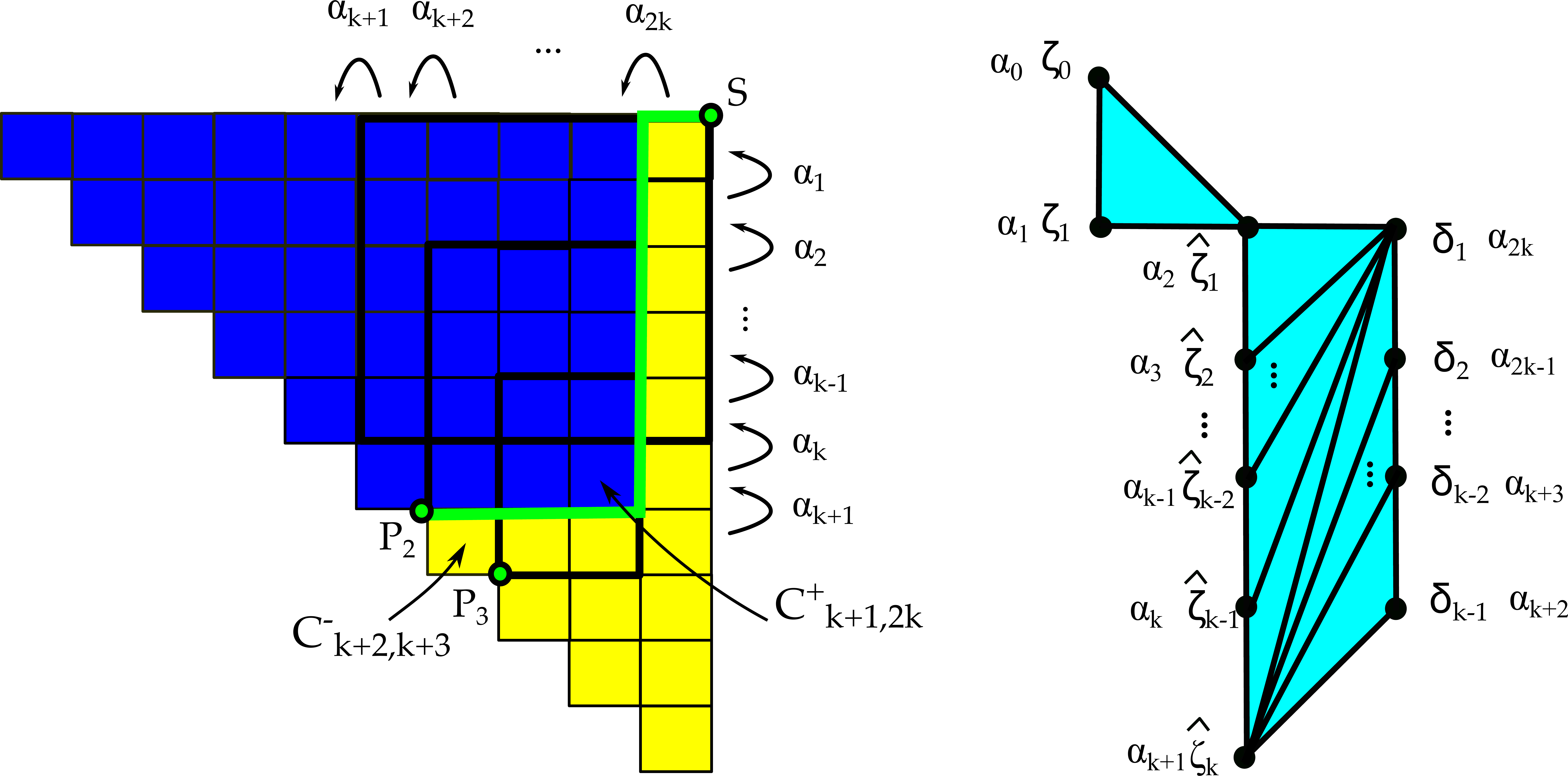}
    \caption{Box Graph and corresponding secondary fiber face  triangulation, from which the next layer in the box graph can be accessed by flopping the curve $C_{k+2, k+3}^-$.    \label{fig:NextLayer}}
\end{figure}

We can then blow down the curve $C_{k+2, k+3}^-$ and blow up by e.g. introducing a new $\mathbb{P}^1$ with projective coordinates $[\xi_1, \xi_2]$ satisfying
\be
s_1 \xi_1  = \xi_2 s_2 \,,\qquad s_3 \xi_1 = \xi_2 s_4 \,.
\ee
The fiber components $F_i$ associated to the roots $\alpha_i$,  that are affected by this flop, split above the codimension two locus $b_1=0$  as follows:
\begin{itemize}
\item $F_{2k}$: this is given by $\delta_1=0$, which has $k+2$ components after the flop
\item $F_{k+1}$: this is $\hat\zeta_k=0$, which looses one component after the flop, and splits into $k-2$ components
\item $F_{k+3}$: this is given by $\delta_{k-2}=0$, which in the new coordinates corresponds to  $s_2=s_4=0$, i.e. $\xi_2=0$ has now two components along $b_1=0$. 
\end{itemize}
This is precisely the splitting that is expected from the box graph analysis after flopping the curve $C^-_{k+2, k+3}$. 
With this flop we have accessed the next `layer' in the box graph, namely, the class of resolutions, which correspond to anti-Dyck paths ending at $P_3$ in figure \ref{fig:NextLayer}.

\begin{figure}
    \centering
    \includegraphics[width=10cm]{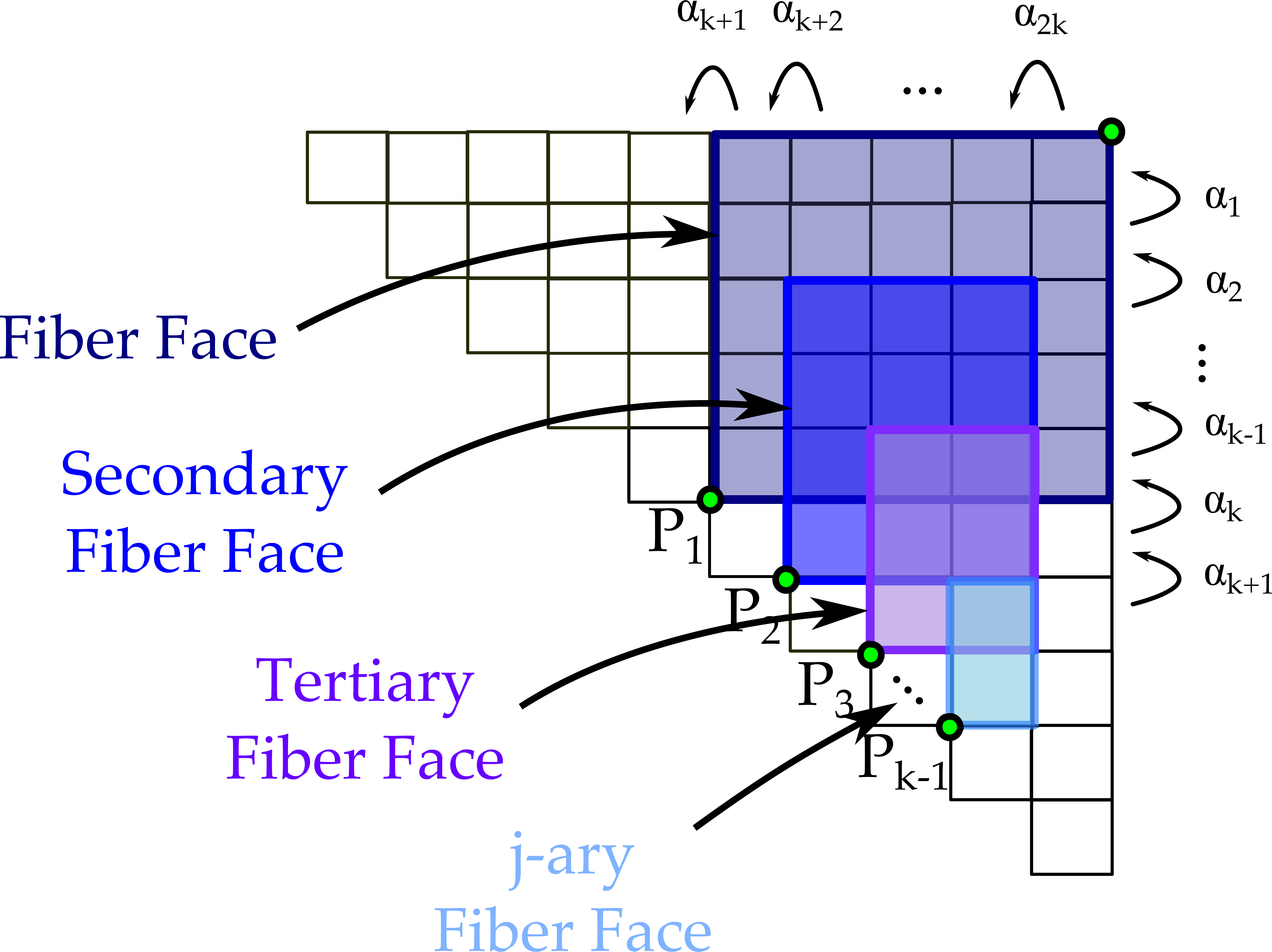}
    \caption{The layer structure of  fiber faces within the box graphs for the anti-symmetric representation of $\mathfrak{su}(2k+1)$.\label{fig:layersboxgraphs}}
\end{figure}

\subsection{Conjecture on Layer Structure}
\label{sec:Layersconj}

The analysis of the last section lends itself to a conjecture about how to construct 
the remaining phases. 
As we have seen in section \ref{sec:BoxToric}, all phases for which the fiber is embedded as a toric hypersurface nicely
organize themselves as anti-Dyck paths inside a square of the box graph, ending at $P_1$ in figure \ref{fig:layersboxgraphs}. In section \ref{sec:CompleteInt} we gained access to another layer of curves by blowing down all of the coordinates $\zeta_i$ for $i = 2, \cdots, k$. 
The crucial point was that the elliptic fiber can be in turn described in the alternative factored form (\ref{eq:su2kp1factforms2}). 
This factorization makes manifest, after blowing down appropriate coordinates, the existence of conifold singularities, which can be used to pass to an alternative resolution. These have a characterization in terms of the triangulation of secondary fiber faces. We have shown that these are precisely the flops, which in the box graph language correspond to 
 the phases for which the anti-Dyck path ends at $P_2$ in figure \ref{fig:layersboxgraphs}. 

A completely analogous structure becomes apparent in \eqref{eq:defsvariables}. To achieve the flop of the curve $C_{k+2, k+3}^-$, we have essentially
factored out $\delta_{k-2}$ from the terms contained in $s_3$ in \eqref{eq:defsvariables}. However, note that $s_3$ contains a factor of $\prod_{i=1}^{k-3} \delta_i$ as well. It is hence possible to introduce a similar birational map to the one defined in \eqref{eq:defsvariables} by employing any of the coordinates $\delta_i$ for $i = 1 \cdots k-3$. 
Correspondingly, after blowdown, we expect there to be conifold  singularities in \eqref{eq:conifoldinsvars}, whereby we reach the set of phases for which the anti-Dyck paths end at $P_3$ of figure \ref{fig:layersboxgraphs}.
Concretely, this will require all of the blowdowns associated with the  $\delta_i$ for $i = 1 \cdots k-2$ at once, followed by the alternative 
small resolutions. This is expected to introduce $k-2$ new coordinates, forming a fiber face corresponding to $P_3$. 

We conjecture that this structure prevails for \emph{all} of the anti-Dyck paths, ending on the points $P_i$, i.e. there is a fiber face which is a strip with sides of length $k-i+2$ and $k-i+1$ associated to each class of paths, which end at one of the points 
$P_i$ such that triangulations of the fiber face are in one-to-one correspondence with anti-Dyck paths ending at $P_i$ of the box graph: 
\be
\hbox{Anti-Dyck Paths Ending at $P_j$} \quad \stackrel{1:1}{\longleftrightarrow} \quad \hbox{$j$-ary Fiber Face Triangulation}
\ee
It is not hard to see that a generalization of the splitting rules over $b_1=0$ observed in sections \ref{sec:BoxToric} and \ref{sec:CompleteInt}
perfectly match the behaviour of the fiber components predicted by the associated box graphs.
%
%
%


\section{Discussion and Outlook}
\label{sec:Discussion}

In this paper we studied the correspondence between resolutions of singular elliptic fibrations and box graphs (or equivalently, Coulomb phases of 3d supersymmetric gauge theories). We have proven the equivalence between a subclass of box graphs and a specific class of resolutions of the elliptic fibration. Each box graph has a unique identification with so-called anti-Dyck paths, and we showed that each resolution type is  characterized in terms of paths ending at one fixed point on the diagonal. 
Moreover, we determined the network of flop transitions and showed the equivalence to the flops predicted by the box graphs. 

More precisely, we have proven a one-to-one correspondence between resolutions obtained by toric methods (of triangulating the fiber face) and a class of box graphs. These have a unique characterization as anti-Dyck paths all ending in one fixed point on the diagonal (in this case, they end at the point $P_1$ in figure \ref{fig:layersboxgraphs}). Furthermore, we have shown that there is a secondary fiber face, which corresponds to another subclass of box graphs, characterized in terms of anti-Dyck paths ending at the point $P_2$ in figure \ref{fig:layersboxgraphs}. For these two classes we have shown in sections \ref{sec:BoxToric} and \ref{sec:CompleteInt} that the triangulation of the fiber faces and box graph phases are in complete agreement.

Beyond these, we do not at present know how the class of resolutions has to be extended in order to account for the phases that are given in terms of box graphs. 
From our analysis, starting with the tops and then passing on to the secondary fiber faces, it seems rather suggestive that the box graphs can be somewhat ``foliated" by generalized fiber face diagrams and their triangulations, as shown in figure \ref{fig:layersboxgraphs}. In other words, we expect each class of anti-Dyck paths with a fixed endpoint on the diagonal to give rise to a specific class of resolutions, as shown in figure \ref{fig:layersboxgraphs}.

As already observed in the companion paper \cite{Braun:2014kla} for $\mathfrak{su}(5)$, the resolutions cease to be of simple hypersurface or complete intersection type, and require for instance determinantal blowups. 
One direction to extend this would be to develop the connection to matrix factorization and resolutions as discussed  in \cite{MR3084719} as well as the more recent developments in \cite{Grassi:2013kha, Collinucci:2014taa} addressing alternative ways of studying F-theory on singular spaces, or their deformations.  Additionally, it would be interesting to extend our analysis 
to (combinations of)  different matter representation and gauge algebras, such as the ones considered in \cite{Hayashi:2014kca}. 

Perhaps most thought-provokingly, one could anticipate to define  a geometric structure  starting from the box graphs, which is constructed from the data of the extremal generators and the knowledge of the splitting of rational curves in the fibers from codimension one to two. We leave these intriguing questions for future work.

Our results are also amenable to applications in mirror symmetry. 
In string theory, the K\"ahler moduli space of a Calabi-Yau variety is not confined to a single K\"ahler cone. 
In fact, it is natural to consider the union of all K\"ahler cones, that are related by flop transitions \cite{Aspinwall:1993nu, Cox:2000vi}. From this point of view, the box graphs yield the structure of the so-called enlarged K\"ahler cone for the K\"ahler moduli, which control the volumes of the fiber components (whilst keeping the K\"ahler moduli of the base fixed). Our results indicate that different phases of the same Calabi-Yau can have very different geometric realizations.  The resolved elliptic fibers can for instance be embedded as hypersurfaces, complete intersections or more general algebraic varieties, which would in turn also change the geometric realization of the whole Calabi-Yau manifold in question.


\subsection*{Acknowledgements}

We thank Philip Candelas, Xenia de la Ossa, Craig Lawrie,  Dave Morrison and Jenny Wong for discussions on related matters. We thank the Aspen Center for Physics and the Galileo Galilei Institute in Florence for hospitality during the completion of this work. 
The research of APB was supported by the STFC grant ST/L000474/1 and the EPSCR grant EP/J010790/1. The work of SSN is supported in part by STFC grant ST/J002798/1. This work was in part performed at the Aspen Center for Physics, which is supported by National Science Foundation grant PHY-1066293. We furthermore thank the Aspen Asie fortune cookies for inspiration during the completion of this work.


\appendix

\section{Number of Triangulations of a Strip}
\label{sect:triang}

In this appendix, we derive an expression for the number of fine triangulations of the point configuration
\bea
P_{n,m} = \{(k,0)| [k=1\cdots n] \, , \quad\,  (l,1)| [l=1 \cdots m]  \}  \, ,
\eea
i.e. we want to triangulate a strip which has $n$ points on one side and $m$ points on the other.

Let us denote the number of fine triangulations of $P_{m,n}$ by $T_{m,n}$. 
We now claim that 
\begin{equation}\label{eq:Tmn}
 T_{n,m} = \binom{n+m-2}{n-1} \,,
\end{equation}
which we are going to prove by induction. Note that this expression is symmetric under the exchange of $n \leftrightarrow m$. The first few terms are easy to check: by inspection one finds that e.g. $T_{1,m}=1$, $T_{2,2}=2$.

To proceed, we decompose the triangulations of $P_{n+1,m}$ in the following way. Let us single out the first point on
the $n$-plane, i.e. $(1,0)$. It will necessarily have a 1-simplex connecting it to one of the points on the $m$-plane. Let us now assume 
the while it shares a 1-simplex with the point $(k_0,1)$, there is no point $(k,1)$ for $k> k_0$ with this property. Note that
$k_0$ can be $1$, in which case $(1,0)$ only meets $(1,1)$ along the boundary of the polytope spanned by the $P_{n,m}$. The crucial observation
is now that for any fixed $k_0$, the triangulation ``to the left'' of the connecting one-simplex is uniquely fixed, whereas there are still
$T_{n,m+1-k_0}$ ways to triangulations the part ``to the right''. Hence we have the recursion relation
\begin{equation}
 T_{n+1,m} = \sum_{k_0 = 1}^{m} T_{n,m+1-k_0} \,.
\end{equation}

To perform the induction step, we assume that the above holds for all $\hat{n} \leq n$ and $\hat{m} \leq m$ and wish to show that this implies that $T_{n+1,m}$ also satisfies \eqref{eq:Tmn}. This is seen by writing
\be\ba
T_{n+1,m} 	&= \sum_{k_0 = 1}^{m} T_{n,m+1-k_0} \nn \\
			&= \sum_{k_0=1}^{m} \binom{n+m-k_0-1}{n-1} \nn\\
			&=\sum_{k=0}^{m-1} \binom{n-1+k}{n-1} \nn\\
			&= \binom{(n+1) + m -2 }{m-1} \,.
\ea\ee
We have used that
\begin{equation}
 \sum_{x=0}^{b}\binom{a+x}{a} =\binom{a+b+1}{b} \, .
\end{equation}
Due to the symmetry between $n$ and $m$ this is sufficient to establish $\eqref{eq:Tmn}$ for all $n,m$.



\section{Fibers and Phases for  $\mathfrak{su}(7)$}
\label{app:SU7}

As  a concrete example, we consider phases of the $\mathfrak{su}(7)$ theory with anti-symmetric representation and construct all the phases geometrically. 
In $\mathfrak{su}(5)$ some features are less transparent due to the small rank, and the general structure becomes apparent only in the case of $\mathfrak{su}(7)$. The Tate form for an $I_7$ Kodaira fiber 
is
\begin{equation}\label{eq:su7}
 y^2  + b_1 x y + b_3 \zeta_0^3 y = x^3 + b_2  \zeta_0 x^2 + b_4 \zeta_0^4x + b_6 \zeta_0^7 \,.
\end{equation}
Using the Weyl group quotient and trace condition, or equivalently the Box Graphs, one can determine the complete network of phases for $\mathfrak{su}(7)$ with ${\bf 21}= \Lambda^2{\bf 7}$. The codimension two locus, where this matter is localized in the Tate model is  $\zeta_0=b_1=0$. 

\begin{figure}
    \centering
    \includegraphics[width=15cm]{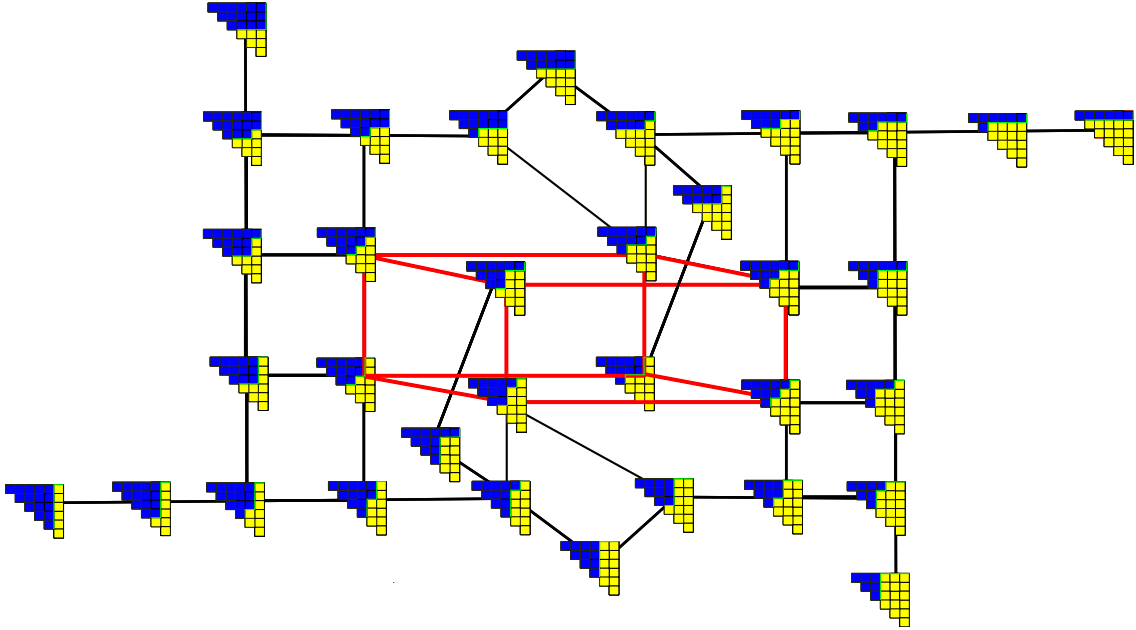}
    \caption{Box graphs for $\mathfrak{su}(7)$ with ${\bf 21}$ matter, with lines connecting the box graphs corresponding to flop transitions. The cube shown with red connections corresponds to the standard algebraic resolutions discussed in more detail in section \ref{sec:AlgRes}. The green lines, separating the blue/yellow (+/-) boxes correspond to the anti-Dyck paths. The geometric counterpart is shown in figure \ref{fig:ResNet}, where the geometric realization of these resolutions are shown. 
     \label{fig:SU7FlopGraph}}
\end{figure}

\subsection{Box Graphs}

As shown in  \cite{Hayashi:2014kca}, there  are 34 box graphs for $\mathfrak{su}(7)$ with $\Lambda^2 {\bf 7}$ with weights 
\be
w_{i,j}=L_i + L_j, \qquad i<j\,.
\ee 
The signs have to satisfy the flow rules, i.e. $+$ (blue) signs flow from right to left and below to above, and the oppositve for $-$ signs (yellow). We will denote $w_{i,j}^{\pm}= \pm w_{i,j}$.
For $SU(n)$ the tracelessness condition implies that there is a diagonal condition that needs to be satisfied. 
Alternatively, the resolution/phase can be characterized by the path that separates the weights that have a positive sign from those with a negative one. This anti-Dyck path has to cross the diagonal at least once, in order to ensure that the diagonal condition is satisfied. 
Flop transitions in box graphs are single box sign changes, which do not violate the flow rules and diagonal condition. The resulting network of flop transitions is shown in figure \ref{fig:SU7FlopGraph} for $\mathfrak{su}(7)$ with the $\Lambda^2 {\bf 7}$ representation.

\begin{figure}
    \centering
    \includegraphics[width=15cm]{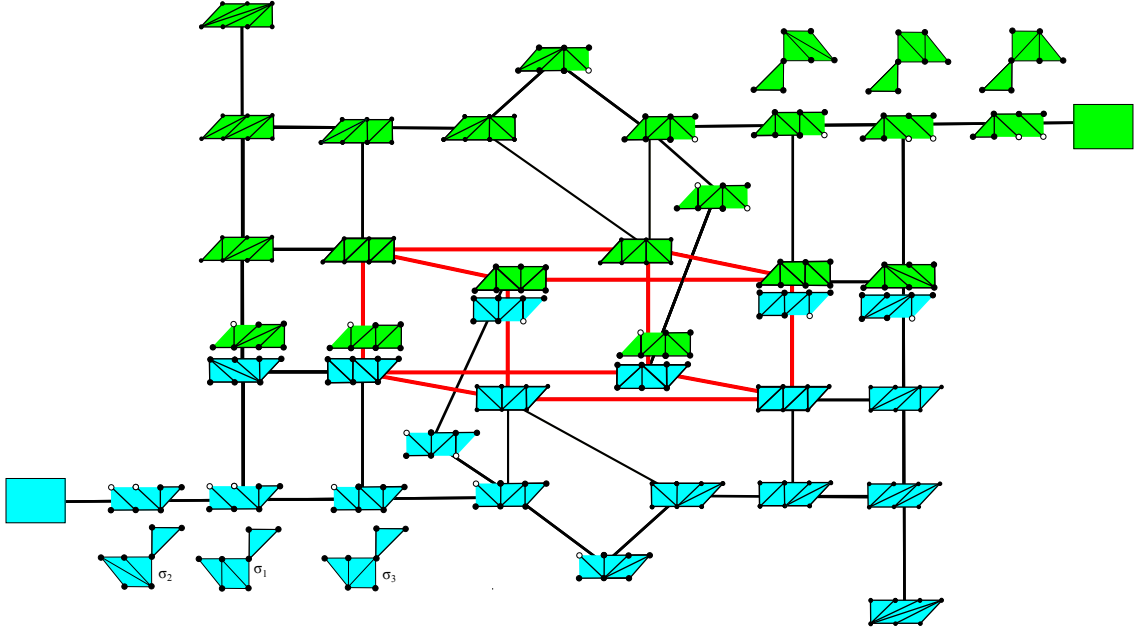}
    \caption{Resolution flop network  for $\mathfrak{su}(7)$ with matter in the $\Lambda^2 {\bf 7}$ representation. This diagram is the geometric counterpart to the flop diagram for the box graphs in figure \ref{fig:SU7FlopGraph}. 
    The turquoise/green differ by reversing the orientation of the assignment between vertices and the fiber face and simple roots. 
    Each diagram corresponds to a triangulation of either the toric top, or a blowdown of this, indicated by the white nodes. 
    In particular the diagrams with multiple nodes blown-down have an alternative description in terms of triangulations of the secondary fiber face  $\varphi$, see figure \ref{fig:SU7Sigma}. Finally, the two empty squares correspond to box graphs, which do not seem to have a straight-forward toric description, however we will determine the corresponding resolution in section \ref{sec:FinFlop}.
    \label{fig:ResNet}}
\end{figure}

\subsection{Fiber Faces and Weighted Blowups}

\subsubsection{Resolution}

It is clear from the general analysis of \cite{Lawrie:2012gg} that for an $I_7$ fiber, 
three successive big resolutions resolve the geometry in codimension one:
\begin{equation}
 y \left(y+ b_1 x + b_3 \zeta_0^3 \zeta_1^2 \zeta_2 \right) = \zeta_1  \zeta_2 \zeta_3 \left(
  x^3 \zeta_2 \zeta_3^2 + b_2 x^2 \zeta_0  + b_4 x \zeta_0^4 \zeta_1^2 \zeta_2  +
 b_6 \zeta_0^7 \zeta_1^4 \zeta_2^2   \right) \, .
\end{equation}
The remaining singularities in higher codimension can all be cured by a small resolutions. This can be realized as
a sequence of three blowups along the divisors $y = \hat{\zeta}_i = 0$ 
\begin{equation}\label{eq:su7restoric}
 y^2 \hat{\zeta}_1  \hat{\zeta}_2  \hat{\zeta}_3 +  b_1 y x + b_3 y \zeta_0^3 \zeta_1^2 \zeta_2 \hat{\zeta}_1^2 \hat{\zeta}_2 =
 x^3 \zeta_1 \zeta_2^2 \zeta_3^3 \hat{\zeta}_2 \hat{\zeta}_3^2 + b_2 x^2 \zeta_0 \zeta_1 \zeta_2 \zeta_3 + b_4 x \zeta_0^4 \zeta_1^3 \zeta_2^2 \zeta_3 \hat{\zeta}_1^2 \hat{\zeta}_2 +
 b_6 \zeta_0^7 \zeta_1^5 \zeta_2^3 \zeta_3 \hat{\zeta}_1^4 \hat{\zeta}_2^2 \, .
\end{equation}

Let us rephrase the resolution process just discussed in terms of toric morphisms of the ambient space. The singular situation
is described by a hypersurface in a toric variety for which the generators of one-dimensional cones are
\be\label{eq:coordsxyz}
v_x=(-1,0,0) \,,\qquad  v_y = (0,-1,0) \,,\qquad  v_{\zeta_0}  = (2,3,1) \,.
\ee
The monomials in \eqref{eq:su7restoric} are assigned to the following points in the M-lattice:
\be \label{eq:su7top}
\begin{array}{c|c|c|c|c|c|c|c}
\hbox{Monomial} &y^2 & b_1 xy & b_3 y \zeta_0^3 & x^3& b_2 \zeta_0 x^2 & b_4 x \zeta_0^{4} &b_6 \zeta_0^{7}  \cr\hline
&&&&&&&\cr 
\begin{array}{c}
\hbox{Lattice} \cr
\hbox{Point}
\end{array}&
 \left(\begin{array}{c}
       -1 \\ 1 \\ 0
      \end{array}\right) &
       \left(\begin{array}{c}
       0 \\ 0 \\ -1
      \end{array}\right) &
       \left(\begin{array}{c}
       1 \\ 0 \\ 0
      \end{array}\right) &
       \left(\begin{array}{c}
       -2 \\ 1 \\ 0
      \end{array}\right) &
       \left(\begin{array}{c}
       -1 \\ 1 \\ -1      \end{array}\right) &
       \left(\begin{array}{c}
       0 \\ 1 \\ 0      \end{array}\right) &
        \left(\begin{array}{c}
       1 \\ 1 \\ 6      \end{array}\right) 
\end{array}
\ee

From the discussion of section \ref{sect:gentoricstuff}, it follows that the singularities are then resolved by refining 
the cone $\langle v_x v_y v_{\zeta_0} \rangle$, by introducing new one-dimensional cones generated by
\be \ba\label{eq:gensu7}
 v_{\zeta_1} &= (1,2,1) \qquad   v_{\hat{\zeta}_1} = (1,1,1) \\
 v_{\zeta_2} &= (0,1,1) \qquad   v_{\hat{\zeta}_2} = (0,0,1) \\
 v_{\zeta_3} &= (-1,0,1) \quad  v_{\hat{\zeta}_3} = (-1,-1,1) \, .
\ea
\ee
These are shown in figure \ref{fig:ToricFan}.
Any triangulation of the polytope spanned by $v_x,v_y,v_{\zeta_0} \cdots v_{\hat{\zeta}_3}$ gives rise to a resolution of \eqref{eq:su7}.
There are ten triangulations of this polytope, nine of which are realized via successive (weighted)
blowups. The power of this point of view is that any toric resolution will introduce the same generators
\eqref{eq:gensu7}, so that the weight system of the ambient space is the same for any resolution:
\be
\begin{array}{ccccccccc}
x & y& \zeta_0 & \zeta_1 & \zeta_2 & \zeta_3 & \hat{\zeta}_1 & \hat{\zeta}_2 & \hat{\zeta}_3 \cr\hline
1 & 1& 1 & -1 & 0&0&0&0&0\cr
1 & 1& 0 & 1 & -1&0&0&0&0\cr
1 & 1& 0 & 0 & 1&-1&0&0&0\cr
0 & 1& 0 & 1 & 0&0&-1&0&0\cr
0 & 1& 0 & 0 & 1&0&0&-1&0\cr
0 & 1& 0 & 0 & 0&1&0&0&-1\cr
\end{array}
\ee
and what discriminates between different resolutions is only the SR-ideal, which is combinatorially equivalent to a
triangulation. Furthermore, it is clear from the above weight system (or, equivalently, the vectors \eqref{eq:gensu7}),
that we will end up with \eqref{eq:su7restoric} for any resolution.

\subsubsection{Weighted Blowups and Triangulations}
\label{sect:buvstriangsu7}

As discussed in sections \ref{sect:gentoricstuff} and \ref{sect:toric_triang}, different sequences of weighted blowups do not necessarily end up with different smooth models, and there are furthermore triangulations which cannot be obtained by any sequence of weighted blowups. In this sections we give
some examples for these phenomena in the context of $\mathfrak{su}(7)$ with ${\bf 21}$.

\begin{figure}
    \centering
    \includegraphics[width=9.5cm]{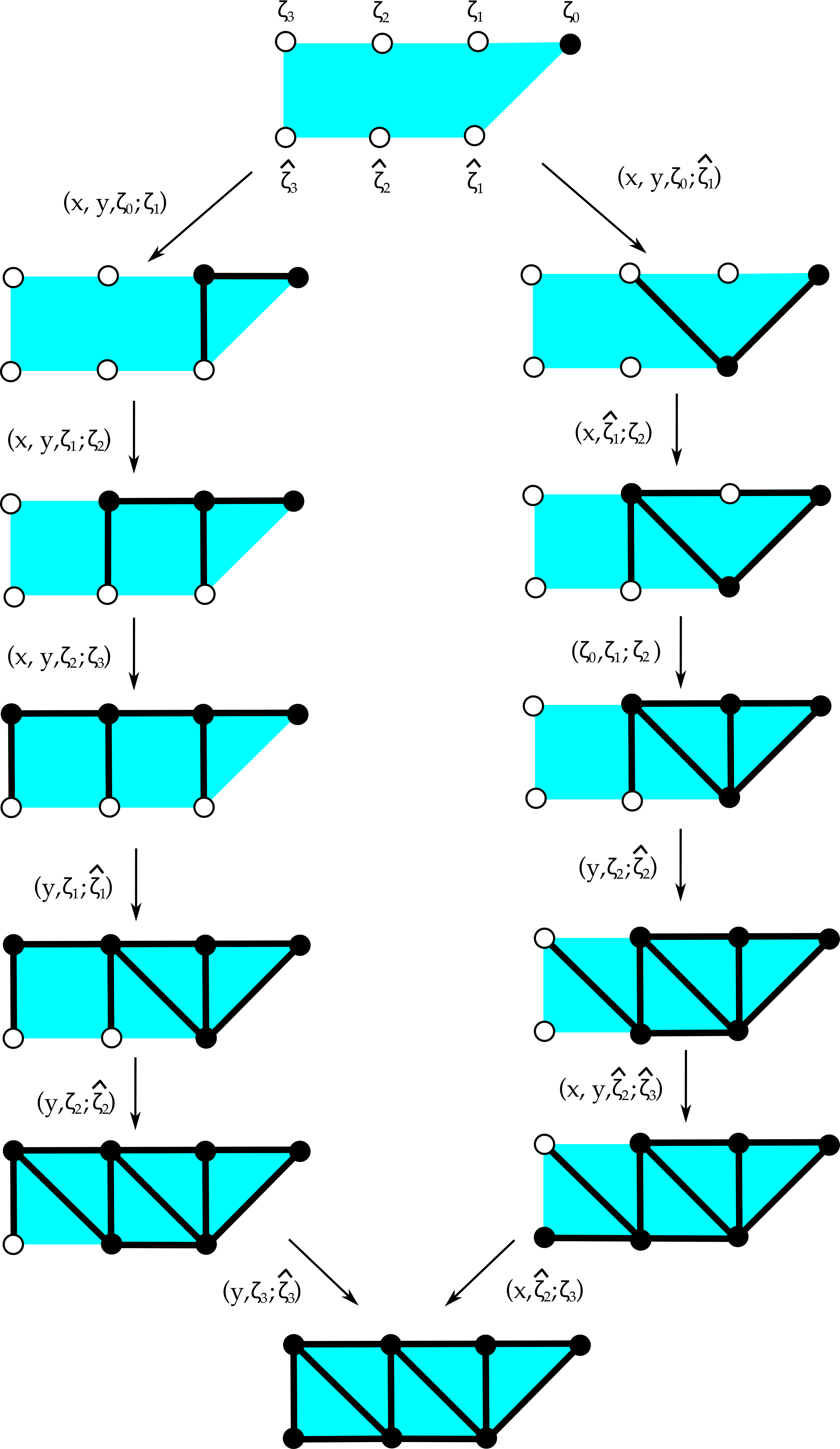}
    \caption{Two different sequences of weighted blowups which end up with the same triangulation and hence the same smooth model. \label{fig:busqvstr}}
\end{figure}

Our first examples concerns two sequences of weighted blowups, which result in the same triangulation and hence in the same phase.
Consider the sequences of blowups shown in figure \ref{fig:busqvstr}. We have only drawn the fiber face part of the fan of the 
toric ambient space and have indicated which blowup is performed in each step. The points drawn in open circles correspond to
homogeneous coordinates that can still be introduced by means of weighted crepant blowups. Note that each $\hat{\zeta}_i$ sits
in the cone spanned by $y$ and $\zeta_i$ (for all $i$) and each $\zeta_i$ sits in the cone spanned by $x$ and $\hat{\zeta}_{i-1}$
(for $i=1..3$). The weights of the individual blowups can be recovered from \eqref{eq:coordsxyz} and \eqref{eq:gensu7} together with \eqref{eq:scalingE}.

\begin{figure}
    \centering
    \includegraphics[width=4cm]{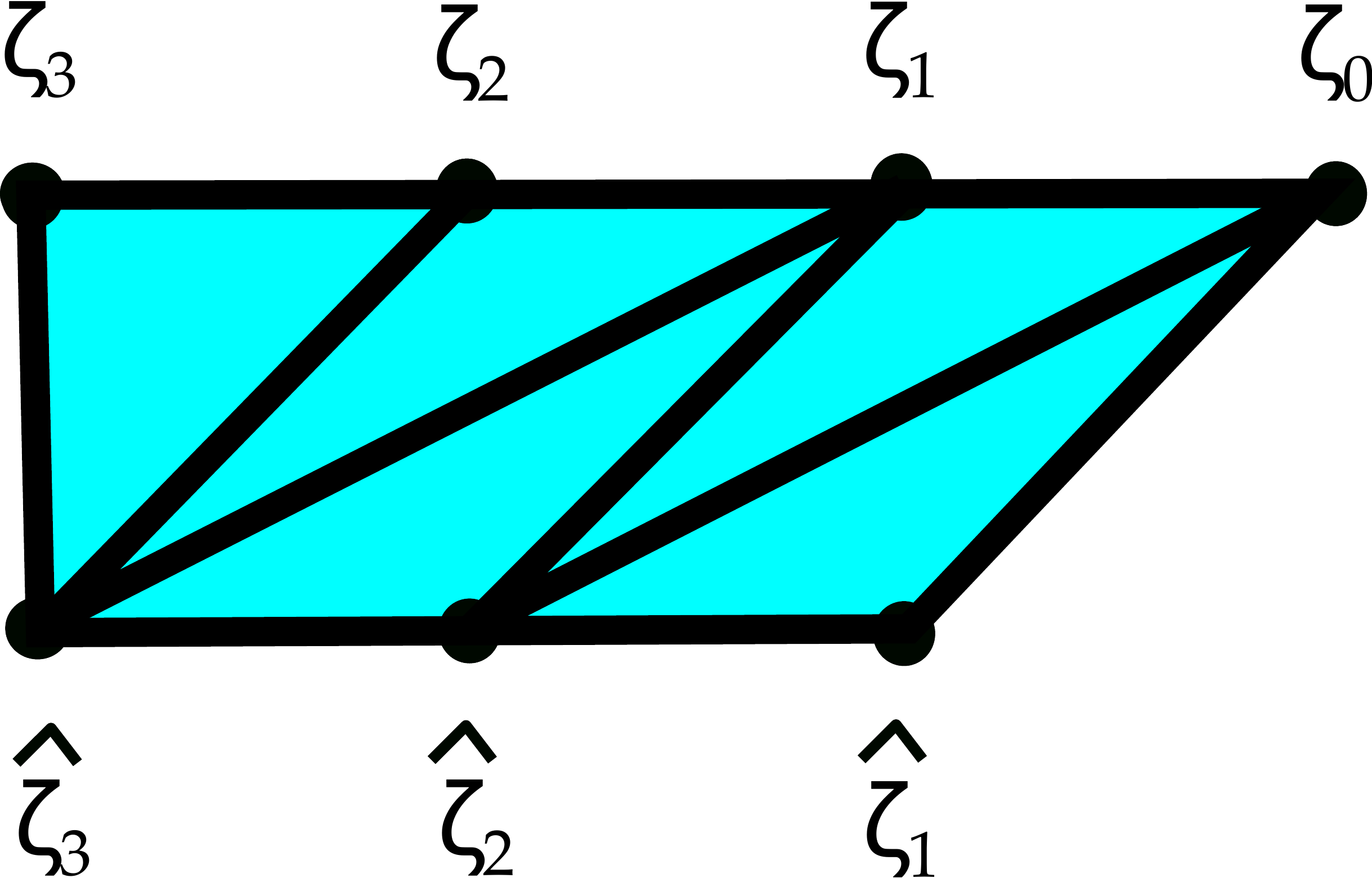}
    \caption{A triangulation which cannot be obtained by a sequence of weighted blowups. \label{fig:nobutriang}}
\end{figure}

\begin{figure}
    \centering
    \includegraphics[width=5cm]{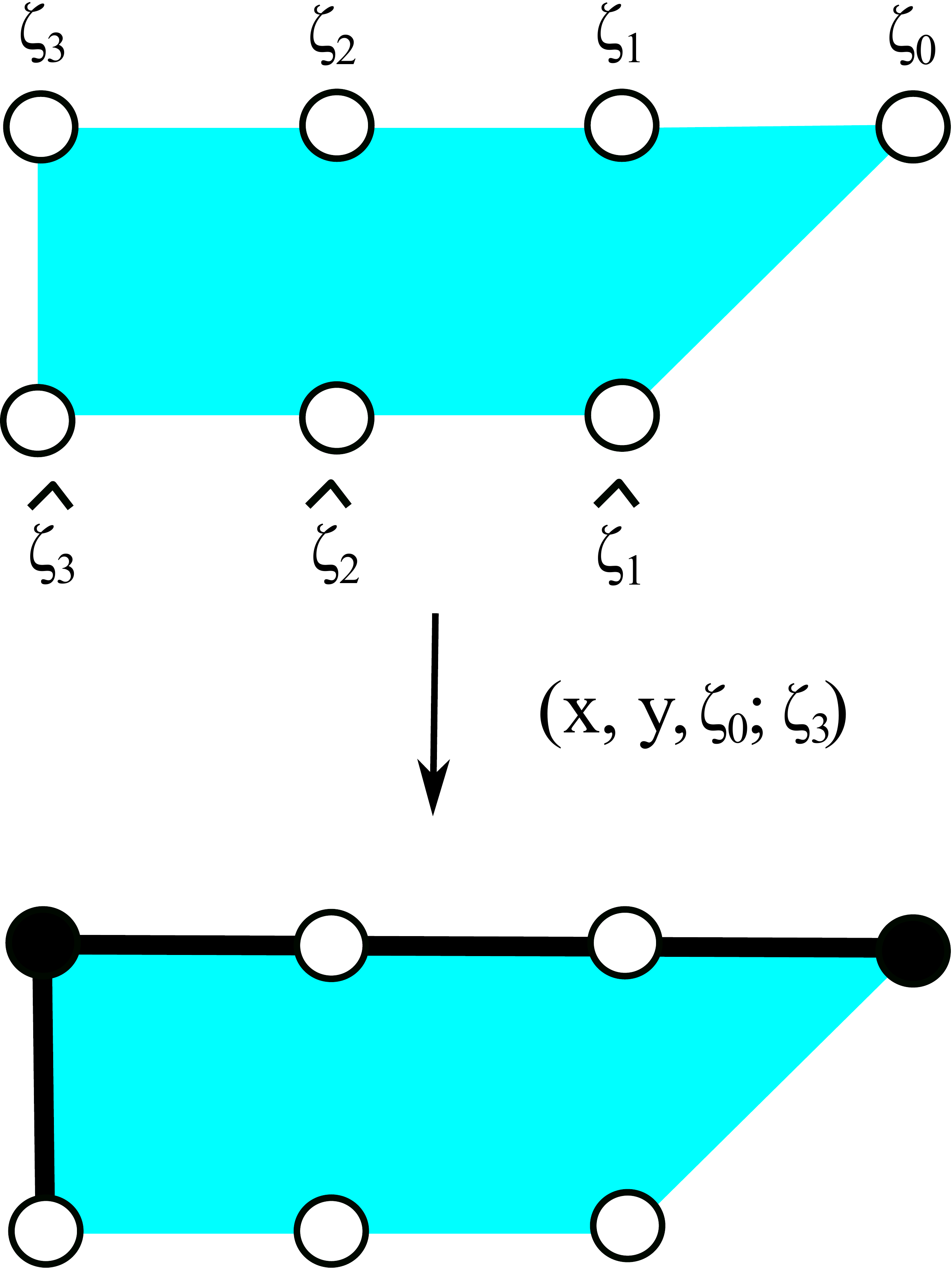}
    \caption{The first step in trying to reach the triangulation shown in figure \ref{fig:nobutriang} by a sequence of blowups. 
    \label{fig:nobutriangstep1}}
\end{figure}

As a second example, consider the triangulation shown in figure \ref{fig:nobutriang}.
It turns out that this phase can never be reached by a sequence of (weighted) blowups. This can be seen by trying
to construct the corresponding blowups. In each step, we have to introduce one of the rays corresponding to the coordinates
$\{\zeta_0,\zeta_1,\zeta_2,\zeta_3, \hat{\zeta}_1,\hat{\zeta}_2,\hat{\zeta}_3\}$. In the first step, the only option we have
is blowing up $(x,y,\zeta_0;\zeta_3)$, and the corresponding cones are shown in figure \ref{fig:nobutriangstep1}.
The reason is that any other choice would necessarily give rise to cones which are not contained
in the triangulation we are aiming for: if we e.g. blow up $(x,y,\zeta_0;\zeta_1)$ we are bound to 
find a 1-simplex connecting $\zeta_1$ with $\hat{\zeta}_1$, whereas blowing up  $(x,y,\zeta_0;\hat{\zeta}_3)$
induces a 1-simplex connecting $\zeta_0$ with $\hat{\zeta}_3$. All other options can be similarly excluded.
As a second step after the blowup $(x,y,\zeta_0;\zeta_3)$, we can still introduce any of $\{\zeta_0,\zeta_1,\zeta_2,\hat{\zeta}_1,\hat{\zeta}_2,\hat{\zeta}_3\}$ by a further blowup. As before, any such blowup will either introduce a 1-simplex between
$\zeta_i$ and $\hat{\zeta}_i$, $\zeta_{i+1}$ and $\hat{\zeta}_i$ ($i=1,2$) or $\hat{\zeta}_3$ and $\zeta_0$, all
of which do not appear in the triangulation in figure \ref{fig:nobutriang}.

Note that even though this triangulation cannot be obtained by a sequence of weighted blowups, there is still a well-defined
morphism corresponding to the whole resolution (which descends from the corresponding morphism of the ambient space).
Furthermore the blown-up ambient space (and hence any algebraic submanifold such as our resolved Calabi-Yau) is still 
projective after the triangulation by the general argument in section \ref{sect:toric_triang}.

\subsubsection{Splitting Rules}

Before discussing how fiber components split over the ${\bf 21}$ matter curve, we identify which divisors
correspond to which Cartan divisors. This is immediate in the present description. 
We can interpret \eqref{eq:su7restoric} as defining a complex two-dimensional variety. In this case, toric divisors 
only have a non-zero intersection if the corresponding points are connected along an edge of the polytope. This means 
we can directly identify
\be
\begin{array}{l|cccccccc}
\hbox{Section} &\zeta_0 & \zeta_1 & \zeta_2 & \zeta_3 & \hat{\zeta}_1 & \hat{\zeta}_2 &\hat{\zeta}_3 \\\hline
\hbox{Cartan Divisor}& D_{\alpha_0} &  D_{\alpha_1} & D_{\alpha_2} & D_{\alpha_3} & D_{\alpha_6} & D_{\alpha_5} & D_{\alpha_4} \cr
\end{array}
\ee

In the cases of interest, where we are considering a Calabi-Yau threefold or fourfold, divisors can also meet
along loci of higher codimension in the base. We can now find a direct map between the triangulations and the splittings of 
the fiber components $\alpha_i$ when we go on top of the ${\bf 10}$ matter curve at $b_1=0$ by using \eqref{eq:su7restoric}. 
The expressions for the different fiber components become
\be
\begin{array}{c|c|l}
\hbox{Root} & \hbox{Section} & \hbox{Equation along }b_1=0 \cr \hline
{\alpha_0} & \ \zeta_0   &0=\hat{\zeta}_2\hat{\zeta}_3 \left(y^2\hat{\zeta}_1 - x^3 \zeta_1 \zeta_2^2 \zeta_3^3 \right) + b_1 yx   \\
 \alpha_1 & \ \zeta_1  & 0= y\left( y \hat{\zeta}_1\hat{\zeta}_2\hat{\zeta}_3 +b_1 x\right)   \\
 \alpha_2 & \ \zeta_2  &0=y\left( y \hat{\zeta}_1\hat{\zeta}_2\hat{\zeta}_3 +b_1 x\right)   \\
 \alpha_3 & \ \zeta_3  &0=y\left( y \hat{\zeta}_1\hat{\zeta}_2\hat{\zeta}_3 + b_3 \zeta_0^3 \zeta_1^2 \zeta_2 \hat{\zeta}_1^2 \hat{\zeta}_2 +b_1 x\right)   \\
 \alpha_4 &\  \hat{\zeta}_3   &0=\zeta_1 \zeta_2 \left(b_3 \zeta_0^3 \zeta_1 \hat{\zeta_1}^2\hat{\zeta}_2- 
 b_2 x^2 \zeta_0 \zeta_3 - b_4 x \zeta_0^4 \zeta_1^2 \zeta_2 \zeta_3 \hat{\zeta}_1^2 \hat{\zeta}_2 - b_6 \zeta_0^7 \zeta_1^4 \zeta_2^2 \zeta_3 \hat{\zeta}_1^4 \hat{\zeta}_2^2 \right) + b_1 xy   \\
 \alpha_5 &\ \hat{\zeta}_2&0=\zeta_0 \zeta_1 \zeta_2 \zeta_3 b_2 x^2 + b_1 xy  \\
 \alpha_6 &\ \hat{\zeta}_1 &0=\zeta_1 \zeta_2 \zeta_3 x^2 \left( b_2 \zeta_0 + x \zeta_2 \zeta_3^2 \hat{\zeta}_2 \hat{\zeta}_3^2  \right) + b_1 xy  
\end{array}
\ee
Note that for any fine triangulation, $\zeta_1,\zeta_2$ and $\zeta_3$ cannot vanish simultaneously with $y$ and $\hat{\zeta}_1,\hat{\zeta}_2$ cannot vanish simultaneously with $x$.

Two divisors $D_{\alpha_i}$ can vanish at the same time if they share a common cone in the fan constructed over simplices of the triangulation. 
To share a common cone, they must hence be connected by a 1-simplex $\sigma_{ij}$ on the face $F_\alpha$. From this, we can read off the following simple rule, already formulated in section \ref{sect:anti-symsplitrules}, Theorem \ref{thm:splittingrules}:

\emph{Let $Z=\{\zeta_0,\zeta_1,\zeta_2,\zeta_3 \}$ and $\hat{Z}= \{\hat{\zeta}_1,\hat{\zeta}_2,\hat{\zeta}_3\}$. Then the number of components each divisor splits into over $b_1=0$ is equal to the number of 1-simplices which connect it to divisors from the set $Z$ or $\hat{Z}$, whichever does not contain the divisor. }

 Note that this means that we will find $4 + 4\cdot 2 = 12$ fiber components of an $I_1^*$ fiber above $b_1=0$, as it should be. Furthermore, it is clear which components of the {\bf 21} matter surface can be obtained as intersections of which divisors in the different phases. 
Under this correspondence, the one-simplices internal to $F_\alpha$ can be associated with weights.
Let us see how this rule works for the first of the two example triangulations discussed above. In the triangulation shown in figure \ref{fig:busqvstr},
there is only a single 1-simplex connecting $\zeta_0,\zeta_1,\hat{\zeta}_3$ to the other side of the fiber face. This means that the three 
fiber components corresponding to $\alpha_0$, $\alpha_1$ and $\alpha_4$ stay irreducible over the locus $b_1=0$. In contrast, there
is more than a single 1-simplex connecting $\zeta_2,\zeta_3$ and $\hat{\zeta}_1,\hat{\zeta}_2$ to the other side, so that 
$\alpha_2,\alpha_3,\alpha_5$ and $\alpha_6$ become reducible over $b_1=0$ into two and three components, respectively. More precisely,
\be
\begin{aligned}
F_{0, 1, 4}&\ \rightarrow\  F_{0, 1, 4}   \\
F_2&\ \rightarrow\   C_a + C_b  \\
F_3&\ \rightarrow\   C_c + F_4  \\
F_5&\ \rightarrow\    C_b + C_c  \\
F_6&\ \rightarrow\    F_0 + F_1 + C_a  \,.
\end{aligned}
\ee
The same splitting is found from the box graph as follows directly from our general analysis in section \ref{fig:ToricBox}.

\subsection{Blowdowns and Flops}
\label{sect:simpflopssu7}

In this section we explore some flops taking us to phases for which the elliptic fiber is no longer embedded as a toric hypersurface.

As already discussed in section \ref{sect:elemtflopssu2kp1} for the general case of $\mathfrak{su}(2k+1)$, we rewrite \eqref{eq:su7restoric} 
in the following two suggestive forms:
\be
\begin{aligned}\label{eq:speciaformeqsu7}
 y \hat{y} &= \zeta_1 \zeta_2 \zeta_3 P   \\
 x W & = \hat{\zeta}_1 \hat{\zeta}_2 S  \, ,
\end{aligned}
\ee
where in this case 
\be
\ba
 \hat{y} & = y\hat{\zeta}_1  \hat{\zeta}_2  \hat{\zeta}_3 +  b_1 x + b_3 \zeta_0^3 \zeta_1^2 \zeta_2 \hat{\zeta}_1^2 \hat{\zeta}_2 \nn \\
 P     & = x^3 \zeta_2 \zeta_3^2 \hat{\zeta}_2 \hat{\zeta}_3^2 + b_2 x^2 \zeta_0 + b_4 x \zeta_0^4 \zeta_1^2 \zeta_2 \hat{\zeta}_1^2 \hat{\zeta}_2 +
 b_6 \zeta_0^7 \zeta_1^4 \zeta_2^2 \hat{\zeta}_1^4 \hat{\zeta}_2^2 \nn \\
 W     & =  x^2 \zeta_1 \zeta_2^2 \zeta_3^3 \hat{\zeta}_2 \hat{\zeta}_3^2 + b_2 x \zeta_0 \zeta_1 \zeta_2 \zeta_3 + b_4 \zeta_0^4 \zeta_1^3 \zeta_2^2 \zeta_3 \hat{\zeta}_1^2 \hat{\zeta}_2 - b_1 y \nn \\
 S     & = y^2 \hat{\zeta}_3 + b_3 y \zeta_0^3 \zeta_1^2 \zeta_2 \hat{\zeta}_1 -  b_6 \zeta_0^7 \zeta_1^5 \zeta_2^3 \zeta_3 \hat{\zeta}_1^3 \hat{\zeta}_2\,.
\ea
\ee
The form is suggestive of conifold singularities, however with a fine (i.e. using all points) triangulation of the fiber
face spanned by \eqref{eq:coordsxyz} and \eqref{eq:gensu7}, the SR ideal always forbids the loci in question. However, for specific
triangulations we may perform blow-downs after which a conifold singularity (sitting over $b_1=0$ in the fiber) 
indeed arises due to the factorizations of \eqref{eq:speciaformeqsu7}. We can then reach the flopped
phase in the obvious way by performing the other small resolution. 
From \eqref{eq:speciaformeqsu7} it is already clear that we should consider blowdowns which 
allow the coordinates $\zeta_1,\zeta_2,\zeta_3$ to vanish with $y$ or $\hat{\zeta}_1, \hat{\zeta}_2$ with $x$. 
In fact, it follows from \eqref{eq:gensu7} that a cone spanned by $y,\zeta_i$ contains 
$\hat{\zeta}_i$ (for all $i$) and a cone over $x,\hat{\zeta}_i$ contains $\zeta_{i+1}$ (for $i=1,2$), which nicely corresponds 
to the factorizations spelled out in \eqref{eq:speciaformeqsu7}.

As discussed in the main text in section \eqref{sect:elemtflopssu2kp1}, not all the corresponding blowdowns give rise to interesting flops.
In the following, we discuss the interesting cases in some more detail.

\subsubsection{$(y, \zeta_1|\hat{\zeta}_1)$}

\begin{figure}[h!]
     \centering
     \includegraphics[width=4cm]{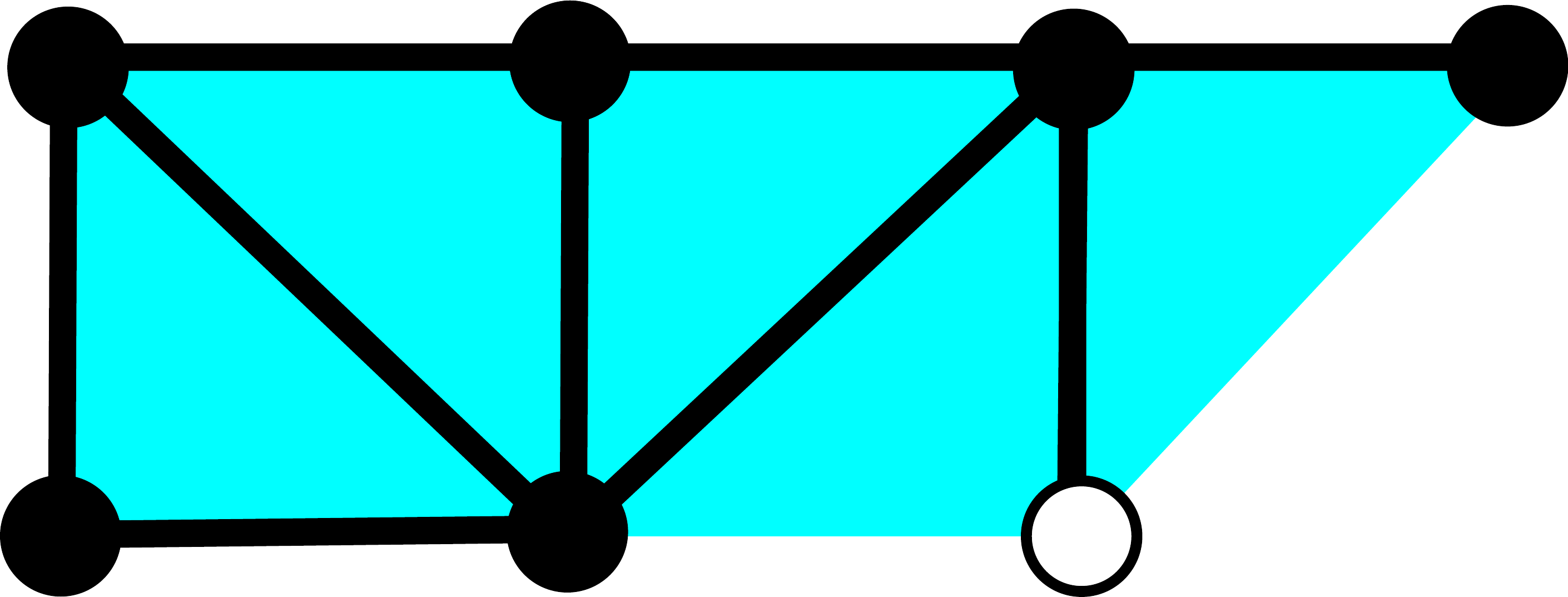} \hspace{2cm}
     \includegraphics[width=4cm]{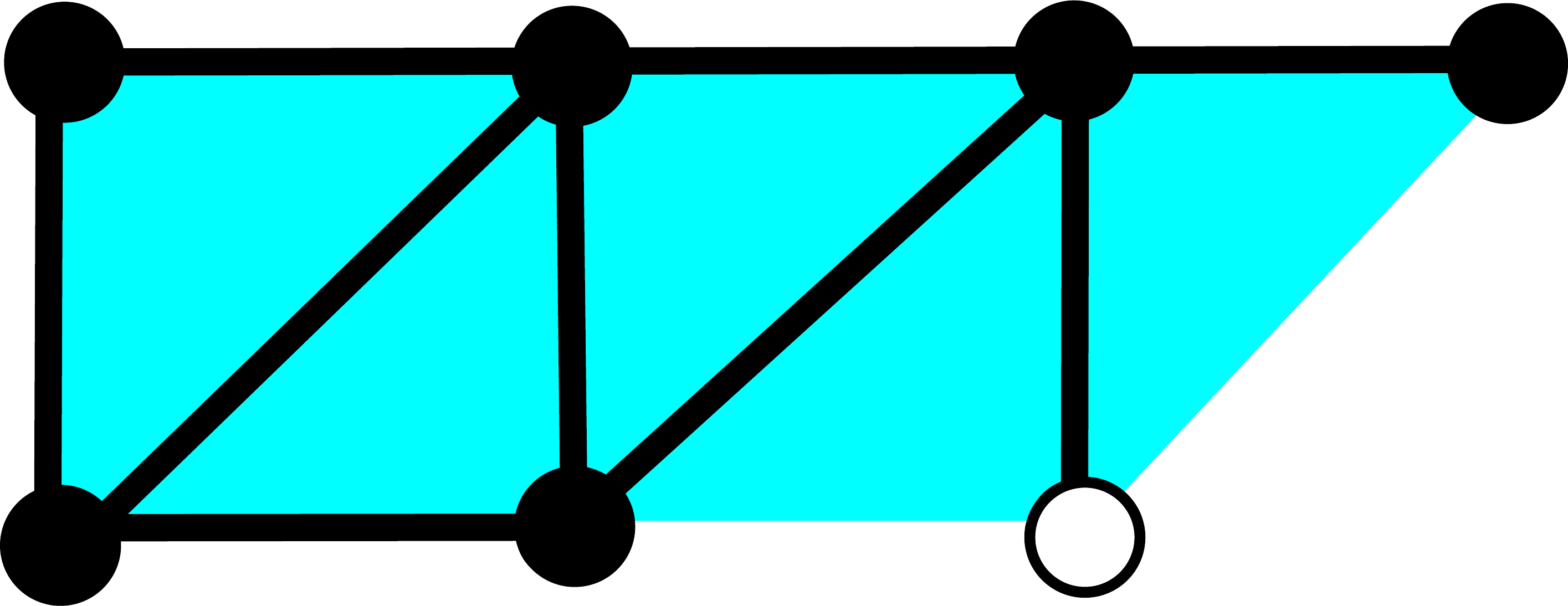}\hspace{2cm}
     \includegraphics[width=4cm]{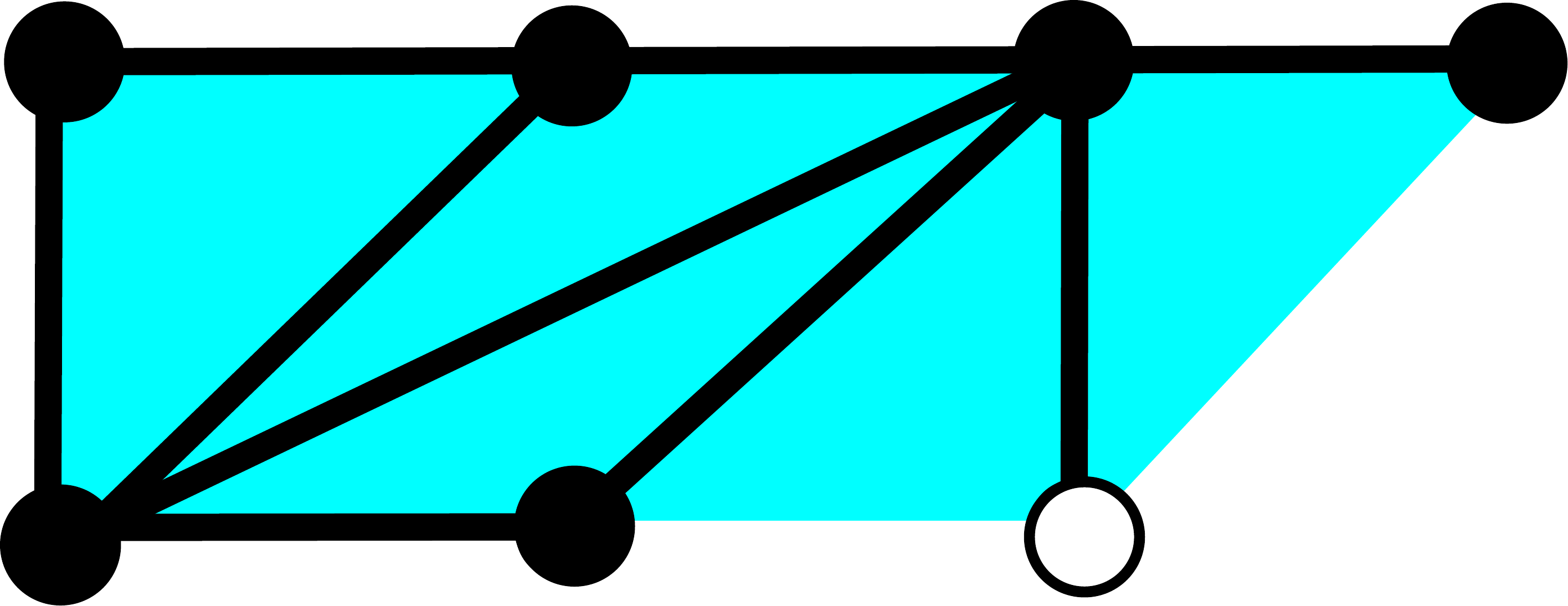}
     \caption{Blowdows of $\hat{\zeta}_1$ for  a resolved $\mathfrak{su}(7)$ model. The points correspond to the same homogeneous coordinates as 
     in figure \ref{fig:ToricFan}. \label{fig:su7bd1}}
 \end{figure}

Let us first consider the blowdowns which result in the fans shown in figure \ref{fig:su7bd1}. 

In both cases, we have fused the cones of the fan such that the ray corresponding to $\hat{\zeta}_1$ is absent. On the level of the toric
ambient space this means that we have blown down the divisor $\hat{\zeta}_1=0$. There is now a shared cone for $y$ and 
$\zeta_1$, so that there is now a conifold singularity on the Calabi-Yau at the locus $y = \zeta_1 = \hat{y} = P = 0$ (which
implies $b_1 = 0$). On the Calabi-Yau, the divisor $\zeta_1 = 0$ becomes reducible and we associate
\be
\begin{aligned}\label{eq:su7a16bd}
D_{\alpha_1}: &\hspace{1cm} \zeta_1 = \hat{y} = 0 \\
D_{\alpha_6}: &\hspace{1cm} \zeta_1 = y = 0 \,.
\end{aligned}
\ee
The reason for this association is that undoing the blowdown again by a blowup $(y,\zeta_1;\hat{\zeta}_1)$, $D_{\alpha_6}$ is mapped to
$\hat{\zeta}_1 = 0 $ and $\zeta_1 = 0$ (which corresponds to $D_{\alpha_1}$) implies $\hat{y} = 0$. Note that in all three cases, the phase
before the blowdown was such that $D_{\alpha_6}$ was splitting in precisely two components over $b_1 = 0$.

On the Calabi-Yau, the singular locus is precisely where $D_{\alpha_1}$ and $D_{\alpha_6}$ meet. At the same time, the curve which was present
at this locus, i.e. at $\zeta_1 = \hat{\zeta}_1 = 0$, is now gone as well. Hence we conclude the we have blown down the component of
$D_{\alpha_6}$ that was shared between $D_{\alpha_1}$ and $D_{\alpha_6}$ over $b_1 = 0$.

We may now reached a flopped phase by performing the second small resolution of the conifold singularity, i.e.
by introducing two new coordinates $\pi$ and $\delta$ satisfying
the equations
\be\ba \label{eq:su7flop1}
 y \hat{y} & = \zeta_1 \zeta_2 \zeta_3 \pi  \cr 
 \delta \pi & = P\,.
\ea\ee
This gives rise to a new smooth space in which $D_{\alpha_1}$ and $D_{\alpha_6}$ are given by intersecting \eqref{eq:su7a16bd} with
\eqref{eq:su7flop1}. Correspondingly, $D_{\alpha_6}$ is now irreducible over $b_1=0$, whereas $D_{\alpha_1}$ receives an extra 
component over this matter curve. Note that all other divisors and curves remain unperturbed under this operation.

\subsubsection{$(x,\hat{\zeta}_2| \zeta_3)$}\label{sect:appsu7xz2bd}

\begin{figure}[h!]
     \centering
     \includegraphics[width=4cm]{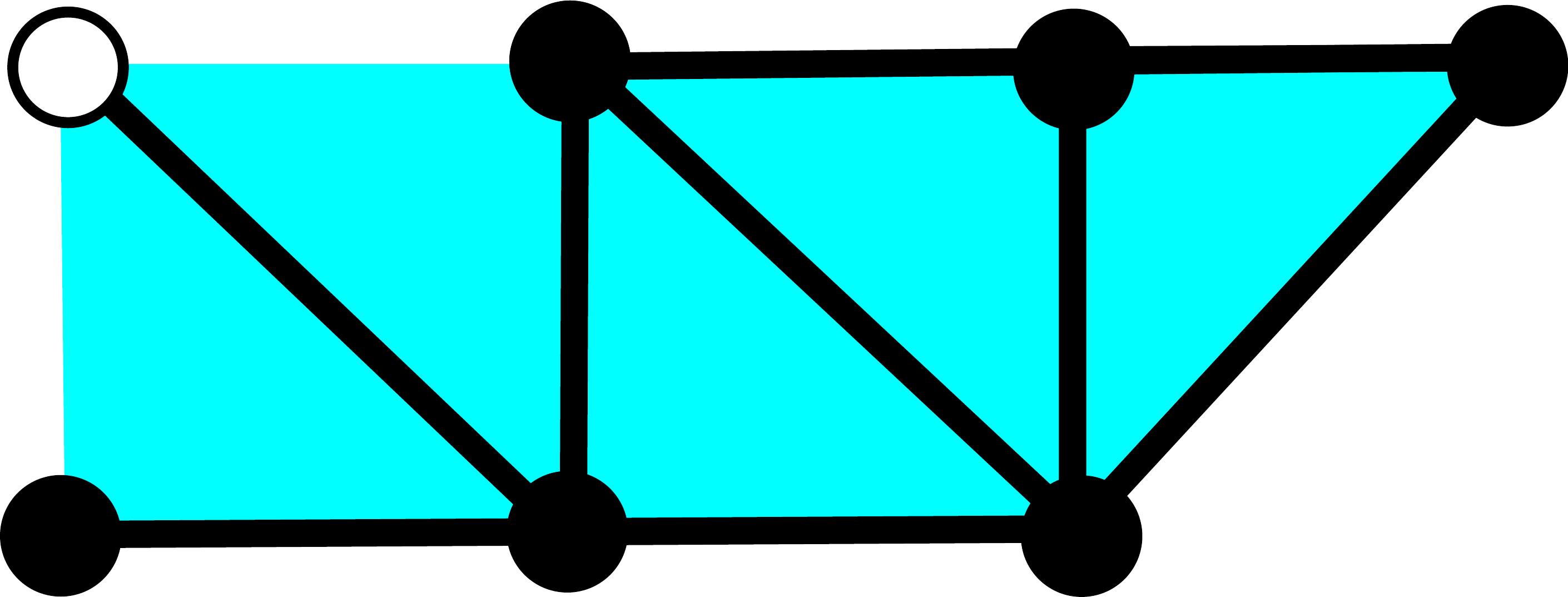} \hspace{2cm}
     \includegraphics[width=4cm]{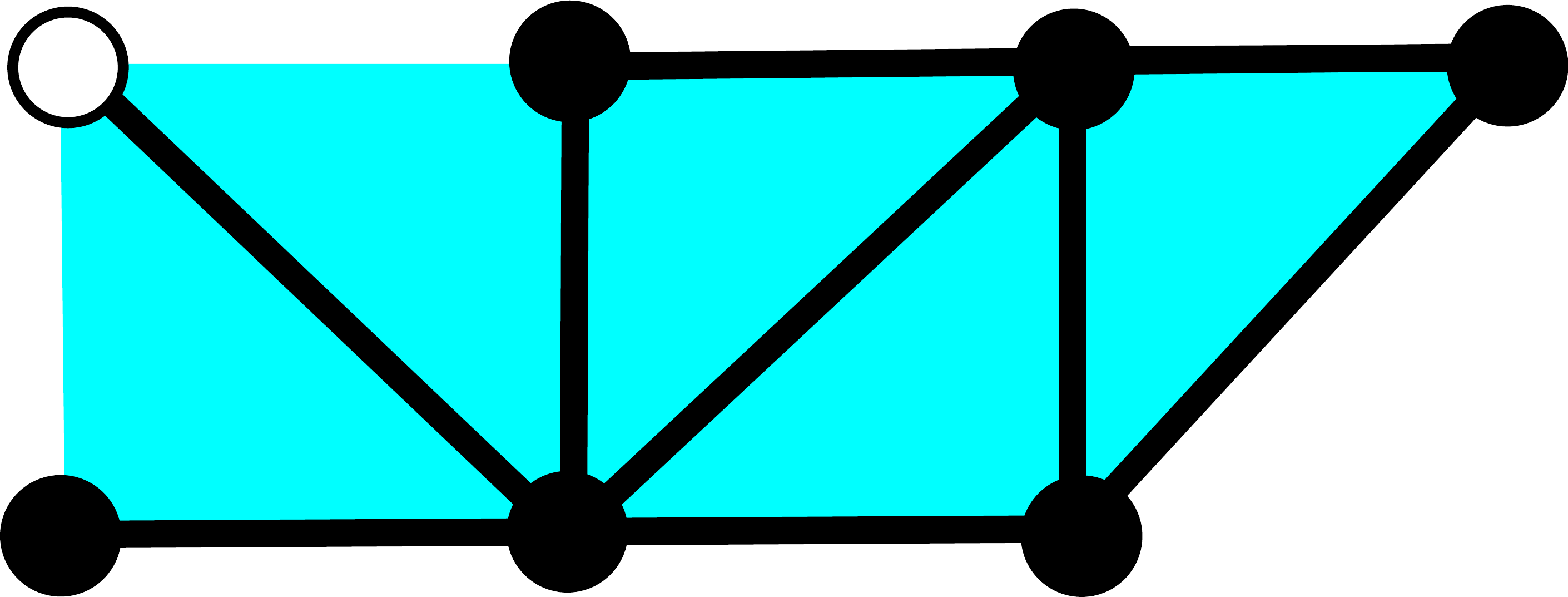}\hspace{2cm}
     \includegraphics[width=4cm]{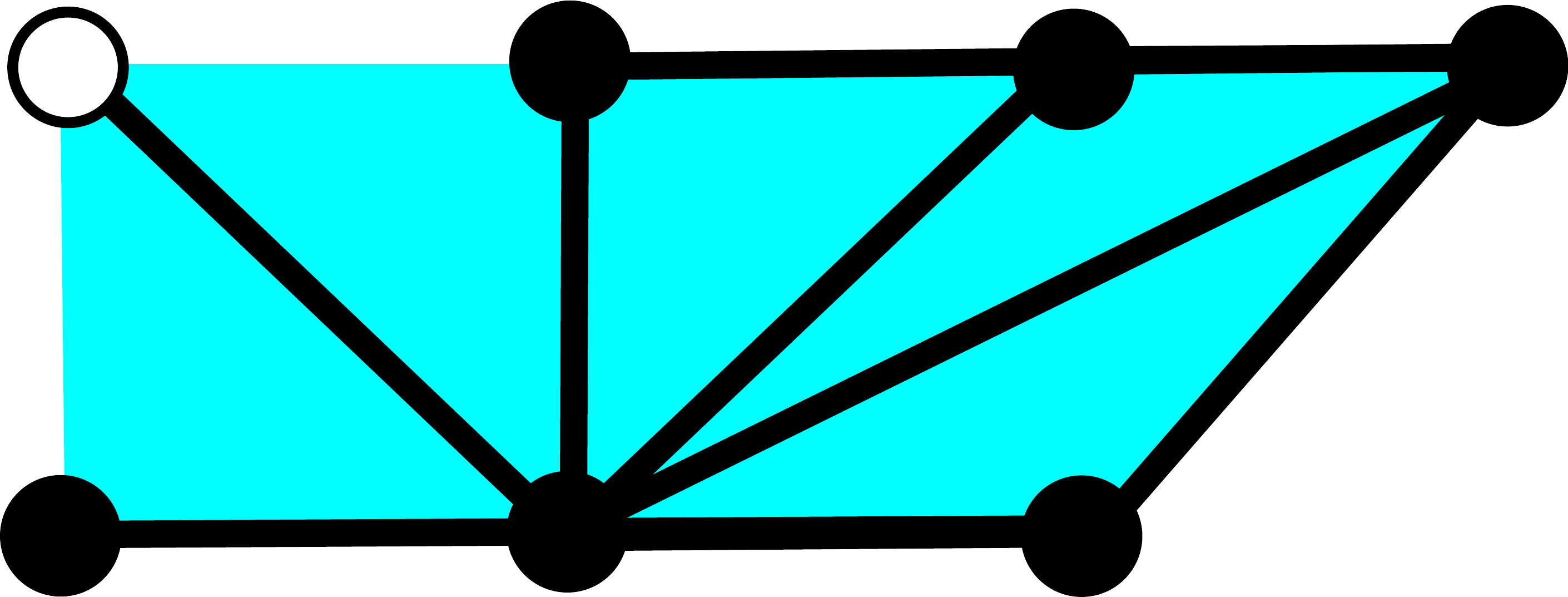}
     \caption{Blowdows of $\zeta_3$ for  a resolved $\mathfrak{su}(7)$ model. The points correspond to the same homogeneous coordinates as
     in figure \ref{fig:ToricFan}.\label{fig:su7bd2}}
 \end{figure}

Similarly, we may blow down $\zeta_3$ reaching the fans shown in figure \ref{fig:su7bd2}.

The blowdown again gives rise to a conifold singularity over the ${\bf 21}$ matter curve and is located at $x=W=\hat{\zeta}_2=S=0$. 
In the blow-down, the divisor $\hat{\zeta}_2$ is reducible and its components are
\be
\begin{aligned}
D_{\alpha_3}:&\hspace{1cm} \hat{\zeta}_2 = x = 0 \cr 
D_{\alpha_5}:&\hspace{1cm}  \hat{\zeta}_2 = W = 0\,.
\end{aligned}
\ee
Again, the conifold singularity is at the locus where these two divisors meet. The triangulation of the corresponding smooth phases before the blowdown
are such that $D_{\alpha_3}$ has precisely two components over $b_1=0$ and the blowdown shrinks the curve which is shared between $D_{\alpha_3}$
and $D_{\alpha_5}$.

A new resolution corresponding to the flopped phase is obtained by setting
\be\ba
 x \omega = \hat{\zeta}_2 S \cr
 \delta \omega = W\,.
\ea\ee
The fiber component $D_{\alpha_3}$ is now at $\hat{\zeta}_2=x=0$, so that it does not split over $b_1=0$ anymore. Similar to
what happened before, $D_{\alpha_5}$ gains another component there, so that their total number stays invariant.

\subsubsection{$(y, \zeta_1|\hat{\zeta}_1)$ and $(x,\hat{\zeta}_2| \zeta_3)$}

\begin{figure}[h!]
     \centering
     \includegraphics[width=4cm]{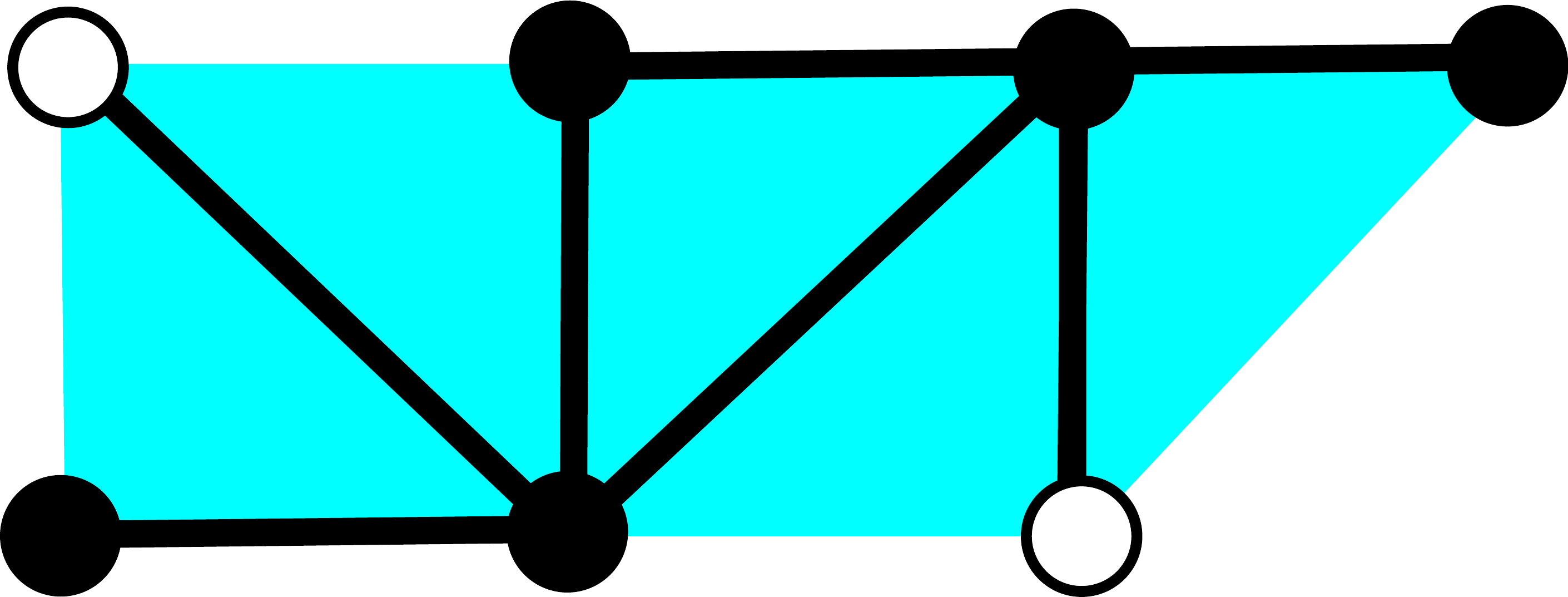} 
     \caption{Blowdown of $\zeta_3$ and $\hat{\zeta}_1$ for  a resolved $\mathfrak{su}(7)$ model. The points correspond to the same homogeneous coordinates as
     in figure \ref{fig:ToricFan}.\label{fig:su7bd3}}
 \end{figure}

As the flops discussed in the last sections were essentially local operations, we can also perform both of 
them independently (if we start with an appropriate smooth model). There is a single fan for which both $\zeta_3$ and
$\hat{\zeta}_1$ can be blown down and flopped. The fan corresponding to the model where both are blown down is shown in figure \ref{fig:su7bd3}.

\subsubsection{Secondary Fiber Face  $\varphi$}

\begin{figure}[h!]
     \centering
     \includegraphics[width=4cm]{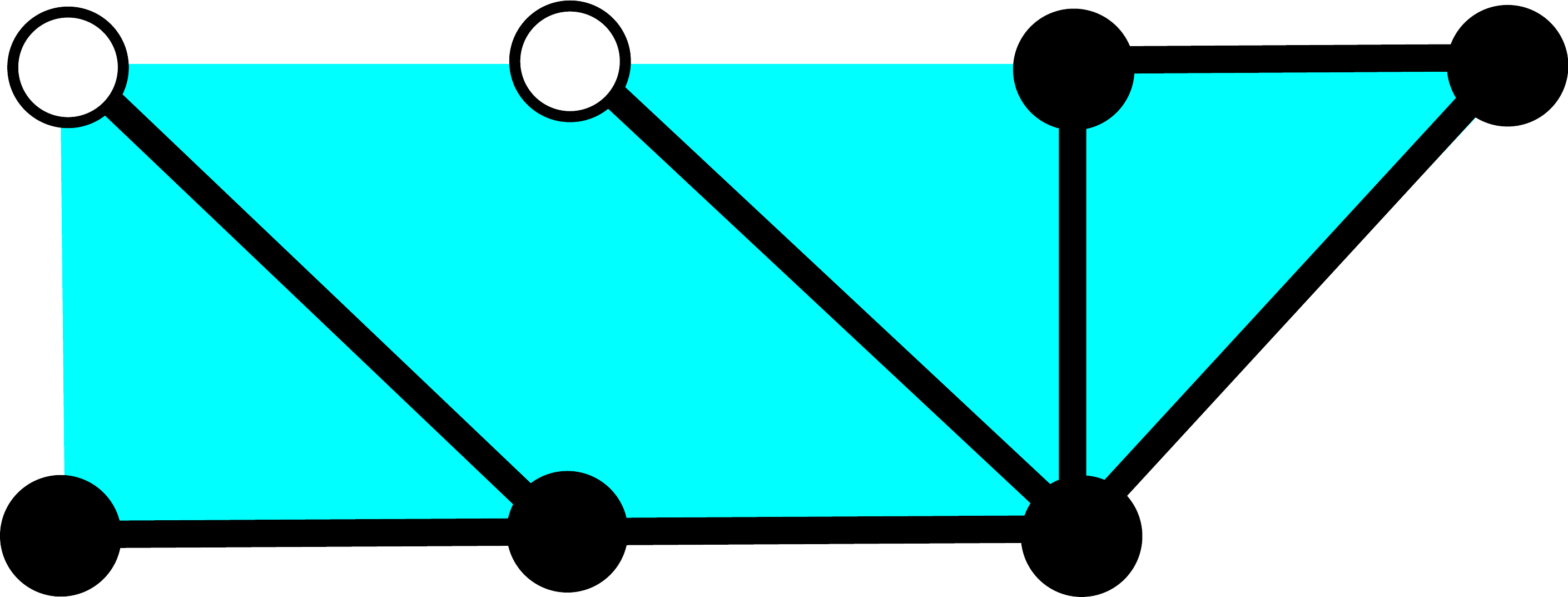} 
     \caption{Blowdown of $\zeta_3$ and $\zeta_2$ for  a resolved $\mathfrak{su}(7)$ model. The points correspond to the same homogeneous coordinates as
     in figure \ref{fig:ToricFan}.\label{fig:su7bd4}}
 \end{figure}

We now describe phases, which can be reached from the partial resolution shown in figure \ref{fig:su7bd4}, which are obtained after blowing down $\zeta_2$ and $\zeta_3$. This can both be constructed
by a sequence of weighted blowups 
\be
(x, y, \zeta_0; \hat{\zeta}_1)\quad
(x, \hat{\zeta}_1, \zeta_0; \zeta_1) \quad 
(x, y, \hat{\zeta}_1; \hat{\zeta}_2)\quad 
(x, y, \hat{\zeta}_2; \hat{\zeta}_3) \,.
\ee
or by subsequently blowing down $\zeta_2$ and $\zeta_3$

Here, both $\hat{\zeta}_1$ and $\hat{\zeta}_2$ are reducible when intersected with the Calabi-Yau and we identify
\be
\begin{aligned}
D_{\alpha_2}:&\hspace{1cm} \hat{\zeta}_1 = x = 0 \cr
D_{\alpha_6}:&\hspace{1cm}  \hat{\zeta}_1 = W = 0 \cr
D_{\alpha_3}:&\hspace{1cm} \hat{\zeta}_2 = x = 0 \cr
D_{\alpha_5}:&\hspace{1cm}  \hat{\zeta}_2 = W = 0 \,.
\end{aligned}
\ee

There are now two\footnote{The point $x = W = \hat{\zeta}_1 = \hat{\zeta}_2 = 0$ is forbidden due to the SR ideal.} 
conifold singularities over $b_1 = 0$, one at $\hat{\zeta}_1 = x = W = S = 0$ and one at $\hat{\zeta}_2 = x = W = S = 0$, which are both apparent from
\eqref{eq:speciaformeqsu7}. In codimension one in the base, our model is already smooth, however.

The different resolutions of this model (besides the ones where we reintroduce the $\zeta_i$) were discussed from a general point of view in section
\ref{sec:CompleteInt}. Instead of repeating the analysis, let us work out the details for $\mathfrak{su}(7)$ more explicitly here.

The most obvious way is to resolve by two blowups, 
\be\label{delta12blowups}
(\hat\zeta_1, \omega; \delta_1)\quad
( \hat\zeta_2, \omega; \delta_2) \,,
\ee
where $\omega$ is the coordinate associated with $W$. Hence that the geometry in question is now described by
\begin{equation}
 \begin{aligned}\label{eq:resdelta12}
  x \omega & = \hat{\zeta}_1 \hat{\zeta}_2 \left( y^2 \hat{\zeta}_3
  + b_3 y \zeta_0^3 \zeta_1^2 \hat{\zeta}_1 \delta_1 -  b_6 \zeta_0^7 \zeta_1^5 \hat{\zeta}_1^3 \hat{\zeta}_2 \delta_1^3 \delta_2 \right) \\
   \delta_1  \delta_2 \omega & = x^2 \zeta_1 \hat{\zeta}_2 \hat{\zeta}_3^2  \delta_2 
   + b_2 x \zeta_0 \zeta_1 + b_4 \zeta_0^4 \zeta_1^3  \hat{\zeta}_1^2 \hat{\zeta}_2  \delta_1^2  \delta_2 - b_1 y \,.
\end{aligned}
\end{equation}
The $\C^*$ actions on the homogeneous coordinates are determined by the weight system
\be
\begin{array}{cccccccccc}
x & y& \zeta_0 & \zeta_1 & \hat{\zeta}_1 & \hat{\zeta}_2 & \hat{\zeta}_3 & \omega & \delta_1 & \delta_2 \cr\hline
1& 2& 1& 0& -1& 0& 0& 2& 0& 0 \cr
1& 1& 1& -1& 0& 0& 0& 1& 0& 0 \cr
1& 1& 0& 0& 1& -1& 0& 1& 0& 0 \cr
1& 1& 0& 0& 0& 1 &-1& 1& 0& 0 \cr
0& 0& 0& 0& 1& 0 &0& 1& -1& 0 \cr
0& 0& 0& 0& 0& 1 &0& 1& 0& -1\cr
\end{array}
\ee
We have chosen a basis of $(\C^*)^6$ reflecting the sequence of blowups that were performed.

Before the two blowups introducing $\delta_1$ and $\delta_2$, the SR ideal of the ambient space can be inferred from the diagram
in figure \ref{fig:su7bd4}. After the blowup, \eqref{delta12blowups} the SR ideal gains the generators
\be
 [ \omega,\hat{\zeta}_1 ]\, , [ \omega,\hat{\zeta}_2 ]\, , [ \delta_2,\hat{\zeta}_1 ] \, .
\ee
Furthermore, any set of coordinates which cannot vanish simultaneously with $\hat{\zeta}_1$ ($\hat{\zeta}_1$) is also forbidden to 
simultaneously vanish with $\delta_1$ ($\delta_2$). In toric language, we may lift the 3-dimensional fan with 
generators \eqref{eq:coordsxyz} and \eqref{eq:gensu7} used above to a four-dimensional fan with the generators
\begin{equation}\label{eq:gen356}
 \ba
v_x = \left(\begin{array}{c}
             -1 \\ 0 \\ 0 \\0  
            \end{array} \right) \, ,
v_y = \left(\begin{array}{c}
             0 \\ -1 \\ 0 \\0  
            \end{array} \right) \, ,            
v_{\zeta_0} = \left(\begin{array}{c}
             2 \\ 3 \\ 1 \\0  
            \end{array} \right) \, ,
v_{\zeta_1} = \left(\begin{array}{c}
             1 \\ 2 \\ 1 \\ {1} 
            \end{array} \right) \, ,
v_{\omega} = \left(\begin{array}{c}
             0 \\ 0 \\ 0 \\ 1  
            \end{array} \right) \, , \\
v_{\hat{\zeta}_1} = \left(\begin{array}{c}
             1 \\ 1 \\ 1 \\ 2  
            \end{array} \right) \, ,
v_{\hat{\zeta}_2} = \left(\begin{array}{c}
             0 \\ 0 \\ 1 \\ 3  
            \end{array} \right) \, ,            
v_{\hat{\zeta}_3} = \left(\begin{array}{c}
             -1 \\ -1 \\ 1 \\ 4  
            \end{array} \right) \, ,
v_{\delta_1} = \left(\begin{array}{c}
             1 \\ 1 \\ 1 \\ 3  
            \end{array} \right) \, ,
v_{\delta_2} = \left(\begin{array}{c}
             0 \\ 0 \\ 1 \\ 4  
            \end{array} \right) \, .            
\ea
\end{equation}

\begin{figure}
    \centering
    \includegraphics[width=15cm]{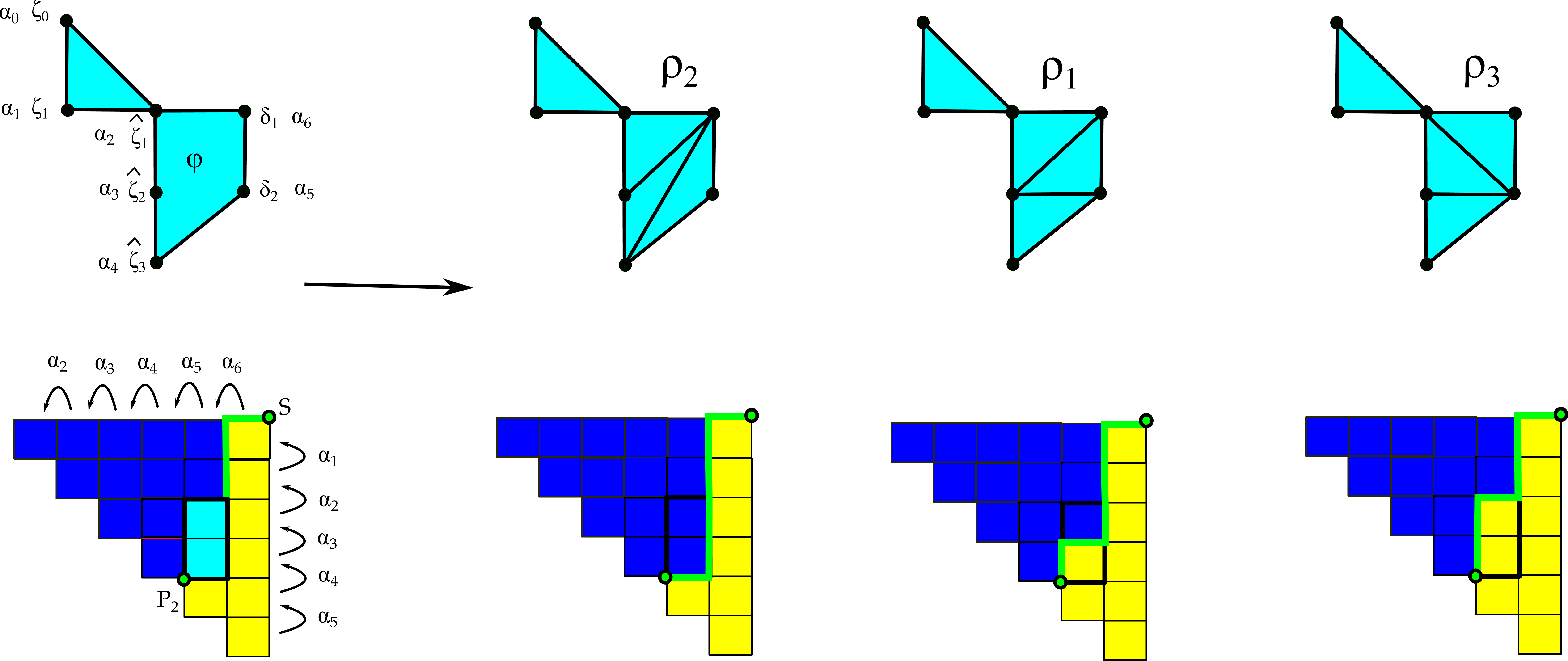}
    \caption{Triangulations and box graphs for the secondary fiber faces of type shown in figure \ref{fig:HatZetaDeltaFF} and \ref{fig:HatZetaDelta}.
    for $\mathfrak{su}(7)$. Each triangulation $\rho_i$ of the secondary fiber face  $\varphi$ corresponds to a sign assignment/colouring of the  turquoise region in the box graph shown in the second line. Note that there are three more colourings of the box graphs, which however correspond to standard toric triangulations of figure \ref{fig:ToricBox}.
   \label{fig:SU7Sigma} }
\end{figure}

The 4-dimensional cones of this fan are spanned by
\begin{equation}\label{eq:cones356}
\begin{aligned}
\langle x \zeta_1 \omega \zeta_0 \rangle,
\langle x y \hat{\zeta}_3 \omega \rangle,
\langle \zeta_1 \hat{\zeta}_1 \zeta_0 \delta_1 \rangle,
\langle \zeta_1 \omega \zeta_0 \delta_1 \rangle,
\langle y \hat{\zeta}_1 \zeta_0 \delta_1 \rangle,
\langle y \omega \zeta_0 \delta_1 \rangle,
\langle x \zeta_1 \hat{\zeta}_1 \delta_1 \rangle,
\langle x \zeta_1 \omega \delta_1 \rangle, 
\langle x \hat{\zeta}_2 \hat{\zeta}_1 \delta_1 \rangle, \\
\langle \hat{\zeta}_2 y \hat{\zeta}_1 \delta_1 \rangle,
\langle x \hat{\zeta}_2 \hat{\zeta}_3 \delta_2 \rangle,
\langle x \hat{\zeta}_3 \omega \delta_2 \rangle,
\langle \hat{\zeta}_2 y \hat{\zeta}_3 \delta_2 \rangle,
\langle y \hat{\zeta}_3 \omega \delta_2 \rangle,
\langle x \hat{\zeta}_2 \delta_1 \delta_2 \rangle,
\langle x \omega \delta_1 \delta_2 \rangle,
\langle \hat{\zeta}_2 y \delta_1 \delta_2 \rangle,
\langle y \omega \delta_1 \delta_2 \rangle\, .
\end{aligned}
\end{equation}
The way these cones fit together can be visualized in the cone diagram shown in figure \ref{fig:SU7Sigma}.
The fiber components over $b_1 = 0$ become:
\be\label{eq:su7deltaphasesdivs}
\begin{array}{c|c|c}
\hbox{Root} & \hbox{Section} & \hbox{over $b_1 = 0$}\cr \hline
{\alpha_0} & \ \zeta_0 &   x \omega  = \hat{\zeta}_1 \hat{\zeta}_2  y^2 \hat{\zeta}_3   \\
&                      &  \delta_1  \delta_2 \omega  = x^2 \zeta_1 \hat{\zeta}_2 \hat{\zeta}_3^2  \delta_2 \\
 \alpha_1 & \ \zeta_1  & x \omega  = \hat{\zeta}_1 \hat{\zeta}_2 y^2 \hat{\zeta}_3  \\
          &            & \delta_1  \delta_2 \omega =  0  \\
 \alpha_2 & \ \hat{\zeta}_1  & x \omega  = 0\\
 &&  \delta_1  \delta_2 \omega  =  b_2 x \zeta_0 \zeta_1 \\
 \alpha_3 & \ \hat{\zeta}_2   & x \omega  = 0  \\
 &&   \delta_1  \delta_2 \omega  = b_2 x \zeta_0 \zeta_1  \\
 \alpha_4 &\  \hat{\zeta}_3  &   x \omega  = \hat{\zeta}_1 \hat{\zeta}_2 \left(
   b_3 y \zeta_0^3 \zeta_1^2 \hat{\zeta}_1 \delta_1 -  b_6 \zeta_0^7 \zeta_1^5 \hat{\zeta}_1^3 \hat{\zeta}_2 \delta_1^3 \delta_2 \right)  \\
  &&    \delta_1  \delta_2 \omega  = b_2 x \zeta_0 \zeta_1 + b_4 \zeta_0^4 \zeta_1^3  \hat{\zeta}_1^2 \hat{\zeta}_2  \delta_1^2  \delta_2 \\
 \alpha_5 &\ \delta_2 &  x \omega  = \hat{\zeta}_1 \hat{\zeta}_2 \left( y^2 \hat{\zeta}_3
  + b_3 y \zeta_0^3 \zeta_1^2 \hat{\zeta}_1 \delta_1  \right) \\
  &&   0 =   b_2 x \zeta_0 \zeta_1 \\
 \alpha_6 &\ \delta_1 &  x \omega  = \hat{\zeta}_1 \hat{\zeta}_2  y^2 \hat{\zeta}_3 \, .\\
  &&   0  = x \zeta_1 \left( \hat{\zeta}_2 \hat{\zeta}_3^2  \delta_2 
   + b_2  \zeta_0 \right)
\end{array}
\ee

Let us now discuss the splitting of the various components in turn. 
\begin{itemize}
\item Even though the expression for $F_0$ can be solved by setting three homogeneous coordinates to zero, all such options are forbidden by the SR ideal.
Correspondingly, $F_0$ stays irreducible.
\item The second equation defining $F_1$ seemingly splits into three components. However, $[\delta_2, \zeta_1]$ is in the SR ideal, as are
$[\zeta_1,\omega,\hat{\zeta}_1]$, $[\zeta_1,\omega,\hat{\zeta}_2]$, $[\zeta_1,\omega,\hat{\zeta}_3]$ and $[\zeta_1,\omega,y]$. Hence $F_1$
sits at $\zeta_1 = \delta_1 = x \omega  - \hat{\zeta}_1 \hat{\zeta}_2 y^2 \hat{\zeta}_3 = 0$ and is irreducible.
\item For $F_2$, it is important that $[\omega, \hat{\zeta}_1]$ and $[\hat{\zeta}_1, \delta_2]$ are in the SR ideal. This forces $x=\delta_1=0$
and makes $F_2$ irreducible over $b_1=0$.
\item The first equation describing the fiber component $F_3$ again forces $x=0$ as $[\omega,\hat{\zeta}_2]$ is in the SR ideal.  Over $b_1=0$, 
this leaves the two components at $\delta_1=0$ and $\delta_2=0$.
\item The component $F_4$ stays irreducible over $b_1=0$ as $[\omega,\hat{\zeta}_1]$ and $[\omega,\hat{\zeta}_2]$ are in the SR ideal.
\item $F_5$ splits into the two components $x=\hat{\zeta_2} = 0$ and  $x= y^2 \hat{\zeta}_3
  + b_3 y \zeta_0^3 \zeta_1^2 \hat{\zeta}_1 \delta_1  = 0$ 
\item $F_6$ splits into the four components $\zeta_1=0$, $ \hat{\zeta}_2 \hat{\zeta}_3^2  \delta_2 
   + b_2  \zeta_0 = 0$ as well as $x = \hat{\zeta}_1 = 0$, $ x = \hat{\zeta}_2 = 0$.
\end{itemize}
In summary, we hence get five extra components over $b_1=0$, as expected, and the  splitting can be summarized as
\begin{equation}
 \begin{aligned}
  F_3 & \rightarrow C_a + C_b \\
  F_5 & \rightarrow C_a + C_c \\
  F_6 & \rightarrow F_1 + F_2 +  C_b + C_d \,.
 \end{aligned}
\end{equation}
The corresponding fiber face and box graph is the case $\rho_1$ shown in figure \ref{fig:SU7Sigma}.

From the general discussion one expects that there should be two more phases described by \eqref{eq:resdelta12}, for which the face containing the
coordinates $\delta_1, \delta_2$ and $\hat{\zeta}_1, \hat{\zeta}_2, \hat{\zeta}_3$ is triangulated differently. These remaining triangulations and box graphs $\rho_2$ and $\rho_3$ are shown in figure \ref{fig:SU7Sigma}.

The case $\rho_2$ is obtained by performing a flop on the resolution associated to $\rho_1$. 
From the box graph, this corresponds to sign-changing the weight $L_{4}+ L_6$, giving rise to the box graph $\rho_2$ in figure  \ref{fig:SU7Sigma}. 
In particular this means that the curve 
$C_a$, which is at $\hat{\zeta}_2 = \delta_2 =0$ (from which $x=0$ follows), and carries weight $L_4 + L_5$, ceases to be extremal. A glance at the corresponding fiber face, 
shows that this curve corresponds to the line connecting $\delta_2$ with $\hat{\zeta}_2$. We can perform a flop in which this 1-simplex is 
replaced by a one-simplex connecting $\delta_1$ and $\hat{\zeta}_3$, as shown in figure \ref{fig:SU7Sigma}, where the fiber face $\rho$ is triangulated now as in $\rho_2$.
The ambient space of the flopped phase hence has the same generators 
\eqref{eq:gen356}, but the 4-dimensional cones are now
\begin{equation}
\begin{aligned}
 \langle x \zeta_1 \omega \zeta_0 \rangle,
\langle x y \hat{\zeta}_3 \omega \rangle,
\langle \zeta_1 \hat{\zeta}_1 \zeta_0 \delta_1 \rangle,
\langle \zeta_1 \omega \zeta_0 \delta_1 \rangle,
\langle y \hat{\zeta}_1 \zeta_0 \delta_1 \rangle,
\langle y \omega \zeta_0 \delta_1 \rangle,
\langle x \zeta_1 \hat{\zeta}_1 \delta_1 \rangle,
\langle x \zeta_1 \omega \delta_1 \rangle,
\langle x \hat{\zeta}_2 \hat{\zeta}_1 \delta_1 \rangle, \\
\langle \hat{\zeta}_2 y \hat{\zeta}_1 \delta_1 \rangle,
\langle x \hat{\zeta}_2 \hat{\zeta}_3 \delta_1 \rangle,
\langle x \hat{\zeta}_3 \omega \delta_2 \rangle,
\langle \hat{\zeta}_2 y \hat{\zeta}_3 \delta_1 \rangle,
\langle y \hat{\zeta}_3 \omega \delta_2 \rangle,
\langle x \hat{\zeta}_3 \delta_1 \delta_2 \rangle,
\langle x \omega \delta_1 \delta_2 \rangle,
\langle y \hat{\zeta}_3 \delta_1 \delta_2 \rangle,
\langle y \omega \delta_1 \delta_2 \rangle\, .
\end{aligned}
\end{equation}
This means that we have replaced $\langle \hat{\zeta}_2, \hat{\zeta}_3, \delta_2 \rangle\, , \langle \hat{\zeta}_2, \delta_1, \delta_2 \rangle$
by  $\langle \hat{\zeta}_2, \hat{\zeta}_3, \delta_1 \rangle\, , \langle \hat{\zeta}_3 \delta_1 \delta_2 \rangle$. 

The elliptic fibration and the fiber components are of course still given by the same equations \eqref{eq:resdelta12}. When we discuss the splitting
over the locus $b_1=0$, however, the components $F_3$ and $F_5$ are still irreducible as now $[\delta_2,\hat{\zeta}_2]$ is in the SR ideal.

The components $F_4$ and $F_6$ now each have an extra component at $\hat{\zeta}_3=\delta_1=0$. The splittings are hence
\begin{equation}\label{PreFinalPhase}
 \begin{aligned}
 F_4 & \rightarrow C_e + C_f \\
 F_6 & \rightarrow F_1 + F_2 + F_3 + C_e + C_d\,,
 \end{aligned}
\end{equation}
which precisely corresponds to the fiber face  $\rho_2$ and associated box graph in figure \ref{fig:SU7Sigma}.

Finally, to describe the resolution $\rho_3$ in figure \ref{fig:SU7Sigma}, consider again $\rho_1$ and flop the 
1-simplex connecting $\delta_1$ with $\hat{\zeta}_2$ by replacing it with a 1-simplex connecting $\delta_2$ with $\hat{\zeta}_2$. Not surprisingly, in this flop the curve shared by $D_{\alpha_3}$ and $D_{\alpha_6}$
is blown down and a new curve, now shared between $D_{\alpha_2}$ and $D_{\alpha_5}$ emerges. Hence we expect this geometry to be identical  to the one described in section \ref{sect:appsu7xz2bd} by the diagram on the left of figure \ref{fig:su7bd2}.

\subsubsection{Flop to Ternary Fiber Face}
\label{sec:FinFlop}

The box graph, which in terms of our nomenclature of fiber faces in the main text, see section \ref{sec:Layers} corresponds to the ternary fiber face (which for $\mathfrak{su}(7)$ is the final layer) is given by
\be
\includegraphics[width=3cm]{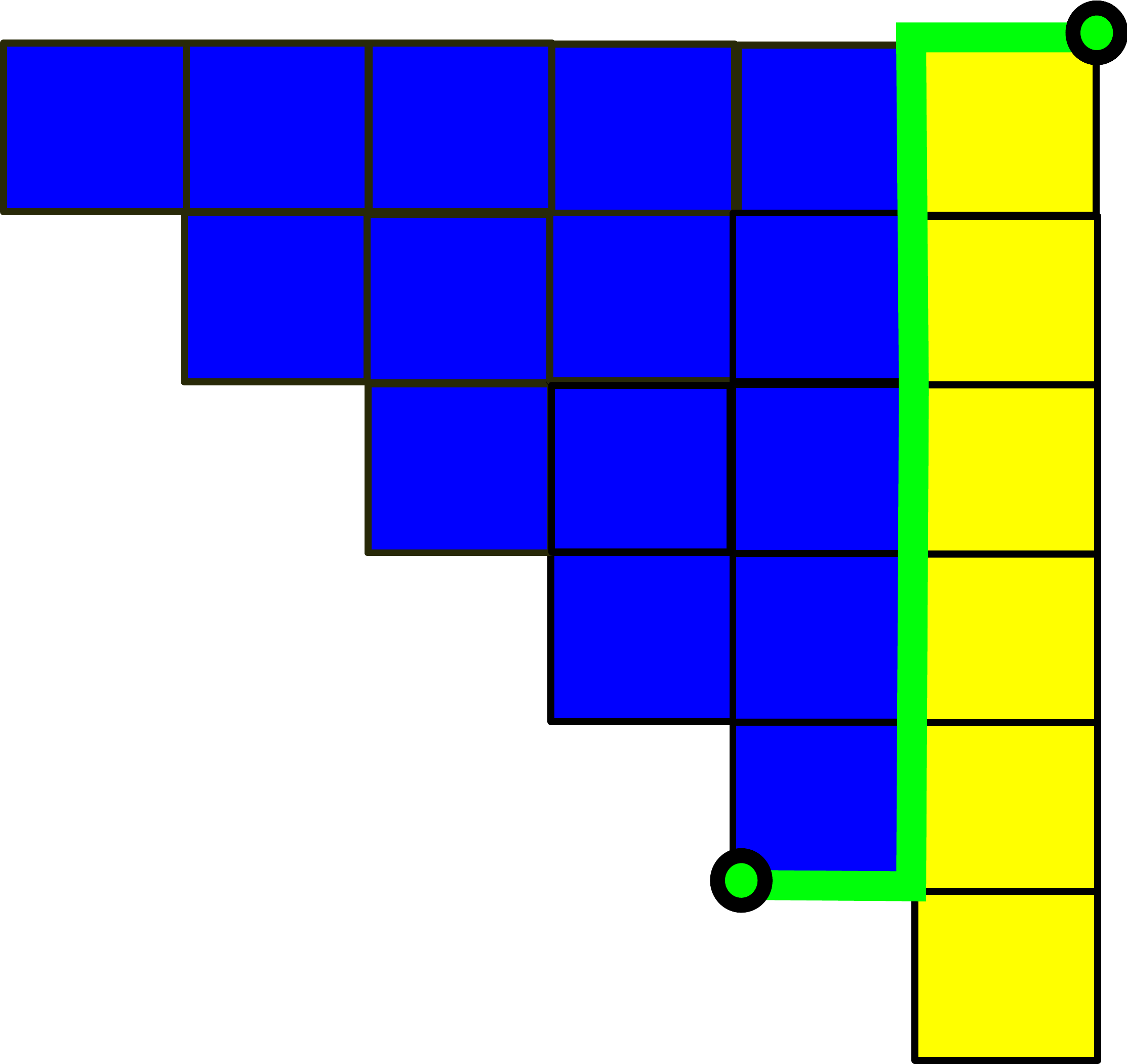}
\ee
Like in section \ref{sec:ThirdLayer}, we can determine this phase   by a flop from the phase associated to $\rho_2$ \ref{sec:CompleteInt}.

Consider  the equation of the phase (\ref{PreFinalPhase}) corresponding to the secondary fiber face triangulation $\rho_2$
\be
\ba
\hat\zeta_1 \hat\zeta _{2} \left(-b_6 \delta _1^3 \delta _2 \zeta_1^5 \zeta _0^7 \hat\zeta _1^3
   \hat\zeta _{2}+b_3 \delta _1 \zeta _1^2 \zeta _0^3 y \hat\zeta_1+y^2 \hat\zeta _{3}\right)-x
   \omega &=0 \cr
   b_4 \delta _1^2 \delta _2 \zeta _1^3 \zeta _0^4 \hat\zeta _1^2 \hat\zeta _{2}+b_2 \zeta _1 \zeta
   _0 x-b_1 y-\delta _1 \delta _2 \omega +\delta _2 \zeta _1 x^2 \hat\zeta _{2} \hat\zeta _{3}^2 &=0 \,.
\ea
\ee
with the projective relations 
\be
\ba
\left\{\zeta _0,\delta _2 \hat\zeta _{2} \hat\zeta _{3}\right\},
\left\{y,\zeta_1\right\},
\left\{\zeta _1,\delta _2 \hat\zeta _{2} \hat\zeta _{3}\right\},
\left\{\omega   ,\hat\zeta _1 \hat\zeta _{2}\right\},
\left\{\hat\zeta _1,\delta _2 \hat \zeta _{3 }\right\},\cr 
\left\{\hat\zeta _{2},\delta _2\right\},
\left\{x,\zeta _0,\delta _1 \hat\zeta_1\right\},
\left\{x,y,\delta _1 \delta _2 \zeta _0 \hat\zeta _1 \hat\zeta _{2}\right\},\left\{\omega ,\hat \zeta _{3 },\delta _1\right\}\,.
\ea\ee
As $\omega=0$ intersects the exceptional curves in points, and is not going to play any role in the flop.

We thus assume that $
\omega\not=0$ and  solve the first equation for $x$ and insert it into the second equation, 
so that the geometry is now the hypersurface
\be\ba\label{NoPatchEq}
0&=b_2 \zeta _0 \zeta _1 \omega 
 \hat\zeta_1 \hat\zeta_2 \left(-b_6 \delta _1^3 \delta _2 \zeta
   _1^5 \zeta _0^7 \hat\zeta_1^3 \hat\zeta_2+b_3 \delta _1 \zeta _1^2 \zeta _0^3 y \hat\zeta_1+y^2 \hat\zeta_3\right)\cr   
&+   \delta _2 \zeta _1 \hat\zeta_1^2 \hat\zeta_2^3 \hat\zeta_3^2
   \left(-b_6 \delta _1^3 \delta _2 \zeta _1^5 \zeta _0^7 \hat\zeta_1^3 \hat\zeta_2+b_3 \delta
   _1 \zeta _1^2 \zeta _0^3 y \hat\zeta_1+y^2 \hat\zeta_3\right)^2+\omega ^2 (b_4
   \delta _1^2 \delta _2 \zeta _0^4 \zeta _1^3 \hat\zeta_1^2 \hat\zeta_2-b_1 y -\delta
   _1 \delta _2 \omega)\,.
   \ea\ee
One can easily check that the curve that has to be flopped to reach the final phase is given by
\be
\hat\zeta_3 = b_1=0 \, :\qquad \delta_1 s_3 =0\,,
\ee
where $\delta_1=0$ is the component that needs to be retained, and $s_3$ is defined in (\ref{s1s2s3s4}). We will now rewrite the equations in new coordinates, 
and show explicitly how a conifold equation emerges. This is very similar to the flops in \cite{Hayashi:2013lra}. Define the new coordinates
\be\label{s1s2s3s4}
\ba
s_1 &=  \hat\zeta_2\cr 
s_2 &= \delta_1 s_1 \cr
s_3 &= -b_2 b_6 \delta _1^2 \delta _2 \zeta _1^6 \zeta _0^8 \omega  \hat\zeta_1^4 \hat\zeta_2^2+b_6^2
   \delta _1^5 \delta _2^3 \zeta _1^{11} \zeta _0^{14} \hat\zeta_1^8 \hat\zeta_2^5 \hat\zeta _{3}^2+\omega  \left(\delta _2 \omega  \left(b_4 \delta _1 \zeta _0^4 \zeta _1^3 \hat\zeta_1^2
   \hat\zeta_2-\omega \right)+b_2 b_3 \zeta _1^3 \zeta _0^4 y \hat\zeta_1^2 \hat\zeta _{2}\right) \cr 
   s_4 & = \delta_1 s_3 \,.
\ea
\ee
The  equation (\ref{NoPatchEq}) can then be rewritten in terms of solely these coordinates (in particular no explicit dependence on $\delta_1$)
\be\ba
s_4= &   -b_2
   \zeta _0 \zeta _1 s_1 y^2 \omega  \hat\zeta_1 \hat\zeta_3+b_1 y  \omega ^2 \cr
 &+  y \delta _2 \zeta _1 \hat\zeta_1^2 \hat\zeta_3^2 \left(b_3 \zeta _1^2 \zeta _0^3 s_2
   \hat\zeta_1+s_1 y \hat\zeta_3\right) \left(2 b_6 \delta _2 \zeta _1^5 \zeta _0^7 s_2^3
   \hat\zeta_1^3-b_3 \zeta _1^2 \zeta _0^3 s_1 s_2 y \hat\zeta_1-s_1^2 y^2 \hat\zeta_3\right) \,,
\ea
\ee
with the extra condition, rewritten in terms of $s_1$ and $s_2$,
\be
s_3 = b_6^2 \delta _2^3 \zeta _1^{11} \zeta _0^{14} s_2^5 \hat\zeta_1^8 \hat\zeta_3^2
-b_2 b_6 \delta_2 \zeta _1^6 \zeta _0^8 s_2^2 \hat\zeta_1^4\omega
+b_4 \delta _2 \zeta _1^3 \zeta _0^4 s_2 \hat\zeta_1^2\omega^2
+b_2 b_3 \zeta _1^3 \zeta _0^4 s_1 y \hat\zeta_1^2\omega
-\delta _2 \omega^3\,,
\ee
and this is equivalent to the initial equation in the patch by imposing the condition, which makes manifest the relation among the new coordinates
\be
s_1 s_4 = s_2  s_3 \,,
\ee
which is exactly the conifold equation. 
Note that $\delta_1$ does not appear in these equations any longer. 

We can think of this equation as the resolution of the conifold with 
\be
s_1 \rho_1 = s_2 \rho_2 \,,\qquad s_3 \rho_1 = s_4 \rho_2 \,,
\ee
where $[\rho_1, \rho_2]$ are projective coordinates on a $\mathbb{P}^1$ and we considered the patch $\rho_1\not=0$ where 
$\delta_1 = \rho_2/\rho_1$. 

The flop of the conifold is now obtained by
\be
s_1 \xi_1 = s_3 \xi_2 \,,\qquad 
s_4 \xi_2 = s_2 \xi_1 \,.
\ee
Consider the patch $\xi_1\not=0$ and introduce $\xi = \xi_2 / \xi_1$, and replacing $s_1$ and $s_2$ accordingly yields
\be
\ba
s_4&= -y \delta _2 \zeta _1 \xi ^3 \hat\zeta_1^2 \hat\zeta_3^2 \left(b_3 \zeta _1^2 \zeta _0^3
   s_4 \hat\zeta_1+s_3 y \hat\zeta_3\right) \left(s_3 y \left(b_3 \zeta _1^2 \zeta _0^3 s_4
   \hat\zeta_1+s_3 y \hat\zeta_3\right)-2 b_6 \delta _2 \zeta _0^7 \zeta _1^5 \xi  s_4^3 \hat\zeta_1^3\right) \cr 
&   +b_2 \zeta _0 \zeta _1 \xi  s_3 y^2 \omega  \hat\zeta_1 \hat\zeta_3-b_1y  \omega
   ^2
\cr 
s_3 &= 
b_6^2 \delta _2^3 \zeta _1^{11} \zeta _0^{14} \xi ^5 s_4^5 \hat\zeta_1^8 \hat\zeta_3^2
-b_2 b_6   \delta _2 \zeta _1^6 \zeta _0^8 \xi ^2 s_4^2 \hat\zeta_1^4\omega
+b_4 \delta _2 \zeta _1^3 \zeta_0^4 \xi  s_4 \hat\zeta_1^2\omega^2
+b_2 b_3 \zeta _1^3 \zeta _0^4 \xi  s_3 y \hat\zeta_1^2\omega
-\delta_2 \omega^3 \,.
\ea
\ee
Let us consider the various components along $b_1=0$ -- we have used the SR ideal to eliminate coordinates that cannot vanish at the same time: 
\be
\ba
\zeta_1=0:\qquad & s_4= \delta_2 + s_3=0 \cr 
\hat\zeta_1=0:\qquad & s_4 = \delta_2 + s_3=0\cr 
\xi=0: \qquad & s_4 = \delta_2+s_3 =0 \cr 
s_3=0 :\qquad &s_4 \left(2 b_3 b_6 \delta _2^2 \zeta _0^{10} \zeta _1^8 \xi ^4 s_4^3 y \hat\zeta_1^6
   \hat\zeta_3^2-1\right)=0 \cr 
   &\delta _2 \left(b_6^2 \delta _2^2 \zeta _1^{11} \zeta _0^{14} \xi
   ^5 s_4^5 \hat\zeta_1^8 \hat\zeta_3^2-b_2 b_6 \zeta _1^6 \zeta _0^8 \xi ^2 s_4^2 \hat\zeta_1^4+b_4 \zeta _1^3 \zeta _0^4 \xi  s_4 \hat\zeta_1^2-1\right) =0
\cr
\hat\zeta_3=0:\qquad &  s_4= s_3 \left(b_2 b_3 \zeta _0^4 \zeta _1^3 \xi  y \hat\zeta_1^2-1\right)-\delta _2 =0\cr
\delta_2=0 :\qquad & b_2 \zeta _0 \zeta _1 \xi  s_3 \omega  y^2 \hat\zeta_1 \hat\zeta_3+s_4=0 \cr 
&s_3 \left(b_2 b_3 \zeta _0^4 \zeta _1^3 \xi  y \hat\zeta_1^2-1\right) =0\cr
s_4=0 :\qquad & 
\zeta _1 \xi  s_3 \hat\zeta_1 \hat\zeta_3 \left(b_2 \zeta _0 \omega +\xi ^2 s_3^3 \hat\zeta_1 \hat\zeta_3^3 \left(b_2 b_3 \zeta _0^4 \zeta _1^3 \xi  \hat\zeta_1^2-1\right)\right) =0 \,.
\ea
\ee
Note that $s_3=\delta_2=0$, which naively looks like an additional component, is in fact not allowed because
it implies $\hat{\zeta}_2=0$ from the definition of $s_3$, which is however not consistent with the projective relations. 
All components are irreducible except $s_4=0$, which corresponds to $F_6$. It can be traced back through the flop to $\delta_1=0$ and 
splits into six components,  which is exactly as required from the final phase. 
This completes the correspondence between the geometric realizations of resolutions and box graphs.




\providecommand{\href}[2]{#2}\begingroup\raggedright\endgroup


\end{document}